\documentclass[useAMS,usenatbib]{mn2e}
\usepackage[dvips]{graphicx}
\usepackage{amsmath}
\usepackage{amsfonts}
\usepackage{amssymb}

\title[An Analytic Model for Rotational Modulations]{An Analytic Model for Rotational Modulations\\
in the Photometry of Spotted Stars}
\author[David M. Kipping]{David M. Kipping$^{1,2}$\thanks{E-mail:
dkipping@cfa.harvard.edu}\footnotemark[1]\\
$^{1}$Harvard-Smithsonian Center for Astrophysics, 60, Garden Street, Cambridge, MA 02138 \\
$^{2}$Carl Sagan Fellow}
\begin{document}

\date{Accepted 2012 September 12. Received 2012 September 12; in original form 2012 August 6}

\pagerange{\pageref{firstpage}--\pageref{lastpage}} \pubyear{2012}

\maketitle

\newcommand{\luna}{{\tt LUNA}}
\newcommand{\multi}{{\sc MultiNest}}
\newcommand{\macula}{{\tt macula}}

\label{firstpage}

\begin{abstract}

Photometric rotational modulations due to starspots remain the most common and 
accessible way to study stellar activity. In the \emph{Kepler}-era, there now
exists precise, continuous photometry of $\sim$150,000 stars presenting an
unprecedented opportunity for statistical analyses of these modulations.
Modelling rotational modulations allows one to invert the observations into
several basic parameters, such as the rotation period, spot coverage, stellar
inclination and differential rotation rate. The most widely used analytic model
for this inversion comes from \citet{budding:1977} and \citet{dorren:1987},
who considered circular, grey starspots for a linearly limb darkened star. In
this work, we extend the model to be more suitable in the analysis of high
precision photometry, such as that by \emph{Kepler}. Our new freely available
Fortran code, \macula, provides several improvements, such as non-linear limb 
darkening of the star and spot, a single-domain analytic function, partial 
derivatives for all input parameters, temporal partial derivatives, diluted 
light compensation, instrumental offset normalisations, differential rotation, 
starspot evolution and predictions of transit depth variations due to unocculted 
spots. Through numerical testing, we find that the inclusion of non-linear limb 
darkening means \macula\ has a maximum photometric error an order-of-magnitude 
less than that of \citet{dorren:1987}, for Sun-like stars observed in the 
\emph{Kepler}-bandpass. The code executes three orders-of-magnitude faster than 
comparable numerical codes making it well-suited for inference problems.

\end{abstract}


\section{Introduction}
\label{sec:intro}

%%% STELLAR ACTIVITY INTRO
\subsection{Stellar Activity}
\label{sub:stellaractivity}

A variety of cool stars with external convection envelopes have been observed
to exhibit magnetic activity similar to that of the Sun (e.g. 
\citealt{kron:1947, mullan:1974, vogt:1975}). These magnetic fields,
generated by cyclonic turbulence in the outer convection zone, penetrate the
stellar atmosphere forming starspots, plages, networks, etc 
\citep{berdyugina:2005}. The study of these manifestations on other stars allows 
for crucial tests of stellar dynamo theory. For example, \citet{skumanich:1972} 
first suggested that rotation plays a key role in generating stellar activity.

Since the discovery of rotationally modulated brightness variations due to
starspots, photometry remains the most common technique for studying stellar
activity. In particular, space-based photometric instruments have 
provided many high-cadence, precise light curves (e.g. HIPPARCOS; 
\citealt{leeuwen:1997}). Recently, the detection of transiting extrasolar 
planets \citep{charbonneau:2000,henry:2000} has led to a surge in the design and 
construction of precise photometric instruments. Notably, the \emph{Kepler 
Mission} has detected 2165 eclipsing binaries \citep{slawson:2011} and 2321 
planetary candidates \citep{batalha:2012} with nearly continuous photometry at a 
precision of $\sim50$\,ppm (for $V\simeq12$) per long-cadence exposure 
(29.4244\,minutes). Preliminary analysis of the \emph{Kepler} target stars (over
150,000) has revealed that those which exhibit periodic modulation generally 
have a much higher amplitude of variability \citep{basri:2011}. One of the 
legacy products of the \emph{Kepler Mission} will be a vast database of 
precise continuous photometry and the effective exploitation of this database
will surely lead to deep insights into stellar activity.

%%% STARSPOTS
\subsection{Starspots}
\label{sub:starspots}

Starspots are a very common source of photometric variability and have a diverse 
value to astronomers, varying from friend to foe. The presence of dark starspots 
on the surface of a rotating star induces periodic photometric variability due 
to the stellar rotation. An analysis of these rotational modulations allows for 
a determination of several basic properties of the star. The most accessible of 
these properties is the rotation period, which can often be inferred using a 
simple Lomb-Scargle periodogram or autocovariance analysis, and has several 
astrophysical uses. For example, the rotation period may be used with 
gyrochronology to estimate the age of the star \citep{barnes:2009}. Another
example is demonstrated in the recent work of \citet{hirano:2012} who show
how a spectroscopic $V\sin I_*$, an estimate of the stellar radius ($R_*$) and 
the rotation period allows one to infer the stellar inclination angle, 
$I_*$.

Employing spot-modelling codes allows for more information than just the 
rotation period to be derived. For example, \citet{walker:2007} used
rotational modulations alone to infer the stellar inclination angle for
$\kappa^1$ Ceti. Here the authors also showed how their measurement could be 
used to predict $V\sin I_*$ and verified their solution was consistent with a
spectroscopic determination. Further more, the authors were also able to
estimate the differential rotation rate of $\kappa^1$ Ceti, which was found to
be reasonably close to Solar.

If a star hosts a transiting planet which passes over a dark starspot, the 
transit light curve appears to increase over the duration of the spot crossing 
event (e.g. \citealt{rabus:2009}). Detecting the same spot-crossing event in two 
consecutive transits, which will have migrated along in longitude, allows the 
observer to infer a nearly coplanar spin-orbit angle. This technique has so far 
been successfully applied to several cases including CoRoT-2b 
\citep{nutzman:2011}, WASP-4b \citep{sanchis:2011} and Kepler-17b 
\citep{desert:2011}. In the case of WASP-4b, the result was verified using
the more traditional spectroscopic technique known as the Rossiter-McLaughlin 
effect \citep{rossiter:1924,mclaughlin:1924}. Modelling spot-crossing events
remains outside of the scope of this work, but exploiting such phenomena
is greatly aided by including information encoded in the out-of-transit
photometry too, as pointed out by \citet{nutzman:2011}.

In contrast to the examples given so far, other authors consider
starspots to be a nuisance rather than a tool, due to their differing goals.
For example, the ``Hunt for Exomoons with Kepler'' (HEK) project 
\citep{hek:2012} anticipates that starspot crossings will be a source of 
false-positives for exomoon identification due to the morphological similarities 
with planet-moon mutual events. Cross-referencing the transit anomaly with
rotational modulations may be used to test whether the event is consistent
with a starspot or not \citep{sanchis:2012}.

Finally, even non-occulted starspots are a source of frustration in some
arenas. In particular, these spots subtly change the perceived transit depth.
Since spots vary in both time and wavelength, they are therefore capable of
producing transit depth variations. This point is highly salient for those
studying the atmospheres of exoplanets, who seek small chromatic variations in 
the depth. The study of rotational modulations can be used to correct the
resulting transmission spectra, as shown in \citet{desert:2009} for example.

It is therefore clear that the study and interpretation of rotation modulations
due to starspots is crucial to several areas of modern astronomical research.
In this work, we aim to provide a revised model for such modulations which
can account for several previously ignored effects.

%%% JUSTIFYING ANALYTIC OVER NUMERICAL
\subsection{Numerical vs Analytic Models}
\label{sub:numericalvsanalytical}

Inverting the brightness modulation of a star into a physical map of the
starspot coverage can be broached in several ways. The most successful technique
is Doppler imaging augmented by precise photometry (e.g. see
\citealt{vogt:1983,cameron:1994,tuominen:2002}). However, using rotational 
modulation alone, as is the case for the \emph{Kepler Mission}, is a 
more challenging problem.

Eclipse mapping of eclipsing binaries (EBs) allows for greatly improved 
inversions of the spot coverage through photometry alone. This is usually done 
by numerically pixelating the star and applying the maximum-entropy-method (MEM) 
to invert the map (e.g. \citealt{cameron:1997}). In principle, a transiting 
planet offers the same opportunity as was discussed earlier for measuring 
spin-orbit alignments. However, eclipse-mapping using planets has several
drawbacks. For example, the much smaller ratio-of-radii means only a thin-strip 
of the star is actually sampled. This means that inferences about the 
non-eclipsed portion of the star, which affect the perceived transit depth for 
example, must be made using the rotational modulations. Further, the 
eclipse-mapping technique only provides a snapshot of the spot-coverage at the 
instant of the transit. For long-period planets, or even stars without transits 
at all, this technique cannot uniquely infer even basic stellar properties, such 
as the rotation period.

Like so many problems in astrophysics, modelling rotation modulations due to 
starspots can be approached using either numerical or analytic techniques. 
Numerical techniques are more diverse, allowing one to compute any starspot 
shape, flux profile, limb darkening law, etc one wishes. However, they come at 
the expense of much higher computation times. Analytic models are extremely 
quick to execute, often outpacing their numerical counterparts by 
orders-of-magnitude. Such models are challenging to derive though and are
limited in that they assume fixed properties of each spot; for example, their
shape is often assumed to be circular. However, rotational modulations represent
a disc-integrated snapshot of a star with unresolved surface features. For
this reason, the size of a spot and the flux-contrast are highly correlated
since altering either will change the amplitude of the resulting rotational
modulations. In this paradigm, modelling elaborate shapes for the starspots
with dozens of free parameters is unlikely to generate a more meaningful model
than a simple circular assumption. Additionally, in the circular model, one
should interpret the spots as really representing a dark patch or 
cluster of small spots on the surface rather than a perfectly circular starspot.

The diverse range of spots which can be modelled using numerical techniques is
therefore not a practical advantage over analytic models. With this advantage
lost, we consider analytic models to be invariably the preferential tool for 
modelling starspots since they will execute with dramatically quicker 
computation times.

%%% Current Analytic Models
\subsection{Current Analytic Models}
\label{sub:currentanalytics}

Analytic models of rotational modulations due to starspots have existed for 
decades. The foundational paper comes from \citet{budding:1977}, who describe an 
analytic model for the rotational modulation due to multiple non-overlapping 
circular starspots. An alternative derivation of an essentially identical model 
is presented in \citet{dorren:1987}. 

These models have been successfully applied to numerous studies of starspots.
We highlight the recent work of the \emph{MOST} space-telescope in detecting
differential rotation on $\epsilon$ Eridani \citep{croll:2006} and 
$\kappa^1$ Ceti \citep{walker:2007}. In both of these cases, the authors make
use of a Markov Chain Monte Carlo (MCMC) algorithm to regress the data, which
yields Bayesian inferences of the parameter posteriors and correlations. The
code, called StarSpotz \citep{starspotz:2006}, includes parallel tempering
to locate the global minimum in the inevitably complex parameter landscape.
Bayesian inference techniques, such as MCMC, offer significant improvements in
the statistical interpretation of modelling starspots, but come at the cost of
being inherently computationally expensive. For this reason, analytic models
are highly prized due to their unmatched computational efficiency.

Despite the successes of the \citet{budding:1977} and \citet{dorren:1987}
model, there are several areas for improvement. Firstly and perhaps most
critically, the models are limited to a linear limb darkening law which is
generally a poor description stellar specific intensity profiles. 
\citet{claret:2000} remark that the most accurate limb-darkening functions
are the quadratic and ``non-linear'' laws, both of which are widely used in
the exoplanet community for example. Indeed, recently \citet{nutzman:2011}, who 
also made use of the \citet{dorren:1987} model, remarked on how extending the 
model to non-linear laws would be a significant improvement. Given the dramatic 
increase in photometric precision since the era of \citet{budding:1977} to the 
\emph{Kepler}-era, this point is not just pertinent but imperative to address. 

Secondly, there currently exists no partial derivatives for the model, which 
would lead to a further significant improvement in regression analysis. One
of the obstacles in achieving this goal is that the \citet{budding:1977}
and \citet{dorren:1987} equation for the model flux is not a single-domain
function and has multiple cases. This means partial derivatives would have
to be tediously derived for all cases individually. Finally, if one sets the
goal of deriving partial derivatives, then it would be advantageous to
observers to include numerous intrinsic effects in the model, for which
their respective free parameters could also have partial derivatives computed.
Examples include differential rotation, starspot evolution, diluted light
and instrumental offsets.

\subsection{Goal of This Work}
\label{sub:goals}

For reasons described in \S\ref{sub:starspots} \& 
\S\ref{sub:numericalvsanalytical}, the principal goal of this paper
is to provide a revised analytic model for rotational modulations due to
starspots. This new model has wide applications for both stars with and without 
orbiting planets offering numerous advantages over the \citet{budding:1977} and
\citet{dorren:1987} algorithms. Specifically, we highlight the following key
features of the model presented in this paper:

\begin{itemize}
\item[{\tiny$\blacksquare$}] Allows for $N_S$ non-overlapping small starspots, 
assumed to be small relative to the stellar radius.
\item[{\tiny$\blacksquare$}] Full non-linear limb darkening of the stellar and 
spot surface is included with vectors $\mathbf{c}$ and $\mathbf{d}$ 
respectively.
\item[{\tiny$\blacksquare$}] Differential rotation is included via a 
latitude-dependency including terms in $\sin^2\Phi$ and $\sin^4\Phi$.
\item[{\tiny$\blacksquare$}] Starspot evolution permitted using a linear 
model.
\item[{\tiny$\blacksquare$}] Umbra/penumbra effect may be generated.
\item[{\tiny$\blacksquare$}] $M$ instrumental offsets are allowed for (e.g.
quarter-to-quarter offsets in \emph{Kepler} data)
\item[{\tiny$\blacksquare$}] $M$ blended light dilution factors are allowed for 
(e.g. quarter-to-quarter contamination in \emph{Kepler} data)
\item[{\tiny$\blacksquare$}] Our solution may be expressed as a single-domain 
analytic function.
\item[{\tiny$\blacksquare$}] Consequently, we are able to provide single-domain 
analytic expressions for the partial derivatives of the model flux with 
respect to all input parameters, as well as time ($F'$).
\item[{\tiny$\blacksquare$}] We also show how the model may be used to predict
transit depth variations (T$\delta$V) due to non-occulted spots.
\item[{\tiny$\blacksquare$}] We make freely available the new algorithm in 
Fortran 90 code, \macula\ (see www.cfa.harvard.edu/$\sim$dkipping/macula.html).
\end{itemize}

In \S\ref{sec:model}, we introduce the model and discuss the various definitions
and how differential evolution and starspot evolution is accounted for. In
this section, we also provide a way to use our model to compute the transit
depth variations due to unocculted spots. In \S\ref{sec:comparison}, we compare 
our results to that of the \citet{dorren:1987} model and show that the 
small-spot approximation used in this work is accurate to within $\sim$100\,ppm 
for spots of angular size $\lesssim 10^{\circ}$. Further, the model has a 
maximum error which is an order-of-magnitude less than that of 
\citet{budding:1977} and \citet{dorren:1987} for a non-linearly limb darkened
Sun-like star observed in the \emph{Kepler} bandpass, with spots of angular size 
below $10^{\circ}$. In \S\ref{sec:example}, we demonstrate an application to a 
previously-studied example, \emph{MOST} observations of $\kappa^1$ Ceti by 
\citet{walker:2007} using a multimodal nested sampling algorithm, \multi\ 
\citep{feroz:2008,feroz:2009}. Discussion of the key highlights is provided in 
\S\ref{sec:discussion}.

\section{The Model}
\label{sec:model}

\subsection{Assumptions}

For clarity, we list the assumptions made throughout this work below:

\begin{enumerate}
\item All starspots are circular and lie on the plane of the stellar surface
\item Starspots never overlap one another
\item Starspots are small relative to the stellar radius
\item Each starspot is grey and has a uniform temperature
\item The star is a sphere with projected circular symmetry (i.e. no gravity
darkening)
\end{enumerate}

We do not claim that these are necessarily physically true statements. The
function of these assumptions is that they are broadly reasonable and allow
for a self-consistent analytic solution for modelling both in- and 
out-of-transit starspots. 

Out of all our assumptions, the one which is most likely to impose practical 
restrictions on the application of our model is the small-spot approximation. 
One may reasonably question why such an assumption is indeed required. By
using the small-spot approximation, we may treat the surface brightness of
the star to be constant under the disk of the starspot. This assumption, 
inspired by the small-planet approximation used for modelling transit light 
curves in \citet{mandel:2002}, leads to dramatically simpler expressions. For
example, the non-linear limb darkening model of \citet{mandel:2002} requires
hypergeometric functions whereas after the authors apply the small-planet
approximation the most computationally expensive function is arc cosine. In
the case of a transiting planet, one can see that the circular symmetry of the 
planet is a simpler problem than that of the foreshortened starspot, suggesting 
an analytic non-linear limb darkening solution without the small-spot 
approximation would certainly not be computationally cheap.

The small-planet approximation is later shown to be dramatically more accurate 
than the linear limb darkening assumption of previous works for almost all 
feasible spot sizes. We estimate a maximum error of $100$\,ppm for spots of
angular sizes $\lesssim10^{\circ}$, and typically the error is much smaller than 
this. More detailed estimators for the accuracy of our small-spot model are 
provided in \S\ref{sec:comparison}.

\subsection{Definitions}

We define the host star to have $N_S$ starspots on its surface labelled by
$k=1,2,...,N_S-1,N_S$. Each spot has a fixed angular radius $\alpha_k$, which
represents the solid-angle of the cone swept out from the stellar centre to 
the stellar surface. Further, each starspot has a flux-per-unit-area contrast
ratio, relative to the star, defined by $f_{\mathrm{spot},k}$. This is
essentially a proxy for the temperature of the spot and is assumed to be
uniform within each starspot (but variable between spots). Setting 
$f_{\mathrm{spot},k}>1$ reproduces a bright facula, rather than a dark starspot.

\begin{figure}
\begin{center}
\includegraphics[width=8.4 cm]{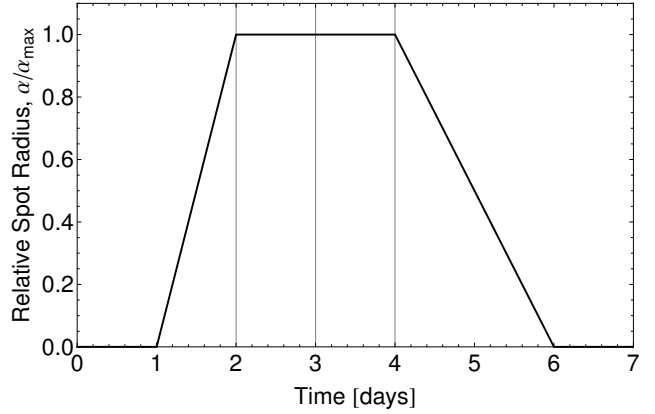}
\caption{\emph{An example of our linear starspot evolution model. We plot
the size of the starspot in units of $\alpha_{\mathrm{max}}$ as a function
of time. The gridlines (left-to-right) mark the end of ingress, the instant
$t_{\mathrm{max}}$, and the start of egress.}}
\label{fig:linear}
\end{center}
\end{figure}

The centre of a starspot has a longitude $\Lambda_k$ and latitude $\Phi_k$.
These two angles may be combined into the auxiliary angle, $\beta_k$, defined
as

\begin{align}
\beta_k &= \cos^{-1}\Big[ \cos I_* \sin\Phi_k + \sin I_* \cos\Phi_k\cos\Lambda_k \Big],
\end{align}

where $I_*$ is the inclination of the star. Starspots may migrate over time from
a reference location due to the stellar rotation. We assume that no migration
occurs in latitude but linear migration is permitted in longitude via

\begin{align}
\Lambda_k &= \Lambda_{\mathrm{ref},k} + \frac{2\pi (t-t_{\mathrm{ref},k})}{P_{*,k}}, \\
\Phi_k &= \Phi_{\mathrm{ref},k},
\end{align}

where $P_{*,k}$ is the time for the $k^{\mathrm{th}}$ spot to undergo a
change of $2\pi$ radians in longitude and $t_{\mathrm{ref},k}$ is an arbitrary
reference time when $\Lambda = \Lambda_{\mathrm{ref},k}$. We stress here that 
modifying our code to include latitude migration is trivial but the partial 
derivatives returned by the algorithm are only valid under the above assumption. 
Due to differential rotation, this period varies for each spot and here we 
assume a simple latitude-dependence for the differential rotation of

\begin{align}
P_{*,k} &= \frac{P_{\mathrm{EQ}}}{1 - \kappa_2 \sin^2\Phi_k - \kappa_4 \sin^4\Phi_k},
\end{align}

where $P_{\mathrm{EQ}}$ is the rotation period for the equator of the star and
$\kappa_2$ and $\kappa_4$ are coefficients of the differential rotation profile.
Additionally, a starspot may evolve via a linear growth/decay model of the 
angular size via

\begin{align}
\frac{\alpha_k(t_i)}{\alpha_{\mathrm{max},k}} &= \mathcal{I}_k^{-1} [\Delta t_1 \mathsf{H}(\Delta t_1) - \Delta t_2 \mathsf{H}(\Delta t_2)] \nonumber\\
\qquad& - \mathcal{E}_k^{-1} [\Delta t_3 \mathsf{H}(\Delta t_3) - \Delta t_4 \mathsf{H}(\Delta t_4)].
\label{eqn:evolution}
\end{align}

and using

\begin{align}
\Delta t_1 &= t_i-t_{\mathrm{max},k}+\frac{L_k}{2}+\mathcal{I}_k,\\
\Delta t_2 &= t_i-t_{\mathrm{max},k}+\frac{L_k}{2},\\
\Delta t_3 &= t_i-t_{\mathrm{max},k}-\frac{L_k}{2},\\
\Delta t_4 &= t_i-t_{\mathrm{max},k}-\frac{L_k}{2}-\mathcal{E}_k,
\end{align}

where $\alpha_{\mathrm{max},k}$ is the angular size of the $k^{\mathrm{th}}$ 
spot at a reference time $t_{\mathrm{max},k}$, $L_k$ is the ``lifetime'' of the 
spot (technically the full-width-full-maximum) and $\mathcal{I}_k$ \& 
$\mathcal{E}_k$ are the ingress \& egress durations of the spot's growth
profile. $\mathsf{H}(x)$ is the Heaviside Theta step-function. An illustrative 
example of our starspot growth/decay model is shown in Figure~\ref{fig:linear}.

%%% LINEAR EXAMPLE

For simplicity, \macula\ defines the reference times 
$t_{\mathrm{ref},k}$ to be equal to $t_{\mathrm{max},k}$, although one may 
change this definition without affecting the validity of the returned partial 
derivatives. Finally, we employ the four-coefficient non-linear limb darkening 
law of \citet{claret:2000}, where the specific intensity of the star is given by

\begin{align}
I_*(r) &= 1 - \sum_{n=1}^4 c_n (1-\mu^{n/2}),
\label{eqn:LD4}
\end{align}

where $c_n$ are the limb darkening coefficients, $\mu=\cos\Theta=\sqrt{1-r^2}$,
$0\leq r\leq1$ is the normalised radial coordinate on the disk of the star.
We employ the definition of a normalised limb darkening coefficient, $c_0$,
as utilised by \citet{mandel:2002} where $c_0 = 1 - c_1 - c_2 - c_3 - c_4$.

\subsection{Solution}

A detailed derivation of the model presented here is provided in Appendices 
\ref{appendixA}, \ref{appendixB} and \ref{appendixC}. To summarise,
the analytic solution for the model flux from $N_S$ non-overlapping circular
starspots may be expressed as

\begin{align}
F_{\mathrm{mod}} &= \sum_{m=1}^M U_m \Pi_{m} \Bigg(\frac{F(\boldsymbol{\alpha},\boldsymbol{\beta})}{B_m F(\boldsymbol{\alpha}=\mathbf{0},\boldsymbol{\beta})} + \frac{B_m-1}{B_m}\Bigg),
\label{eqn:Fmod}
\end{align}

where $U_m$ is the instrumental offset of the $m^{\mathrm{th}}$ data set (a
normalisation factor for each \emph{Kepler} quarter, for example), $B_m$ is
a blending factor for each data set (or quarter) and $\Pi_{m}$ is a box-car 
function defined by

\begin{align}
\Pi_{m}(t_i;T_{\mathrm{start},m},T_{\mathrm{end},m}) &= \mathsf{H}(t-T_{\mathrm{start},m}) - \mathsf{H}(t-T_{\mathrm{end},m}).
\end{align}

In the above, $T_{\mathrm{start},m}$ is the start of the $m^{\mathrm{th}}$ data 
set and $T_{\mathrm{end},m}$ is the end of the $m^{\mathrm{th}}$ data set. The 
$F(\boldsymbol{\alpha},\boldsymbol{\beta})$ function is given by

\begin{align}
F(\boldsymbol{\alpha},\boldsymbol{\beta}) &= 1 - \sum_{n=0}^4 \Big(\frac{n c_n}{n+4}\Big) - \sum_{k=1}^{N_S} \frac{A_k}{\pi} \Bigg[ \nonumber \\
\qquad& \Bigg(\sum_{n=0}^4 \frac{4 (c_n-d_n f_{\mathrm{spot},k})}{n+4} \frac{ \zeta_{-,k}^{\frac{n+4}{2}} - \zeta_{+,k}^{\frac{n+4}{2}} }{ \zeta_{-,k}^2 - \zeta_{+,k}^2 + \delta_{\zeta_{+,k},\zeta_{-,k}} } \Bigg) \Bigg],
\label{eqn:Fsub}
\end{align}

where $\delta_{x,y}$ is the Kronecker delta fucntion and 
$\mathbf{c}=\{c_0,c_1,c_2,c_3,c_4\}^T$ and $\mathbf{d}=\{d_0,d_1,d_2,d_3,d_4\}^T$ 
describe the non-linear limb darkening coefficients of the stellar surface and 
spot surface respectively. The function $A_k$ defines the sky-projected area of 
the $k^{\mathrm{th}}$ starspot and is given by:

\begin{align}
A_k(\alpha_k,\beta_k) &= \mathbb{R}\Big[ \cos^{-1}[\cos\alpha_k\csc\beta_k]\nonumber \\
\qquad& + \cos\beta_k\sin\alpha_k \Xi_k - \cos\alpha_k\sin\beta_k \Psi_k \Big],
\end{align}

where we use

\begin{align}
\Xi_k &= \sin\alpha_{k}\cos^{-1}[-\cot\alpha_{k}\cot\beta_{k}],\\
\Psi_k &= \sqrt{1-\cos^2\alpha_k\csc^2\beta_k}.
\end{align}

Finally, Equation~\ref{eqn:Fsub} includes the $\zeta$ function, which we
define as:

\begin{align}
\zeta(x) &= \cos x \mathsf{H}(x) \mathsf{H}(\frac{\pi}{2}-x) + \mathsf{H}(-x),
\end{align}

and we further define $\zeta_{-,k} = \zeta(\beta_k-\alpha_k)$ and 
$\zeta_{+,k} = \zeta(\beta_k+\alpha_k)$. We provide some typical examples
generated by \macula\ using random input parameters for four realisations
of a 5-spot model in Figure~\ref{fig:examples}.

\subsection{Generating Umbra/Penumbra}
\label{sub:umbra}

Sunspots can manifest with umbra and penumbra, or simply as ``naked'' spots.
Formally, our model assumes naked non-overlapping spots. However, our
algorithm \macula\ does allow one to place spots on top of one another too.
Doing so allows one to generate umbra/penumbra effects via a superposition.

To generate a single starspot with an umbra and penumbra of angular radii 
$\alpha_{\mathrm{umbra}}$ and $\alpha_{\mathrm{penumbra}}$ respectively, one 
simply generates two spots of these sizes. If the umbra has a flux contrast of 
$f_{\mathrm{umbra}}$ and the penumbra has $f_{\mathrm{penumbra}}$, then the two 
spots generated will have $\{\alpha,f_{\mathrm{spot}}\}$ equal to 
$\{\alpha_p,f_{\mathrm{penumbra}}\}$ and 
$\{\alpha_u,f_{\mathrm{penumbra}}-f_{\mathrm{umbra}}\}$.

\subsection{Transit Depth Variations from Unocculted Spots}
\label{sub:TdeltaV}

It is well-known that an eclipsing body which occults a starspot leaves a 
significant imprint on the transit profile \citep{rabus:2009}. Non-occulted dark 
spots also affect the transit indirectly via an amplification of the apparent
transit depth \citep{czesla:2009}; so-called transit depth variations 
(T$\delta$V). This occurs because the planet transits a non-spotty region where 
more flux is concentrated and thus more of the total flux is blocked out by the 
eclipse. The observed transit depth is defined as:

\begin{align}
\delta_{\mathrm{obs}} &= \frac{F_{\mathrm{out-of-transit}} - F_{\mathrm{in-transit}}}{F_{\mathrm{out-of-transit}}}.
\end{align}

For an unspotted star, this yields:

\begin{align}
\lim_{\boldsymbol{\alpha}\rightarrow\mathbf{0}} \delta_{\mathrm{obs}} = p^2 = \delta,
\end{align}

where $p$ is the ratio of the planet to star radius, $R_P/R_*$. To derive the
T$\delta$V effect, let us first consider that the spot is a bright facula. In 
this case, the spot actually increases the total amount of flux emitted by the 
star - it provides some extra flux $F_{\mathrm{extra}}$. This extra flux must be 
given by

\begin{align}
F_{\mathrm{extra}} = F(\boldsymbol{\alpha},\boldsymbol{\beta}) - F(\boldsymbol{\alpha}=\mathbf{0},\boldsymbol{\beta}).
\end{align}

An observationally equivalent scenario would be to consider this extra flux as 
originating from a spatially unresolved background star. This well-known 
scenario is often dubbed a ``blend'' because the extra flux source is uneclipsed 
and thus the total eclipse depth is diluted due to the blend source. 
\citet{kippingtinetti:2010} showed that a blend source changes the transit depth 
via:

\begin{align}
\delta_{\mathrm{obs}} &= \frac{\delta}{\mathcal{B}},\\
\mathcal{B} &= \frac{F_* + F_{\mathrm{extra}}}{F_*}.
\end{align}

The above allows for a simple calculation of the T$\delta$V effect. One 
additional effect we can include at this point is genuine background/foreground
blend sources with a blend factor $B_m$. For the bright facula then, the 
observed transit depth becomes

\begin{align}
\frac{\delta_{\mathrm{obs}}}{\delta} &= \frac{F(\boldsymbol{\alpha}=\mathbf{0},\boldsymbol{\beta})}{F(\boldsymbol{\alpha},\boldsymbol{\beta})} \frac{1}{\sum_{m=1}^M \Pi_m B_m}.
\label{eqn:TdeltaV}
\end{align}

For a bright facula, $F(\boldsymbol{\alpha},\boldsymbol{\beta}) >
 F(\boldsymbol{\alpha}=\mathbf{0},\boldsymbol{\beta})$ and thus
$\delta_{\mathrm{obs}} < \delta$ (for $\mathbf{B}=\mathbf{1}$) i.e. the transit 
depth becomes shallower due to the ``blend'' source, as expected. This logic is 
easily extended to dark starspots and accurately predicts the T$\delta$V effect, 
except that dark uncocculted spots cause deeper transits. The analogy of a 
spatially unresolved background star becomes unphysical in that the background 
star now emits negative flux, but this is beside the point. One key conclusion 
is that unocculted facula behave as ``blends'' and unocculted spots behave as 
``anti-blends''.

Another subtlety is that the above that the effect is purely due to blindly 
normalising the data by the local baseline, which is affected by rotational 
modulation. ``Astrophysical detrending'' of a continuous photometric time 
series, in this case using a spot-model to detrend the data, would eliminate any 
apparent depth variations. However, ground-based observers are usually only able 
to obtain a small amount of data either side of a transit event leaving no
alternative except to blindly normalise the data. In fact, even space-based 
transit observations are almost always blindly normalised using polynomials, 
running medians or linear trends. \macula\ therefore offers the opportunity the
astrophysically detrend photometry.

\begin{figure*}
\begin{center}
\includegraphics[width=18.0 cm]{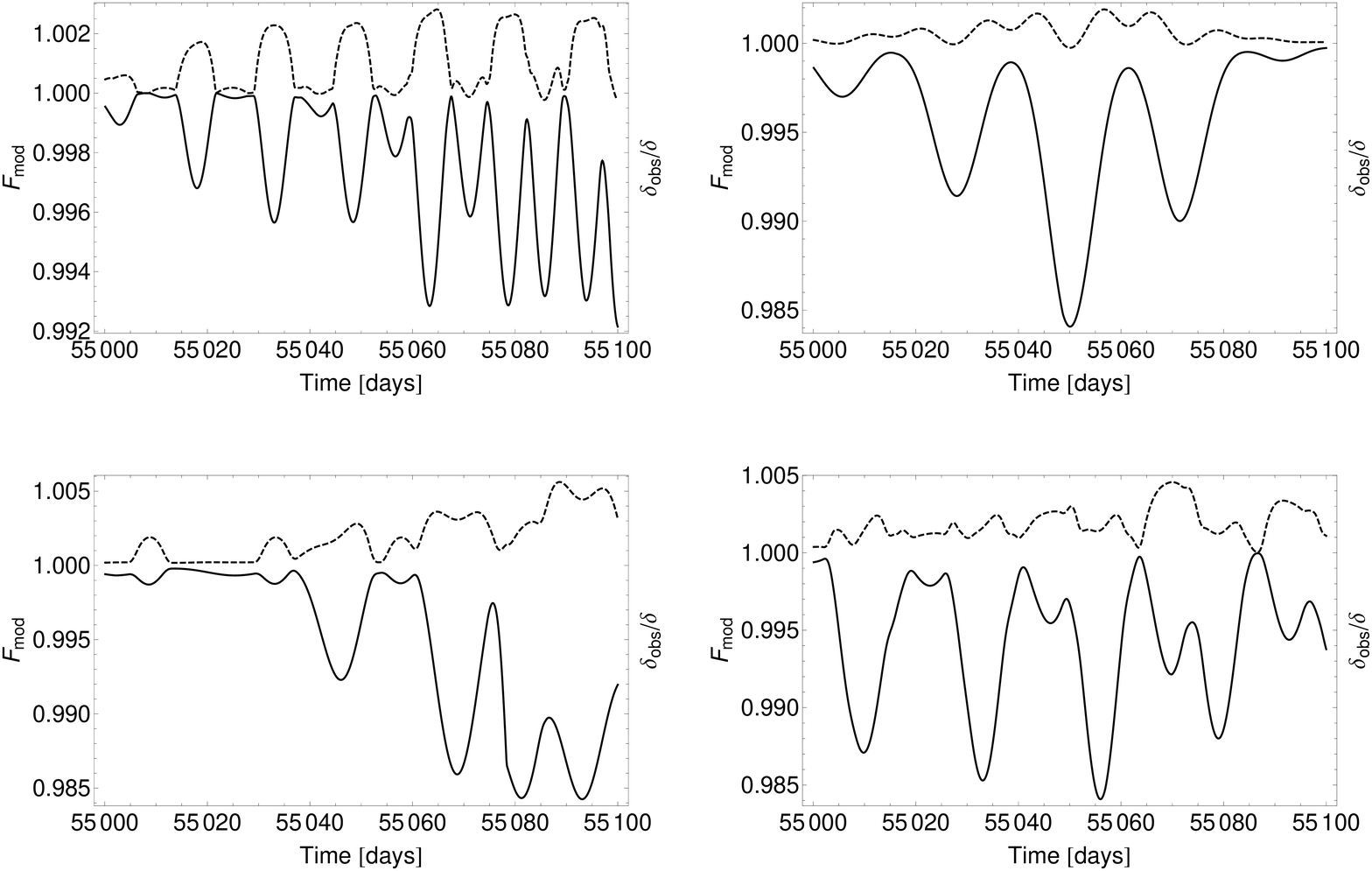}
\caption{\emph{Examples of light curves (solid) and T$\delta$Vs (dashed) for 
four randomly generated scenarios using \macula. Each simulation assumes 5
spots with random properties, including random angular sizes between $0^{\circ}$ 
and $10^{\circ}$ and Sun-like non-linear limb darkening. Even with 5 spots, the 
photometric behaviour can be highly complex due to spot evolution, which is also 
randomly generated in all four cases.}} 
\label{fig:examples}
\end{center}
\end{figure*}

In the typical case of blindly normalised data, Equation~\ref{eqn:TdeltaV} 
allows one to fit a set of transit depth variations with a spot model using 
\macula. Alternatively, one may wish to make causal predictions of the 
T$\delta$V effect based upon out-of-transit rotational modulations. We stress 
that the equation is only valid if the spots are unocculted. These T$\delta$Vs 
may occur in time due to rotational modulation (see examples in 
Fig~\ref{fig:examples}), or even in wavelength due to the chromatic nature of 
spots. Indeed, correcting for the chromatic T$\delta$V effect is crucial in 
accurate interpretation of exoplanet transmission spectra (e.g. 
\citealt{desert:2009}). A more detailed discussion of the applications
of T$\delta$Vs is presented in \S\ref{sub:apps}

\macula\ directly returns the function $(\delta_{\mathrm{obs}}/\delta)$ at
all times, $t_i$, which are inputted. This feature is on/off switchable so that 
one may either choose to not use the feature or perhaps just input one (or a 
few) value(s) of $t_i$, such as the time(s) of transit minimum. Partial
derivatives of this function are not provided although may be easily computed
since \macula\ evaluates the partial derivatives of $A_k$ and 
$F_{\mathrm{mod}}$.

%%% EXAMPLES

\section{Comparison to the Dorren (1987) Model}
\label{sec:comparison}

\subsection{Overview}

As discussed earlier in \S\ref{sub:currentanalytics}, the most widely used
analytic models for starspot modelling come from \citet{budding:1977} and
\citet{dorren:1987}. The models are essentially identical and so we will
refer to comparison to the \citet{dorren:1987} model only from here-on-in
for brevity.

The main difference between our model and that of \citet{dorren:1987} is
that our derivation assumes the starspot is small i.e.
$\sin\alpha\lesssim0.1$, whereas \citet{dorren:1987} did not. By making
this assumption, we have derived a full non-linear limb darkening treatment
for starspots, whereas the \citet{dorren:1987} model is limited to a simple
linear limb darkening law only. Further, our model is amenable to the inclusion
of spot-crossing events due the parametric form of the expressions describing
the arcs along the starspot rim and bulge.

Since the two models essentially only differ in their treatment of limb
darkening, one should expect them to be exactly equivalent for the case of a
uniform brightness star, which we easily verified through numerical
experiments. However, for the case of a limb darkened star, one might ask, for 
what spot size does our model significantly deviate away
from that of \citet{dorren:1987}? We will investigate this question in the
following subsections.

\subsection{Model Error due to our Small-Spot Approximation}

Let us define the model flux, as predicted by this work, as 
$F_{\mathrm{mod}}$ (as used throughout). Let us further define the model flux
as predicted by \citet{dorren:1987} as $F_{\mathrm{mod,D87}}$. If we assume
a linearly limb darkened star, then the only difference between the model of
\citet{dorren:1987} and that of this work is that we assume a small-spot and
\citet{dorren:1987} do not. Therefore, direct comparison between the two models 
for linear limb darkened stars yields the error in our model of assuming a
small-spot. Accordingly, one may write that the relative error in our model is:

\begin{align}
\Delta F_{\mathrm{mod}} &= \Big|\frac{ F_{\mathrm{mod,D87}} - F_{\mathrm{mod}}(c_1=c_3=c_4=0;c_2=u_L)}{ F_{\mathrm{mod,D87}} }\Big|
\end{align}

For simplicity, we assume the limb darkening of the spot and star are equivalent
and set the spot-star contrast, $f_{\mathrm{spot}}$, to be zero (a black spot). 
Numerically evaluating $\Delta F_{\mathrm{mod}}$ over the domain of interest 
reveals the error is maximised when $\beta=\alpha$. Therefore, we define 
$\Delta F_{\mathrm{mod}}^{\mathrm{max}}=\Delta F_{\mathrm{mod}}(\beta=\alpha)$.

The function $\Delta F_{\mathrm{mod}}^{\mathrm{max}}$ grows with both $u_1$ 
and $\alpha$, tending to zero when they both equal zero, as expected. For a 
Sun-like star ($T_{\mathrm{eff}} = 6000$\,K, $\log g = 4.5$\,dex, [M/H] = 0), 
\citet{claret:2011} estimate that the best fitting linear limb darkening 
coefficient in the \emph{Kepler} bandpass is $u_L = 0.5733$. One may now
set $\Delta F_{\mathrm{mod}}^{\mathrm{max}}$ to some desired tolerance level
(e.g. the noise level of the data) and solve for $\alpha$ i.e. the maximum
spot size which leads to model errors below the tolerance level. We plot
this function in Figure~\ref{fig:modelerrors} (solid-line).

Since a $M_{\mathrm{Kep}}=12$ star has a typical noise of $\sim50$\,ppm per 
long-cadence measurement (note that most \emph{Kepler} targets are fainter than 
this), we estimate that the maximal error of our small-spot approximation is 
below that of a typical \emph{Kepler} measurement error for starspots of angular 
radius $\alpha \lesssim 7.6^{\circ}$. This corresponds to a spot coverage of 
$\lesssim 1.7$\%. Note that the modal spot coverage of stars in the 
\emph{Kepler} sample is $\simeq1$\% \citep{basri:2011}. A 1.7\% spot coverage 
roughly corresponds to $V_{\mathrm{rng}}=0.83$ on Figure~4 of 
\citet{basri:2011}, which can be seen to encompass the majority of periodic 
variables. Given the conservative assumptions of using a relatively bright 
$12^{\mathrm{th}}$ magnitude star (most \emph{Kepler} targets are fainter), the 
fact that we assume only a single spot is responsible for the entire spot 
coverage (making the spot as large as possible) and the fact that the error 
derived is the \emph{maximal} error rather than the typical error, we conclude 
that the large majority of spotted stars within the \emph{Kepler} sample are 
appropriately modelled by the expressions in this work.

\subsection{Model Error due to a Linear Limb Darkening Law Assumption}

For larger spots, the modelling error becomes larger due to our small-spot
assumption and thus an observer may opt to use the \citet{dorren:1987} model
instead. However, we point out that this model assumes a linear limb-darkening
law which is somewhat unphysical in itself. The question therefore arises,
at what point is the error in assuming a large spot with linear limb darkening 
better than assuming a small spot with non-linear limb darkening?

We have already calculated the error due to the small-spot assumption, assuming
the star is perfectly described by a linear limb darkening law 
($\Delta F_{\mathrm{mod}}$). We may similarly define an error in assuming a 
linear limb darkening law when the star is really described by a non-linear
law:

\begin{align}
\Delta F_{\mathrm{mod,D87}} &= \Big|\frac{ F_{\mathrm{mod,D87}} - F_{\mathrm{mod}} }{ F_{\mathrm{mod}} }\Big|
\end{align}

For a star with the same properties as used in the previous example 
($T_{\mathrm{eff}} = 6000$\,K, $\log g = 4.5$\,dex, [M/H] = 0), 
\citet{claret:2011} estimate that the best fitting non-linear limb darkening 
coefficients in the \emph{Kepler} bandpass are 
$\{c_1,c_2,c_3,c_4\}=\{0.3999,0.4269,-0.0227,-0.0839\}$. We also assume
a black spot with the same limb darkening as the star, as was done for the
previous example. Plotting the function $\Delta F_{\mathrm{mod,D87}}$ for several 
realisations of $\alpha$ as a function of $\beta$, we find the maximal error 
occurs at $\beta=\alpha/2$. Thus we define 
$\Delta F_{\mathrm{mod,D87}}^{\mathrm{max}}=\Delta F_{\mathrm{mod,D87}}(\beta=\alpha/2)$.

\begin{figure*}
\begin{center}
\includegraphics[width=18.0 cm]{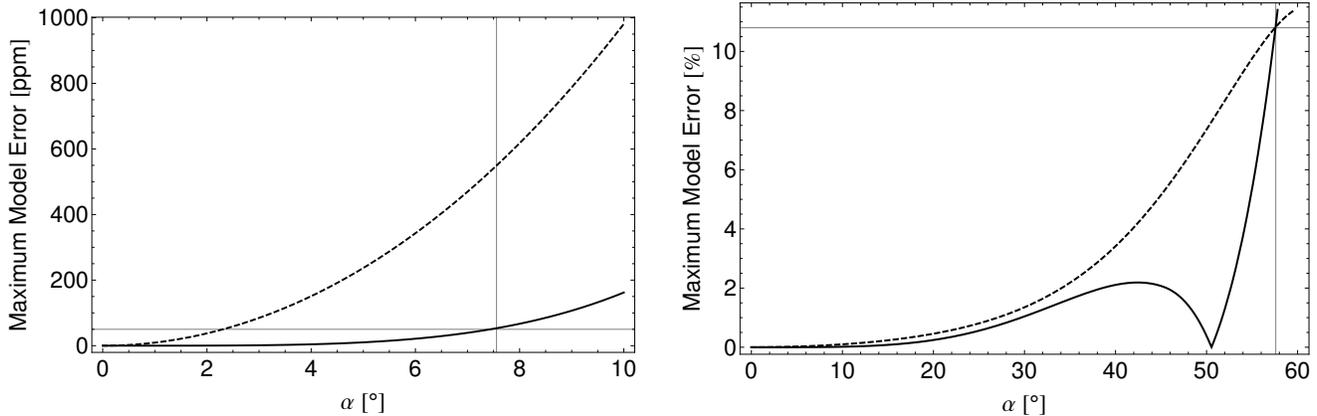}
\caption{\emph{Left: Comparison of the maximum model error of the 
\citet{dorren:1987} linear limb darkening assumption (dashed) versus the 
small-spot approximation of this work (solid), for a Sun-like star. As 
expected, over the range of small-spot sizes, the model presented in this work 
is considerably more accurate. Gridlines mark the point at which our model is
accurare to 50\,ppm at $\alpha=7.6^{\circ}$. Right: Same as left-panel, except
we zoom out to a greater $x$-scale. The \citet{dorren:1987} becomes more
accurate than the model presented in this work for spots larger than 
$57.6^{\circ}$ (marked with gridlines). At this point, the error in both models
is 105\,mmag and is arguably unusable in either case. Note that the location of
the minimum in our model error near $50^{\circ}$ is sensitive to the limb
darkening coefficients used.}} 
\label{fig:modelerrors}
\end{center}
\end{figure*}

In Figure~\ref{fig:modelerrors}, we plot this function along with
$\Delta F_{\mathrm{mod}}^{\mathrm{max}}$ as a function of $\alpha$ for
the Sun-like limb darkening coefficients computed by \citet{claret:2000}.
The figure reveals that the error in assuming a small-spot is substantially
smaller than the error in assuming a linear limb darkening law for 
$\alpha\leq10^{\circ}$, as one should expect. As an example, sunspots typically
have angular sizes $\lesssim 5^{\circ}$ and for $\alpha=5^{\circ}$ we find
$\Delta F_{\mathrm{mod}}^{\mathrm{max}}=10$\,ppm whereas 
$\Delta F_{\mathrm{mod,D87}}^{\mathrm{max}}=237$\,ppm i.e. our model is more 
than an order-of-magnitude more accurate. For spots of size $\alpha=10^{\circ}$
we find $\Delta F_{\mathrm{mod}}^{\mathrm{max}}=162$\,ppm versus
$\Delta F_{\mathrm{mod}}^{\mathrm{max}}=981$\,ppm. 

We find that the significant error in assuming a linear limb darkening law does 
not become a better approximation than the small-spot model until 
$\alpha>57.6^{\circ}$, for Sun-like limb darkening. After this 
point, the error in our model rapidly tends to infinity and becomes untenable.
The exact locations of the various turnovers and minima in 
Figure~\ref{fig:modelerrors} depend upon the limb darkening parameters used.
Also, the reliability of the $\Delta F_{\mathrm{mod,D87}}^{\mathrm{max}}$
function worsens for large $\alpha$ since the ``truth'', assumed to be
$F_{\mathrm{mod}}$ itself starts to become erroneous at high $\alpha$.

Nevertheless, it is clear that our model is more accurate than the 
\citet{dorren:1987} model for even a starspot equivalent to the largest spot 
ever detected ($\alpha\simeq30^{\circ}$; \citealt{strassmeier:1999}). Despite
this, we would urge observers to use a numerical approach for such large spots
since the model errors are significantly greatly than typical measurement 
errors. Spots of size $\alpha\lesssim10^{\circ}$ should be well-described by the 
analytic model presented in this work.

%%% MODEL ERRORS

\section{An Example Application to $\kappa^1$ Ceti}
\label{sec:example}

\subsection{\emph{MOST} observations of $\kappa^1$ Ceti}

$\kappa^1$ Ceti is a relatively nearby G5 dwarf 9.1\,pc from the Solar System.
The star is notable for having a fairly rapid rotational period of $\sim$9\,days
and for being a bright Sun-like star at $V=4.84$. \emph{MOST} observations of
$\kappa^1$ Ceti in 2003 revealed the presence of two starspots with rotation
periods of 8.9\,d and 9.3\,d \citep{rucinski:2004}. However, this single
data set was insufficient to uniquely determine the latitudes of the spots
and thus the differential rotation coefficient, $\kappa_2$ could not be
measured (the authors did not consider the $4^{\mathrm{th}}$-order coefficient
$\kappa_4$).

Subsequently, \emph{MOST} observed $\kappa^1$ Ceti two more times in 2004 and
2005 in order to gather enough data that a unique solution could be inferred.
Indeed, this data, reported by \citet{walker:2007}, was argued by the authors 
to be sufficient to locate a single minimum. The authors made use of the
StarSpotz \citep{starspotz:2006} algorithm to regress the data, which in turn
employs the \citet{budding:1977} model for starspots (note that this is
equivalent to the model of \citealt{dorren:1987}). Starspotz locates the global
minimum using parallel tempering and derives parameter posteriors using the
Markov Chain Monte Carlo (MCMC) technique. 

The photometry span three data sets, exhibit differential rotation and 
seven spots over three years (ranging from $\alpha_k=5.95^{\circ}$ to 
$\alpha_k=16.76^{\circ}$) and thus required considerable computational
effort by \citet{walker:2007}. For these reasons, the data make for an ideal 
test of not only our model here but for an alternative regressing technique.

\subsection{Multimodal Nested Sampling with \multi}

Nested sampling \citep{skilling:2004} is a Monte Carlo method which puts the 
calculation of the Bayesian evidence in a central role, but also produces 
posterior inferences as a by-product. Nested sampling is generally considerably
more efficient than MCMC methods for computing the Bayesian evidence of a model
fit. For example, in cosmological applications, \citet{mukherjee:2006} showed 
that their implementation of the method requires a factor of $\sim100$ fewer 
posterior evaluations than thermodynamic integration with MCMC. A full 
discussion of nested sampling is given in \citet{skilling:2004} and 
\citet{feroz:2008} and for brevity we direct those interested to these
works. 

Multimodal nested sampling is an implementation of the technique to efficiently 
search parameter space under the assumption that one or more modes may exist
in the data. \citet{feroz:2008,feroz:2009} describe multimodal nested
sampling in detail, in particular in regard to their publicly available
algorithm \multi. \multi\ is used by the ``Hunt for Exomoons with Kepler'' (HEK) 
project \citep{hek:2012} to compare the Bayesian evidence of a planet versus
planet-with-moon regression. We will here demonstrate the use of \multi\ with
our starspot model for the $\kappa^1$ Ceti \emph{MOST} photometry. Currently, 
\multi\ does not make use of the likelihood partial derivatives and so the
partial derivatives were turned off in our implementation of \macula. Since we 
only fit a single model through the data, there is no use of the Bayesian 
evidence here and thus we employ the constant efficiency mode of \multi\ at a 
target efficiency of 1\% with 4000 live points.

\subsection{Priors}

In order to make a fair comparison to the \citet{walker:2007} result, we make
the same assumptions as the original authors. Accordingly, we assume
the same number of spots for each data set i.e. 2 spots for 2003, 3 spots for
2004 and 2 spots for 2005. The spots are assumed to be non-evolving over the
course of each data set and have a lifetime which ensures they only exist within
a single data set, as was assumed by \citet{walker:2007}. We also assume
$f_{\mathrm{spot},k} = 0.22$ for all $k$ and that the differential rotation
profile is quadratic is nature (i.e. we fix $\kappa_4=0$).
Finally, limb darkening for the spot and the star are equivalent and follow
a linear law governed by $u_L = 0.6840$. Using these assumptions, we have the 
same number of free parameters (27) as was used by \citet{walker:2007}.

The 27 parameters are 7 reference longitudes, $\Lambda_{0,k}$, 7 reference
latitudes, $\Phi_{0,k}$, 7 angular radii, $\alpha_{0,k}$, one equatorial
rotation period, $P_{\mathrm{EQ}}$, one differential rotation coefficient, 
$\kappa_2$, one stellar inclination angle, $I_*$ and three instrumental
offset terms, $U_m$. Rather than label the offsets by $m=1,2,3$, we use
$m=2003,2004,2005$ for each year. Since each year has unique starspots,
we do use the spot labels $k=1,2,3,4,5,6,7$ but instead use $2003\_1,2003\_2,
2004\_1,2004\_2,2004\_3,2005\_1$\&$2005\_2$. These labels more clearly
identify the spots associated with each year and follow the labelling notation
of \citet{walker:2007}. We adopt the uniform priors for all 27 parameters with 
the same range as that of \citet{walker:2007}.

\subsection{Results}

\begin{figure*}
\begin{center}
\includegraphics[width=18.0 cm]{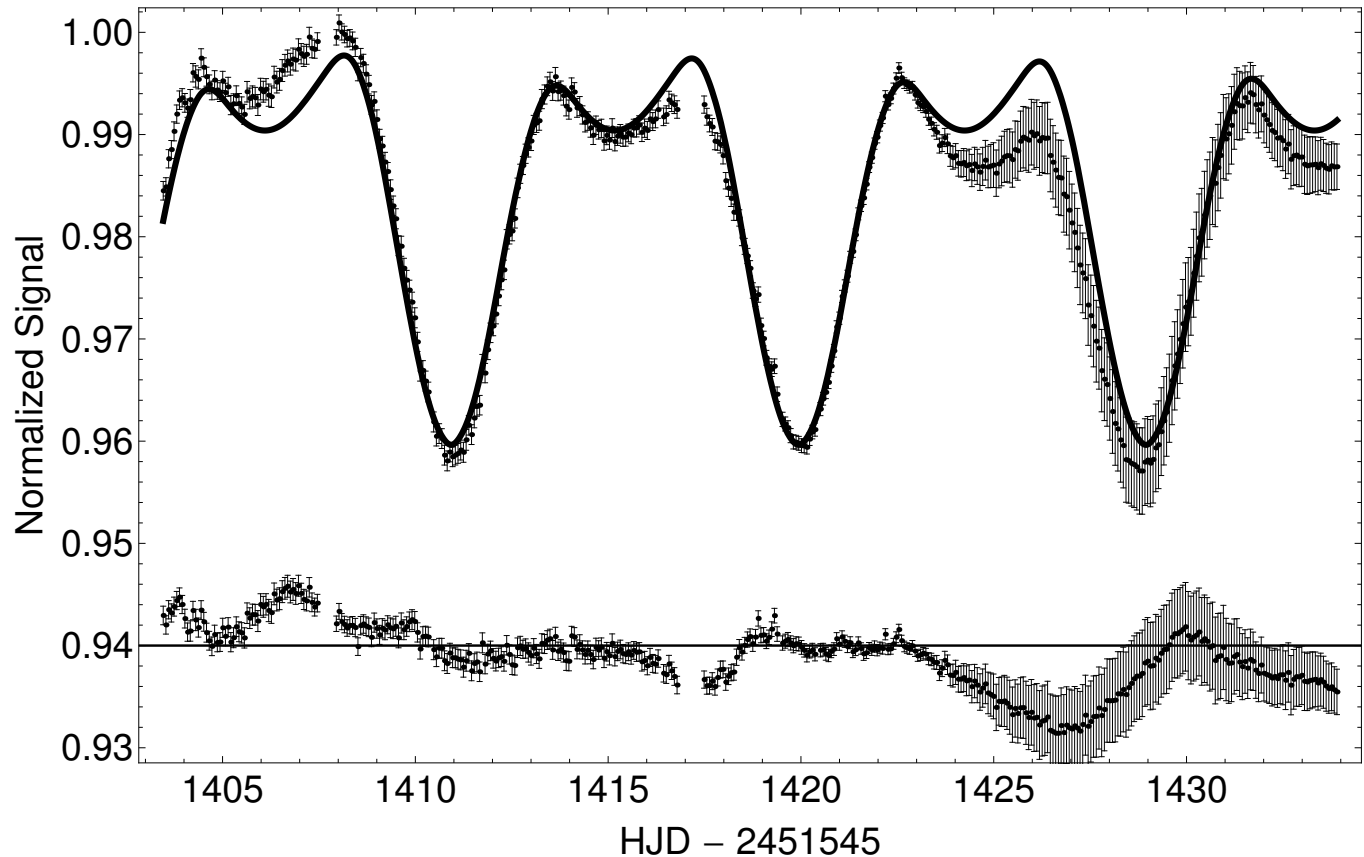}
\caption{\emph{Maximum a-posteriori two-spot model fit to the 2003 MOST data
of $\kappa^1$ Ceti using the analytic model presented in this work. Regression
performed using \multi\ in conjunction with the 2004 \& 2005 data. Residuals to 
the fit are offset by 0.94. Figure may be directly compared to Figure~4 of
\citet{walker:2007}, where one can see an essentially indistinguishable result.}} 
\label{fig:most_2003}
\end{center}
\end{figure*}

\begin{figure*}
\begin{center}
\includegraphics[width=18.0 cm]{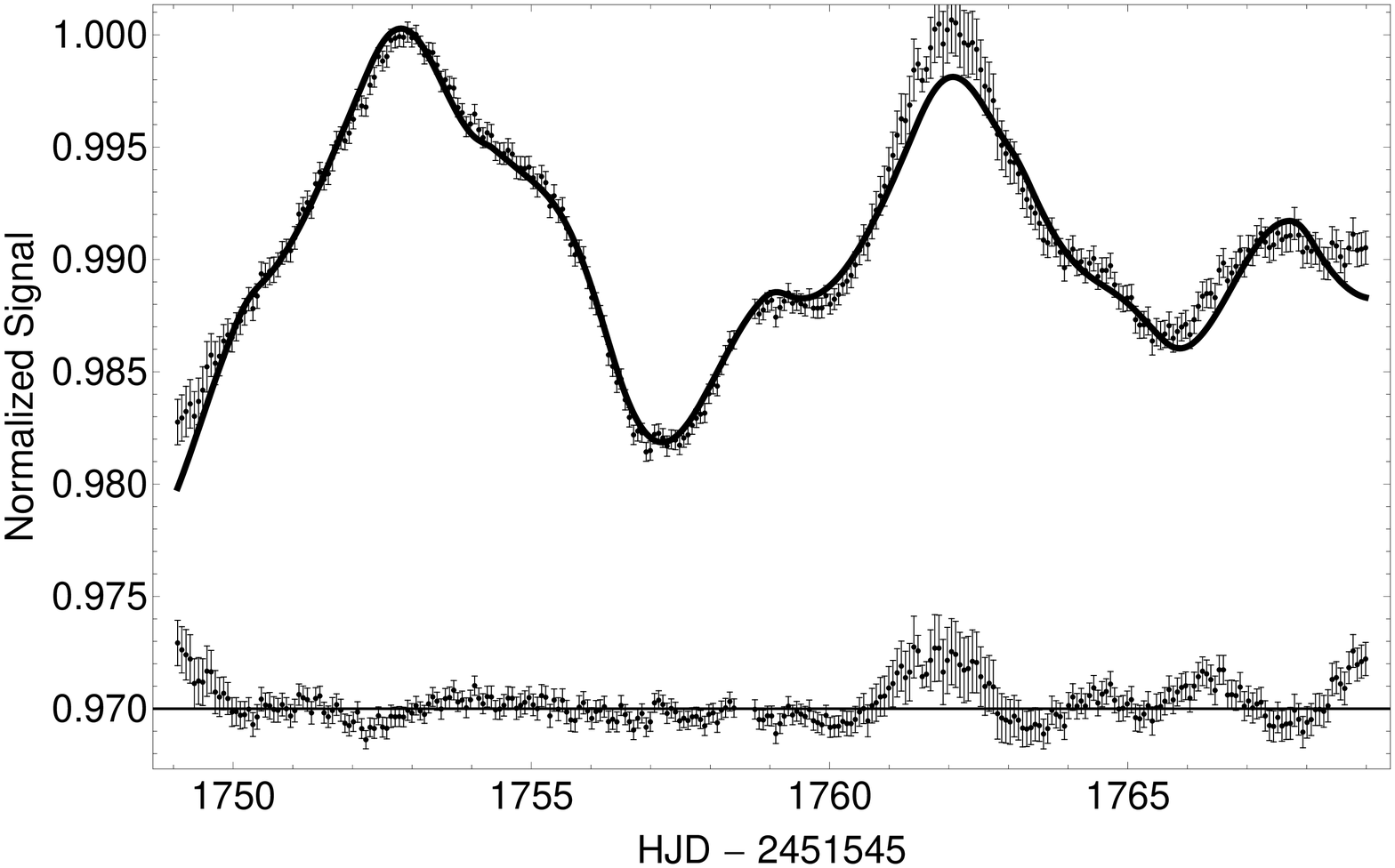}
\caption{\emph{Maximum a-posteriori two-spot model fit to the 2004 MOST data
of $\kappa^1$ Ceti using the analytic model presented in this work. Regression
performed using \multi\ in conjunction with the 2003 \& 2005 data. Residuals to 
the fit are offset by 0.97. Figure may be directly compared to Figure~5 of
\citet{walker:2007}, where one can see an essentially indistinguishable result.}} 
\label{fig:most_2004}
\end{center}
\end{figure*}

\begin{figure*}
\begin{center}
\includegraphics[width=18.0 cm]{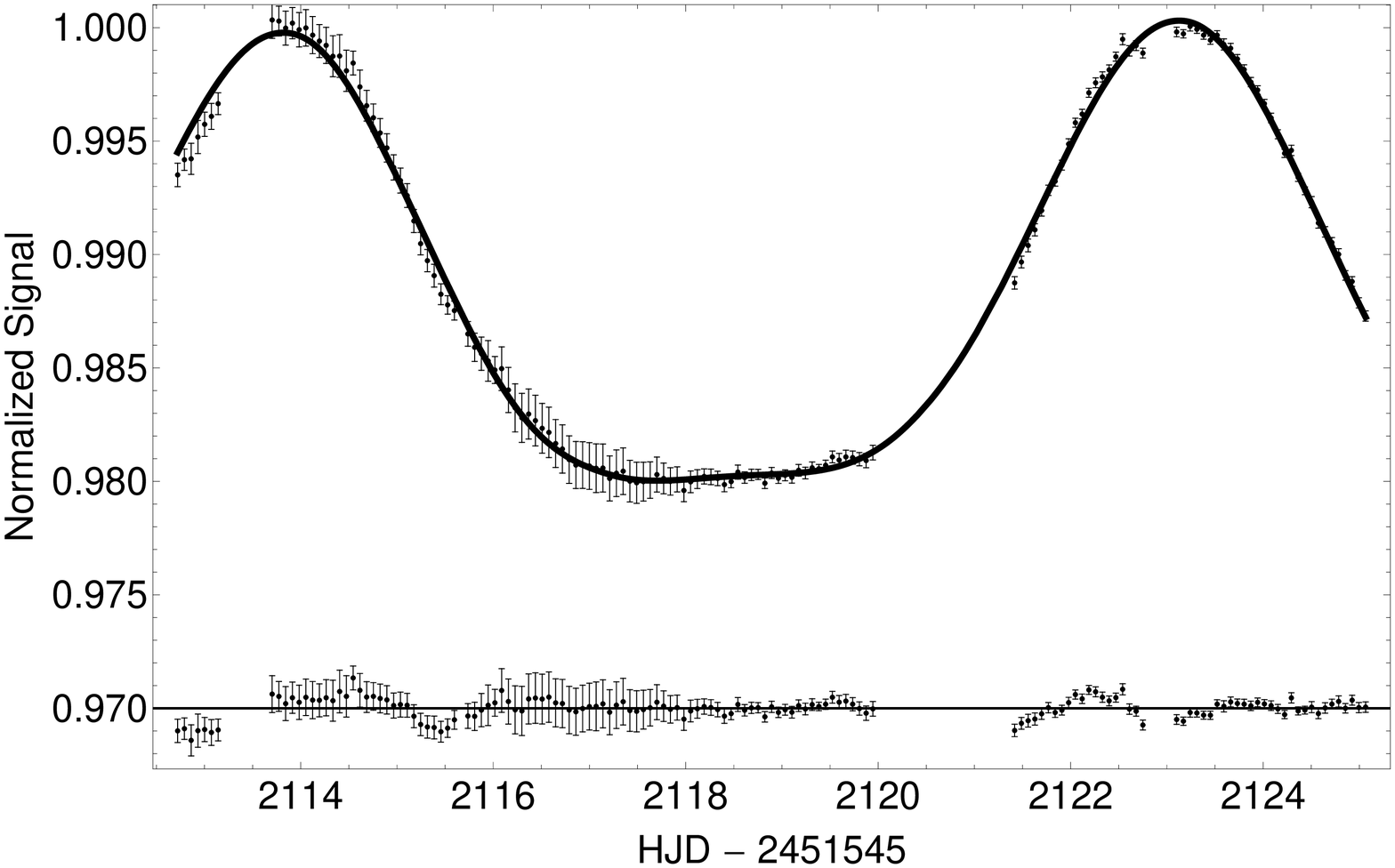}
\caption{\emph{Maximum a-posteriori two-spot model fit to the 2005 MOST data
of $\kappa^1$ Ceti using the analytic model presented in this work. Regression
performed using \multi\ in conjunction with the 2003 \& 2004 data. Residuals to 
the fit are offset by 0.97. Figure may be directly compared to Figure~6 of
\citet{walker:2007}, where one can see an essentially indistinguishable result.}} 
\label{fig:most_2005}
\end{center}
\end{figure*}

The global maximum a-posteriori model fit is presented in 
Figures~\ref{fig:most_2003}, \ref{fig:most_2004} \& \ref{fig:most_2005} for
the data sets in 2003, 2004 and 2005 respectively. Table~\ref{tab:results}
presents the posteriors of the best fitting mode and compares them side-by-side
with the results reported by \citet{walker:2007}.

\begin{table}
\caption{\emph{Results from fitting the MOST data of $\kappa^1$ Ceti using
the \macula\ model presented in this work and the \multi\ algorithm. 
Column~2 shows the 68.3\% credible range derived by \citet{walker:2007}
(taken from column 6 of Table~3 of that work).}} % title of Table
\centering % used for centering table
\begin{tabular}{c c c} % centered columns (3 columns)
\hline\hline %inserts double horizontal lines
\textbf{Parameter} & \textbf{Walker et al. (2007)} & \textbf{This Work} \\ [0.5ex] % inserts table
%heading
\hline
$I_*$ [$^{\circ}$] & $57.8$-$63.5$ & $60.1_{-1.3}^{+1.3}$ \\
$P_{\mathrm{EQ}}$ [days] & $8.74$-$8.81$ & $8.785_{-0.019}^{+0.018}$ \\
$\kappa_2$ & $0.085$-$0.096$ & $0.0868_{-0.0025}^{+0.0025}$ \\
$U_{2003}$ & $1.0003$-$1.0017$ & $1.00105_{-0.00037}^{+0.00040}$\\
$U_{2004}$ & $1.0129$-$1.0150$ & $1.01916_{-0.00113}^{+0.00061}$ \\
$U_{2005}$ & $1.0029$-$1.0051$ & $1.00449_{-0.00054}^{+0.00058}$ \\
$\alpha_{\mathrm{max},2003\_1}$ [$^{\circ}$] & $11.63$-$11.86$ & $11.771_{-0.060}^{+0.062}$ \\
$\Lambda_{\mathrm{ref},2003\_1}$ [$^{\circ}$] & N/A & $61.06_{-0.38}^{+0.38}$ \\
$\Phi_{\mathrm{ref},2003\_1}$ [$^{\circ}$] & $29.5$-$34.8$ & $31.8_{-1.4}^{+1.4}$ \\
$\alpha_{\mathrm{max},2003\_2}$ [$^{\circ}$] & $5.68$-$6.18$ & $5.93_{-0.13}^{+0.14}$ \\
$\Lambda_{\mathrm{ref},2003\_2}$ [$^{\circ}$] & N/A & $-105.7_{-1.3}^{+1.3}$ \\
$\Phi_{\mathrm{ref},2003\_2}$ [$^{\circ}$] & $32.9$-$39.8$ & $35.9_{-2.0}^{+1.9}$ \\
$\alpha_{\mathrm{max},2004\_1}$ [$^{\circ}$] & $7.73$-$8.09$ & $7.92_{-0.10}^{+0.10}$ \\
$\Lambda_{\mathrm{ref},2004\_1}$ [$^{\circ}$] & N/A & $50.53_{-0.98}^{+0.97}$ \\
$\Phi_{\mathrm{ref},2004\_1}$ [$^{\circ}$] & $9.3$-$16.8$ & $13.9_{-2.0}^{+2.0}$ \\
$\alpha_{\mathrm{max},2004\_2}$ [$^{\circ}$] & $14.44$-$17.31$ & $17.12_{-0.86}^{+0.94}$ \\
$\Lambda_{\mathrm{ref},2004\_2}$ [$^{\circ}$] & N/A & $154.8_{-1.0}^{+1.0}$ \\
$\Phi_{\mathrm{ref},2004\_2}$ [$^{\circ}$] & $-47.6$-$-43.2$ & $-46.9_{-1.2}^{+1.2}$ \\
$\alpha_{\mathrm{max},2004\_3}$ [$^{\circ}$] & $11.60$-$13.53$ & $14.26_{-0.51}^{+0.48}$ \\
$\Lambda_{\mathrm{ref},2004\_3}$ [$^{\circ}$] & N/A & $-32.1_{-1.2}^{+1.2}$ \\
$\Phi_{\mathrm{ref},2004\_3}$ [$^{\circ}$] & $74.9$-$78.4$ & $79.26_{-0.71}^{+0.58}$ \\
$\alpha_{\mathrm{max},2005\_1}$ [$^{\circ}$] & $9.24$-$10.28$ & $9.99_{-0.27}^{+0.31}$ \\
$\Lambda_{\mathrm{ref},2005\_1}$ [$^{\circ}$] & N/A & $-163.6_{-1.1}^{+1.2}$ \\
$\Phi_{\mathrm{ref},2005\_1}$ [$^{\circ}$] & $55.4$-$62.1$ & $60.0_{-1.7}^{+1.7}$ \\
$\alpha_{\mathrm{max},2005\_2}$ [$^{\circ}$] & $7.94$-$8.52$ & $8.35_{-0.15}^{+0.17}$ \\
$\Lambda_{\mathrm{ref},2005\_2}$ [$^{\circ}$] & N/A & $64.9_{-1.5}^{+1.5}$ \\
$\Phi_{\mathrm{ref},2005\_2}$ [$^{\circ}$] & $42.9$-$50.1$ & $47.3_{-1.9}^{+1.8}$ \\ [1ex]
\hline\hline %inserts single line
\end{tabular}
\label{tab:results} % is used to refer this table in the text
\end{table}

As revealed in Table~\ref{tab:results}, the agreement between the derived
system and spot parameters of $\kappa^1$ Ceti is excellent with marginal
differences between the estimates. The residuals of the fits in 
Figures~\ref{fig:most_2003}, \ref{fig:most_2004} \& \ref{fig:most_2005}
closely match those of \citet{walker:2007}. This therefore shows that
\macula\ coupled with \multi\ is capable of matching the results of StarSpotz.

Some remaining anomalies in the residuals are evident and one may be tempted to 
input more spots to fit these out. However, \citet{walker:2007} specifically 
caution against such a process arguing that the anomalies correlate to 
moon-light contamination and other instrumental effects.

Although it is not the focus of this work to explore the inter-parameter
correlations and optimal fitting strategies, we here briefly comment on this
issue. Our fits reveal the strongest correlations between $\alpha$ values
associated with the same data set i.e. the amplitudes of the signal
components are correlated. We also find that the equatorial period, the
stellar inclination and the individual latitudes exhibit mutual correlations,
resulting from the uncertainty in the differential rotation determination.

%%% 2003 MOST DATA

%%% 2004 MOST DATA

%%% 2005 MOST DATA

\section{Discussion}
\label{sec:discussion}

\subsection{Performance}

To test the speed of \macula, we generated 1000 time stamps of a single 
synthetic input data set for a random choice of the star's parameters. In all 
cases, full non-linear limb darkening is employed. We generate a single spot 
with random parameters and call \macula\ 100,000 times to evaluate the typical 
execution time. Every call inputted random star and spot parameters in order to 
obtain a reliable average execution time. All simulations are run on a 
single-thread of a Intel Core i7 2.9\,GHz processor with \macula\ compiled in 
g95 using the optimisation flag -O3.

When \macula\ is called, one may instruct the code whether to compute the 
partial derivatives. With derivatives turned off, \macula\ requires 0.59\,$\mu$s 
per data point. Turning derivatives on yields 6.09\,$\mu$s per data point. 
Therefore, the act of turning on derivatives leads to a slowing down of the code 
by a factor of $\simeq10.3$, for a single-spot model. Note that these times 
include the small overhead of generating random system parameters too.

Increasing the spot-number, we find that the no-derivatives call
scales linearly with $N_S$. However, the derivatives call
exhibits super-linear, yet sub-quadratic, scaling of $N_S^{1.74}$,
or roughly $N_S^{7/4}$. For this scaling, doubling the number of 
spots increases the CPU time by a factor of 3.34.

Due to its analytic nature, \macula\ performs significantly faster than 
numerical codes made available in the literature. For example, 
\citet{boisse:2012} presented their numerical algorithm SOAP and report that 
generating 10,000 time stamps of a single synthetic starspot requires less than 
40\,s (but presumably close to this value). This indicates SOAP requires 
$\sim$4\,ms per data point, compared to \macula\ which requires 0.6\,$\mu$s per 
data point i.e. \macula\ is around 6800 times faster than SOAP. There are
several points which make a direct comparison somewhat unfair though. \macula\
does not compute radial velocity variations, whereas SOAP does (although
\S\ref{sub:apps} shows how radial velocity variations are easily generated
from the outputs of \macula). Further, the authors used a slower 2.33\,GHz
Intel Core Duo processor for their benchmark tests. Nevertheless, it is
clear that the difference in computation speeds is three orders-of-magnitude,
making \macula\ a powerful tool in inverse-problems.

\subsection{Benefits}

The analytic model for starspots presented here has several advantages which
we list here:

\begin{itemize}
\item[{\tiny$\blacksquare$}] An analytic algorithm for modelling photometric 
rotational modulation due to multiple circular, grey starspots, performing
three orders-of-magnitude faster than comparable numerical codes.
\item[{\tiny$\blacksquare$}] Reproduces light curves with a maximum model error
an order-of-magnitude less than that of the previous \citet{budding:1977} and
\citet{dorren:1987} for a Sun-like non-linear limb darkened star observed in
the \emph{Kepler} bandpass, for spots of angular size $\lesssim 10^{\circ}$.
\item[{\tiny$\blacksquare$}] Model accounts for spot contrast, non-linear limb 
darkening, differential rotation and starspot evolution.
\item[{\tiny$\blacksquare$}] Includes baseline normalisation parameters for
$M$ data sets, as well as $M$ blended light dilution factors to aid in
\emph{Kepler} analysis.
\item[{\tiny$\blacksquare$}] Computes transit depth variations (T$\delta$Vs) due 
to unocculted spots.
\item[{\tiny$\blacksquare$}] Partial derivatives of the model flux is provided
with respect to all model parameters and time, and may be turned on/off as 
desired (see Appendices~\ref{appendixD}\&\ref{appendixE} for derivations).
\item[{\tiny$\blacksquare$}] Code is freely available as a Fortran routine,
\macula, located at www.cfa.harvard.edu/$\sim$dkipping/macula.html.
\end{itemize}

\subsection{Potential Applications}
\label{sub:apps}

\subsubsection{Rotational Modulation Measurements}

We foresee several possible applications of \macula. Firstly, measuring the
rotational modulation of variable stars may be used to determine the rotation 
period, which may in turn constrain the ages of stars with gyrochronology 
\citep{barnes:2009}. Stars monitored with high signal-to-noise continuous
photometry, such as that from \emph{Kepler}, may also reveal differential
rotation and the stellar inclination angle. An example of this type of
regression is the analysis of \citet{walker:2007} for $\kappa^1$ Ceti,
where the inclination angle derived from rotational modulation alone and an
analytic model for starspots yields a result fully consistent with the
spectroscopic $V\sin I_*$ value. Note that we also reproduce this result
using \macula\ in this work. In addition, it may be possible to measure starspot
evolution using the linear model employed by \macula.

Rotational modulation analyses using \macula\ are not limited to cool stars, 
which have been most commonly observed to exhibit such behaviour. There also 
exists evidence for spots on hot stars, such as the rapidly rotating B star HD 
174648 \citep{degroote:2011}. Indeed, \macula\ will also be applicable for bright 
spots on hot massive stars, such as those proposed by \citet{cantiello:2011}
to explain observations of late O-type and early B-type stars made by CoRoT.

\macula\ also produces predictions for the transit depth variations (T$\delta$V)
at any time stamp inputted. This may permit for the determination of rotational 
periods from T$\delta$Vs alone; highly useful for ground-based observations
lacking the continuous photometry of space-based observatories. It may also
be useful in testing whether observed T$\delta$Vs are consistent with
starspots versus some other hypothesis e.g. planetary oblateness with precession
\citep{carter:2010}.

\subsubsection{Astrophysical Detrending}

Due to the analytic nature of \macula, the code is quick to execute and
therefore many find uses in astrophysical detrending of photometry. For example,
the PDC-MAP algorithm of \emph{Kepler} is designed to remove instrumental
trends but preserve the astrophysical signal, such as rotational modulation
due to starspots. In performing a search for transits, or a detailed modelling
of a known transit, detrending this rotational modulation is required. Whilst
polynomial models or harmonic filtering may be used, an 
astrophysically-grounded model, such as \macula, offers a viable alternative due 
to its fast execution time.

\subsubsection{Radial Velocity Variations due to Starspots}

We briefly remark that \macula\ may be used to predict radial velocity
variations due to starspots via the $F$\,$F'$ method described in 
\citet{aigrain:2012}. Here, the authors propose that radial velocity
variations can be reliably predicted from flux variations ($F$) alone.
Specifically, the authors argue that the flux multiplied by its derivative in
time reveals the radial velocity variations. \macula\ returns both 
$F_{\mathrm{mod}}$ and $\partial F_{\mathrm{mod}}/\partial t$ for all input
times (derivation provided in Appendix~\ref{appendixF}).

\subsubsection{Correcting Transmission Spectra}

If a planet transits across a star, the atmosphere of the planet can reveal
chromatic variations in the transit depth due to molecular absorption. In this
way, transit measurements can reveal the ``transmission spectrum'' of an
exoplanet. If the host star has unocculted starspots, one would expect to
see chromatic variations in the depths purely from spots too, introducing a
confounding signal. Further, transit depth measurements are often scattered
both in wavelength and in time meaning that rotational modulation can also
introduce spurious depth variations. Correcting for starspots is therefore
a major challenge in studying the atmospheres of extrasolar planets. For
example, \citet{desert:2009} found it necessary to use rotational modulation
data of HD 189733 in order to correct \emph{Spitzer} measurements of the
planet's transit depth.

\macula\ offers a self-consistent way of modelling such corrections. 
Observations of rotational modulation may be used to directly infer T$\delta$Vs, 
provided the data are in the same bandpass as that used for the transit 
measurements. If the bandpasses differ (which is practically speaking likely),
then one can estimate the depth variations by assuming a model for the
spectral radiance of the starspots (e.g. blackbody). Nevertheless, we point
out that obtaining several spectra within a few rotation periods of the
intended transit measurement would be the most ideal way to correct for
such activity, obviating the need for spectral radiance modelling.

\subsubsection{Exomoon False-Positive Vetting}

Whilst we leave the issue of modelling planet-spot crossings to future work,
it is worth noting that such events may resemble exomoon mutual transits and are
anticipated to be a source of false-positives in the ``Hunt for Exomoon with
Kepler'' (HEK) project \citep{hek:2012}. Even without detailed spot-crossing
models, \macula\ offers some simple tests to compare these two competing
hypotheses. Firstly, the starspot coverage can be estimated from the 
out-of-transit variability, allowing one to gauge the feasibility of an 
observed anomaly being a spot-crossing event. Secondly, the derived rotation 
period from the rotational modulations can be used to check whether the light 
curve anomalies are consistent or inconsistent with such a period. Finally, 
\citet{sanchis:2012} (see Figure~1) have recently shown that the phase of a 
transit mid-time with respect to the rotational modulations 
($\Phi_{\mathrm{tra}}$) is related to the phase of a starspot with respect to 
the transit mid-time ($\Phi_{\mathrm{spot}}$). Any variations not matching this 
relationship would be difficult to explain as being due to a starspot.

\section*{Acknowledgments}

We are grateful to Pieter Degroote for a very helpful review of our work.
Special thanks to Bryce Croll \& Gordon Walker for providing us with the 
\emph{MOST} data of $\kappa^1$ Ceti. Thanks to Jonathan Irwin, Roberto 
Sanchis-Ojeda amd Joel Hartman for very helpful conversations in preparing this 
work. DMK is funded by the NASA Carl Sagan Fellowships. This research made use 
of the Michael Dodds Computing Facility, courtesy of the HEK project.

%%%%%%%%%%%%%%%%%%%%%%%%%%%%%%%%%%%%%%%%%%%%%%%%%%%%%%%%%%%%%%%%%%%%%%%%%%%%%%%%
%%%%%%%%%%%%%%%%%%%%%%%%%%%%%%%%%%%%%%%%%%%%%%%%%%%%%%%%%%%%%%%%%%%%%%%%%%%%%%%%
%%%%%%%%%%%%%%%%%%%%%%%%%%%%%%%%%%%%%%%%%%%%%%%%%%%%%%%%%%%%%%%%%%%%%%%%%%%%%%%%
%%%%%%%%%%%%%%%%%%%%%%%%%%%%%%%%%%%%%%%%%%%%%%%%%%%%%%%%%%%%%%%%%%%%%%%%%%%%%%%%
%%%%%%%%%%%%%%%%%%%%%%%%%%%%%%%%%%%%%%%%%%%%%%%%%%%%%%%%%%%%%%%%%%%%%%%%%%%%%%%%
%%%%%%%%%%%%%%%%%%%%%%%%%%%%%%%%%%%%%%%%%%%%%%%%%%%%%%%%%%%%%%%%%%%%%%%%%%%%%%%%
%%%%%%%%%%%%%%%%%%%%%%%%%%%%%%%%%%%%%%%%%%%%%%%%%%%%%%%%%%%%%%%%%%%%%%%%%%%%%%%%
%%%%%%%%%%%%%%%%%%%%%%%%%%%%%%%%%%%%%%%%%%%%%%%%%%%%%%%%%%%%%%%%%%%%%%%%%%%%%%%%
%%%%%%%%%%%%%%%%%%%%%%%%%%%%%%%%%%%%%%%%%%%%%%%%%%%%%%%%%%%%%%%%%%%%%%%%%%%%%%%%

\clearpage
\appendix

\section{Geometry of the Spot}
\label{appendixA}

\subsection{Position of the Spot}

The basic unit of our model is a circle, which represents a starspot, on the 
canvas of a star's surface. We begin by considering a single starspot and
show later how the result is generalised to multiple spots. We will assume that 
the star is perfectly spherical in what follows. The geometry of the spot is 
characterised by a position and a size. The position, which we define as the 
position of the starspot's centre relative to the centre of the star, must be a 
two-dimensional vector given that a surface has a two-dimensional topology. An 
appropriate positional vector would be longitude ($\Lambda$) and latitude 
($\Phi$). We define these terms to exist in the range $-\pi<\Lambda<\pi$ and 
$-\pi/2<\Phi<\pi/2$.

We initially consider the centre of the spot to be located in a Cartesian
frame at a location given by the unit vector $\hat{\mathbf{k}} = \{0,0,1\}^T$ 
(where we adopt units of the stellar radius). In all frames of reference, we
consider the observer to be located along the $\hat{z}$-axis at $z=+\infty$.

The centre of the spot can be described at any longitude and/or latitude by 
multiplying the unit vector $\hat{\mathbf{k}}$ by two rotation matrices,
accounting for longitude and latitude. At this stage, we denote the
longitude and latitude using the notation $\tilde{\Lambda}$ and $\tilde{\Phi}$
respectively, which we dub ``apparent longitude'' and ``apparent latitude''. 
This is done in order to reserve the symbols $\Lambda$ and $\Phi$ (the true 
longitude and latitude) for later when we will account for stellar inclination 
as well. The rotation matrices for apparent longitude and latitude are defined 
by the notation $\textbf{M}_{\tilde{\Lambda}}$ then $\textbf{M}_{\tilde{\Phi}}$,
respectively.

Consider that the action of these two rotation matrices leads to a
position for the centre of the spot defined by the vector 
$\mathbf{R}_{\mathrm{centre}}$. Due to the non-commutative nature of 
linear algebra, the order in which one chooses to perform these rotations will 
affect the results. Here we follow historical precedent and define:

\begin{align}
\mathbf{R}_{\mathrm{centre}} &= \mathbf{M}_{\tilde{\Lambda}} \mathbf{M}_{\tilde{\Phi}} \hat{\mathbf{k}}, \\
\mathbf{R}_{\mathrm{centre}} &= \mathbf{M}_{\tilde{\Lambda},\tilde{\Phi}} \hat{\mathbf{k}},
\end{align}

where we have

% NOTE - ALL ROTATION MATRICES ARE CLOCKWISE, EXCEPT WHERE STATED.
% LONGITUDE IS AN EXAMPLE OF AN EXCEPTION, SINCE WE DO ANTICLOCKWISE.
% THIS IS DONE TO GIVE CORRECT X,Y DEPENDENCY EXPECTED.

\begin{eqnarray}
 \mathbf{M}_{\tilde{\Lambda}} = 
  \begin{bmatrix}
    \cos\tilde{\Lambda} & 0 & \sin\tilde{\Lambda} \\
    0 & 1 & 0 \\
    -\sin\tilde{\Lambda} & 0 & \cos\tilde{\Lambda}
  \end{bmatrix},
\label{eqn:MLambda}
\end{eqnarray}

\begin{eqnarray}
 \mathbf{M}_{\tilde{\Phi}} = 
  \begin{bmatrix}
    1 & 0 & 0 \\
    0 & \cos\tilde{\Phi} & \sin\tilde{\Phi} \\
    0 &-\sin\tilde{\Phi} & \cos\tilde{\Phi}
  \end{bmatrix}.
\label{eqn:MPhi}
\end{eqnarray}

One may combine the two matrices into a general transformation matrix, 
$\mathbf{M}_{\tilde{\Lambda},\tilde{\Phi}}$, given by

\begin{eqnarray}
 \mathbf{M}_{\tilde{\Lambda},\tilde{\Phi}} = 
  \begin{bmatrix}
    \cos\tilde{\Lambda} & -\sin\tilde{\Lambda}\sin\tilde{\Phi} & \sin\tilde{\Lambda}\cos\tilde{\Phi} \\
    0 & \cos\tilde{\Phi} & \sin\tilde{\Phi} \\
    -\sin\tilde{\Lambda} & -\cos\tilde{\Lambda}\sin\tilde{\Phi} & \cos\tilde{\Lambda}\cos\tilde{\Phi}
  \end{bmatrix}.
\label{eqn:M}
\end{eqnarray}

We use this matrix to determine

\begin{eqnarray}
 \mathbf{R}_{\mathrm{centre}} = 
  \begin{bmatrix}
    x_{\mathrm{centre}} \\
    y_{\mathrm{centre}} \\
    z_{\mathrm{centre}}
  \end{bmatrix} =
  \begin{bmatrix}
    \sin\tilde{\Lambda}\cos\tilde{\Phi} \\
    \sin\tilde{\Phi} \\
    \cos\tilde{\Lambda}\cos\tilde{\Phi}
  \end{bmatrix}.
\label{eqn:spotposition}
\end{eqnarray}

\subsection{Size of the Spot}

We wish to define the radius of the spot in terms of a solid angle swept out
from the centre of the star. Let us define this solid angle to be given by
$\alpha$. A spot of solid angle radius $\pi/2$ radians would reach from 
pole-to-pole and thus we define $0<\alpha<\pi/2$. Later when we account for
limb darkening effects (\S\ref{sub:LDspots}), we show that it is necessary to 
assume $0<\alpha<\pi/4$ and this should be interpreted as the hard-limit of our 
model, \macula.

For a starspot with a position vector described by
$\mathbf{R}_{\mathrm{centre}} = \hat{\mathbf{k}}$, it is trivial to show 
that the apparent radius of the spot would be $\sin\alpha$. In this frame, the 
spot appears as a perfect circle on the $X$-$Y$ plane.

\subsection{Rim of the Spot}

We define the rim of the spot to be those points which lie along the 
two-dimensional projected perimeter of the starspot, when viewed along a vector 
normal to the stellar surface and passing through centre of the starspot (i.e. 
when $\mathbf{R}_{\mathrm{centre}} = \hat{\mathbf{k}}$). After applying 
the rotation matrices to account a starspot's apparent longitude and/or 
latitude, the position vectors describing the loci of points along the rim are 
transformed too. 

Let us define the position vector of the loci of points along the rim, when
viewed in the frame such that $\mathbf{R}_{\mathrm{centre}} = \hat{\mathbf{k}}$,
by the vector $\mathbf{R}_{\mathrm{rim}}'$. After accounting for the spot's
apparent longitude and latitude, we use the vector $\mathbf{R}_{\mathrm{rim}}$.

For $\mathbf{R}_{\mathrm{rim}}'$, the loci of points may be described using 
parametric equations, taking advantage of the fact the projection of the spot is 
a perfect circle (as described in the previous subsection).

\begin{eqnarray}
 \mathbf{R}_{\mathrm{rim}}' = 
  \begin{bmatrix}
    x_{\mathrm{rim}}' \\
    y_{\mathrm{rim}}' \\
    z_{\mathrm{rim}}'
  \end{bmatrix} =
  \begin{bmatrix}
    \sin\alpha\cos\varphi \\
    \sin\alpha\sin\varphi \\
    \cos\alpha
  \end{bmatrix},
\label{eqn:untransformedrim}
\end{eqnarray}

where $0<\varphi<2\pi$ traces the loci of all points along the starspot rim. We 
may now apply the rotation matrix $\mathbf{M}_{\tilde{\Lambda},\tilde{\Phi}}$ to 
find the parametric expressions describing the rim for any apparent longitude or 
latitude, thereby accounting for the fore-shortening effect.

\begin{align}
\mathbf{R}_{\mathrm{rim}} &= \mathbf{M}_{\tilde{\Lambda},\tilde{\Phi}} \mathbf{R}_{\mathrm{rim}}',
\end{align}

which may be shown to yield

\begin{align}
x_{\mathrm{rim}} &= \sin\alpha\cos\tilde{\Lambda}\cos\varphi + \sin\tilde{\Lambda} (\cos\alpha\cos\tilde{\Phi}-\sin\alpha\sin\tilde{\Phi}\sin\varphi), \\
y_{\mathrm{rim}} &= \sin\alpha\cos\tilde{\Phi}\sin\varphi + \cos\alpha\sin\tilde{\Phi}, \\
z_{\mathrm{rim}} &= \cos\alpha\cos\tilde{\Lambda}\cos\tilde{\Phi} - \sin\alpha(\sin\tilde{\Lambda}\cos\varphi + \cos\tilde{\Lambda}\sin\tilde{\Phi}\sin\varphi).
\end{align}

\subsection{Bulge of the Spot}

Consider again the frame in which one views the spot down the vector normal to
the stellar surface and passing through the spot's centre. In the model
described in this work, the spot lives in three-dimensions in a Cartesian
framework. Notably, the spot exhibits a bulge due to the curvature of the
stellar surface. From the perspective of the star's centre, the
loci of the points on this bulge can be described by two angles; a radial
angle, $\theta$, and an azimuthal angle, $\nu$. We may define these loci
by again starting from the frame in which
$\mathbf{R}_{\mathrm{centre}} = \hat{\mathbf{k}}$, and applying rotation 
matrices appropriately. In this simple frame, we define the position vector for 
the loci of the points existing on the bulge as

\begin{align}
\mathbf{R}_{\mathrm{bulge}}' &= \mathbf{M}_{\nu} \mathbf{M}_{\theta} \hat{\mathbf{k}},
\end{align}

where we have

\begin{eqnarray}
 \mathbf{M}_{\theta} = 
  \begin{bmatrix}
    1 & 0 & 0 \\
    0 & \cos\theta & \sin\theta \\
    0 & -\sin\theta & \cos\theta
  \end{bmatrix},
\label{eqn:Mtheta}
\end{eqnarray}

\begin{eqnarray}
 \mathbf{M}_{\nu} = 
  \begin{bmatrix}
    \cos\nu & \sin\nu & 0 \\
    -\sin\nu & \cos\nu & 0 \\
    0 & 0 & 1
  \end{bmatrix}.
\label{eqn:Mnu}
\end{eqnarray}

Here the radial angle, $\theta$, is bound to be $-\alpha<\theta\leq\alpha$ i.e. 
it cannot subtend an angle greater than the solid angle radius of the spot.
The azimuthal angle has the freedom to be $-\pi<\nu<\pi$. We use these matrices 
to determine:

\begin{eqnarray}
 \mathbf{R}_{\mathrm{bulge}}' = 
  \begin{bmatrix}
    x_{\mathrm{bulge}}' \\
    y_{\mathrm{bulge}}' \\
    z_{\mathrm{bulge}}'
  \end{bmatrix} =
  \begin{bmatrix}
    \sin\theta \sin\nu \\
    \sin\theta \cos\nu \\
    \cos\theta
  \end{bmatrix}.
\label{eqn:untransformedbulge}
\end{eqnarray}

Note the dash, which (as before) is used to denote that this is derived in
the frame not accounting for a spot's apparent longitude and/or latitude. As was 
done earlier, we may now apply the longitude-latitude rotation matrix, 
$\mathbf{M}_{\tilde{\Lambda},\tilde{\Phi}}$, to account for any orientation 
desired:

\begin{align}
\mathbf{R}_{\mathrm{bulge}} &= \mathbf{M}_{\tilde{\Lambda},\tilde{\Phi}} \mathbf{R}_{\mathrm{bulge}}',
\end{align}

which gives

\begin{align}
x_{\mathrm{bulge}} &= \sin\tilde{\Lambda}\cos\tilde{\Phi}\cos\theta + \sin\theta(\cos\tilde{\Lambda}\sin\nu-\sin\tilde{\Lambda}\sin\tilde{\Phi}\cos\nu), \\
y_{\mathrm{bulge}} &= \cos\tilde{\Phi}\sin\theta\cos\nu + \sin\tilde{\Phi}\cos\theta, \\
z_{\mathrm{bulge}} &= \cos\tilde{\Lambda}\cos\tilde{\Phi}\cos\theta - \sin\theta(\sin\tilde{\Lambda}\sin\nu+\cos\tilde{\Lambda}\sin\Phi\cos\nu).
\end{align}

\subsection{A Useful Simplification}

Due to the circular symmetry of the problem, it is actually degenerate to use
two angles to describe the position of the spot. All that matters is how
close to the spot is to the limb, regardless as to the combination of longitude
and latitude responsible. For this reason, we may define any combination of
these terms using a single ``auxiliary angle'' we dub $\beta$ (in-keeping
with the notation of \citealt{dorren:1987}).

The angle of interest is the angle subtended between the vector 
$\hat{\mathbf{k}}$ (pointing towards the observer) and the vector describing the
position of the spot's centre relative to the centre of the star 
$\mathbf{R}_{\mathrm{centre}}$. Let us define this as the auxiliary angle 
$\beta$. $\beta$ can be found by using the dot-product rule of these two
relevant vectors:

\begin{align}
\mathbf{R}_{\mathrm{centre}} \cdot \hat{\mathbf{k}} &= |\mathbf{R}_{\mathrm{centre}}| |\hat{\mathbf{k}}| \cos\beta
\end{align}

Since $\hat{\mathbf{k}}$ is a unit-vector in the $\hat{Z}$-direction, then
this dot-product simply extracts the $\hat{Z}$-component of 
$\mathbf{R}_{\mathrm{centre}}$. Therefore we have:

\begin{align}
\beta &= \cos^{-1}[ z_{\mathrm{centre}} ],
\label{eqn:beta}
\end{align}

which may be evaluated here to be

\begin{align}
\beta &= \cos^{-1}[\cos{\tilde{\Lambda}}\cos{\tilde{\Phi}}].
\end{align}

$\beta$ may also be thought of as being like a net longitude shift at zero 
latitude i.e. ${\Phi}\rightarrow 0$ and ${\Lambda}\rightarrow \beta$.

Due to the mirror symmetry of the problem, we only need consider $0<\beta<\pi$
to derive all possible scenarios. The vectors of interest now become, without
any loss of generality,

\begin{eqnarray}
 \mathbf{R}_{\mathrm{centre}} = 
  \begin{bmatrix}
    x_{\mathrm{centre}} \\
    y_{\mathrm{centre}} \\
    z_{\mathrm{centre}}
  \end{bmatrix} =
  \begin{bmatrix}
    \sin\beta \\
    0 \\
    \cos\beta
  \end{bmatrix},
\label{eqn:etacentre}
\end{eqnarray}

\begin{eqnarray}
 \mathbf{R}_{\mathrm{rim}} = 
  \begin{bmatrix}
    x_{\mathrm{rim}} \\
    y_{\mathrm{rim}} \\
    z_{\mathrm{rim}}
  \end{bmatrix} =
  \begin{bmatrix}
    \sin\alpha\cos\beta\cos\varphi + \cos\alpha \sin\beta \\
    \sin\alpha\sin\varphi \\
    \cos\alpha\cos\beta - \sin\alpha\sin\beta\cos\varphi
  \end{bmatrix},
\label{eqn:etarim}
\end{eqnarray}

\begin{eqnarray}
 \mathbf{R}_{\mathrm{bulge}} = 
  \begin{bmatrix}
    x_{\mathrm{bulge}} \\
    y_{\mathrm{bulge}} \\
    z_{\mathrm{bulge}}
  \end{bmatrix} =
  \begin{bmatrix}
    \sin\beta\cos\theta + \cos\beta\sin\theta\sin\nu \\
    \cos\nu\sin\theta \\
    \cos\theta\cos\beta - \sin\theta\sin\beta\sin\nu
  \end{bmatrix}.
\label{eqn:etabulge}
\end{eqnarray}

\section{Four Cases}
\label{appendixB}

\subsection{Overview}

\subsubsection{Case I}

In order to compute the flux from a starspot, we need to compute the projected
area in the $X$-$Y$ plane. It can be easily seen that four distinct cases exist 
for the geometry of the rim and bulge. The most obvious case is the dominant 
source of flux variations since the spot is nearly face-on. For $\beta$ between 
0 and some angle close to the limb of the star, the loci of points on the bulge 
lie fully inside the rim of the starspot, as seen in the projected $X$-$Y$ 
plane. This case is trivial to model and the rim expressions may be used alone 
to compute the area of the starspot. Case I is valid for 
$0<\beta<\beta_{\mathrm{crit}}$ where we are yet to define 
$\beta_{\mathrm{crit}}$ but it can be understood to be angle close
to the limb of the star.

\subsubsection{Case II}

Case II occurs as $\beta$ approaches $\pi/2$ from $0$. It is defined as when the
loci of points on the bulge are no longer contained within the projected rim of
the starspot. Since the bulge has a $Z$-component, as we rotate round in
longitude, this $Z$-component will be transferred into an ever-increasing
$X$-component. Eventually, this $X$-component exceeds the rim's maximal 
$X$-value at which point ``the bulge pokes out of the rim''. Case II is valid 
for $\beta_{\mathrm{crit}}<\beta<\pi/2$.

\subsubsection{Case III}

Case III occurs as $\beta$ increases beyond $\pi/2$ i.e. the centre of the spot
is behind the star. However, a portion of the starspot is still in view and
causes a flux decrement. It can be easily understood that once $\beta$
exceeds $\pi/2 + \alpha$ then the spot has fully disappeared behind the back
of the star. Thus, case III is valid for $\pi/2<\beta<\pi/2+\alpha$.

\subsubsection{Case IV}

Case IV is simply the case of the spot fully behind the star and thus there is
no contribution to the model flux. This is valid for $\pi/2+\alpha<\beta<\pi$
(recalling that $\beta$ is defined only within the range $0<\beta<\pi$ due to 
the mirror symmetry of the problem).

\subsection{Case II}

\subsubsection{Optimal Bulge Curve}

We here devote a section to case II alone, due to the non-trivial nature of
solving for its parametric equations. We first start by defining the ``optimal
bulge curve''. For any $Y$ co-ordinate of a point lying within the bulge, the
optimal bulge curve is the corresponding $X$ co-ordinate which maximises $X$.
It is the curve which seems to extend furthest to the limb of the star, as
seen in the transformed frame. Since all loci on the 2D surface of the bulge
are defined by two parametric terms ($\theta$ and $\nu$), it should be clear
that the parametric equation of the optimal bulge curve will require only
one term; either $\theta$ or $\nu$, but not both. We arbitrarily choose here to 
define our optimal bulge curve purely in terms of $\nu$.

The optimal bulge curve also exhibits the greatest separation from 
$\{X,Y\}=\{0,0\}$, relative to all other loci on the bulge. Thus we expect that
$x_{\mathrm{bulge}}^2+y_{\mathrm{bulge}}^2$ is maximised and so:

\begin{align}
\frac{\partial(x_{\mathrm{bulge}}^2+y_{\mathrm{bulge}}^2)}{\partial\theta} = 0.
\end{align}

Solving the above for $\cos\theta$ yields two solutions, only one of which is
the maximum:

\begin{align}
\cos\theta_{\mathrm{optimal}} &= \frac{2 \sin\beta\sin\nu }{\sqrt{3+\cos2\beta-2\cos2\nu\sin^2\beta}},
\end{align}

where we only consider the range $0<\nu<\pi/2$ and $0<\beta<\pi/2$ here (the
latter due to the case II conditions and the former due to symmetry about the
$X$-axis). This yields the following parametric expression in the $X$-$Y$
plane:

\begin{align}
x_{\mathrm{optimum}} &= \frac{ 2\sin\nu }{ \sqrt{3+\cos2\beta-2\cos2\nu\sin^2\beta} }, \\
y_{\mathrm{optimum}} &= \frac{ 2\cos\nu\cos\beta }{ \sqrt{3+\cos2\beta-2\cos2\nu\sin^2\beta} }.
\end{align}

Evaluating the equation for $x_{\mathrm{optimum}}$ at $\nu=0$ reveals 
$x_{\mathrm{optimum}}=0$. Thus, when $\nu=0$ the optimal bulge 
curve intersects the $x$-axis, although we note that at this point the
corresponding $\theta_{\mathrm{optimum}}$ point may be exceed $\alpha$ and thus
may not truly exist on the bulge. However, it reveals that $x$ increases as
$\nu$ increases form 0 to $\pi/2$.

\subsubsection{Intersection of Optimal Bulge Curve and the Rim}

For case II, where $\beta_{\mathrm{crit}}<\beta<\pi/2$, there exists a certain 
point where the optimal bulge curve intersects the starspot rim. We denote this 
location as $\{x_{\mathrm{intersection}},y_{\mathrm{intersection}}\}$. The 
location corresponds to a unique parametric location along the rim, 
$\varphi_{\mathrm{intersection}}$. Similarly, there exists a unique parametric
location along the optimal bulge curve, $\nu_{\mathrm{intersection}}$.

Let us deal with $\varphi_{\mathrm{intersection}}$ first. Since the optimal 
bulge curve extends outside the rim, this location can be shown to occur when
the rim's $X$-$Y$ distance from the origin is maximised i.e. when
$x_{\mathrm{rim}}^2+y_{\mathrm{rim}}^2$ is maximised. We therefore must solve
the following expression for $\nu$:

\begin{align}
\frac{\partial(x_{\mathrm{rim}}^2+y_{\mathrm{rim}}^2)}{\partial \varphi} = 0,
\end{align}

which may be shown to yield:

\begin{align}
\cos \varphi_{\mathrm{intersection}} &= \cot\alpha\cot\beta
\end{align}

The intersection point along the optimal bulge curve can be found by minimising
the distance on the $X$-$Y$ plane between the optimal bulge curve and the rim.
Therefore, we must solve the following expression for $\nu$:

\begin{align}
\frac{\partial}{\partial\nu} \Big(& [x_{\mathrm{optimal}} - x_{\mathrm{rim}}(\varphi=\varphi_{\mathrm{intersection}})]^2 \nonumber\\
\qquad& + [y_{\mathrm{optimal}} - y_{\mathrm{rim}}(\varphi=\varphi_{\mathrm{intersection}})]^2 \Big) = 0,
\end{align}

which yields the following solution:

\begin{align}
\cos^2\nu_{\mathrm{intersection}} &= 1 - \cot^2\alpha\cot^2\beta.
\end{align}

Feeding this back into the expressions for the optimal bulge curve, we locate
the Cartesian co-ordinates of the intersection point:

\begin{align}
x_{\mathrm{intersection}} &= \cos\alpha\csc\beta, \\
y_{\mathrm{intersection}} &= \sin\alpha\sqrt{1-\cot^2\alpha\cot^2\beta},
\end{align}

where it is again understood this is for the range $0<\nu<\pi/2$ only.

\subsubsection{The Critical Angle, $\beta_{\mathrm{crit}}$}

As discussed earlier, $x_{\mathrm{optimal}}=0$ for $\nu=0$ and increases up to
a maximum at $\nu=\pi/2$. Similarly, by definition the parametric expression
for $x_{\mathrm{rim}}$ is maximised for $\varphi=0$. Case II is only valid for a 
bulge which pokes out of the rim and its boundary will occur for 
$x_{\mathrm{optimal}}(\nu=\pi/2)=x_{\mathrm{rim}}(\varphi=0)$. Solving for 
$\beta$, we find:

\begin{align}
\cos\beta_{\mathrm{crit}} &= \sin\alpha, \nonumber \\
\beta_{\mathrm{crit}} &= \pi/2 - \alpha.
\end{align}

This therefore proves an intuitive point. The optimal bulge pokes out of the rim
when the rim hits the edge of the star. At this instant, the starspot rim makes 
contact with the projected rim of the star and the optimal bulge is just a 
single point at $\{x,y\} = \{1,0\}$. As $\beta$ becomes larger, the starspot rim
gradually disappears behind the back of the star and the optimal bulge curve
spreads out along the projected rim of the star.

\subsubsection{Projected Area of the Starspot: I. The Rim}

The projected area of the starspot, for case II, can be thought of as the sum
of the projected area bound by the rim and that of the bulge poking out, with
the transition occurring at the intersection points derived. The rim therefore
bounds an area between $\varphi_{\mathrm{intersection}}<\varphi<(2\pi-
\varphi_{\mathrm{intersection}})$. We consider here the area above the $x$-axis 
only, which can later be simply doubled due to symmetry about the $x$-axis.

We start by re-writing the expression for $x_{\mathrm{rim}}(\varphi)$ to make 
$\varphi$ the subject:

\begin{align}
\varphi(x_{\mathrm{rim}}) &= \cos^{-1}\Big(\csc\alpha\sec\beta (x_{\mathrm{rim}}-\cos\alpha\sin\beta)\Big)
\end{align}

We may now replace the $\varphi$ in $y_{\mathrm{rim}}(\varphi)$ to obtain 
$y_{\mathrm{rim}}(x_{\mathrm{rim}})$:

\begin{align}
y_{\mathrm{rim}}(x_{\mathrm{rim}}) &= \sin\alpha\sqrt{1-(x_{\mathrm{rim}}\csc\alpha\sec\beta - \cot\alpha\tan\beta)^2}
\end{align}

The area bounded by the rim is therefore given by:

\begin{align}
A_{\mathrm{rim}} &= 2 \int_{x_{\mathrm{rim}}(\varphi=\pi)}^{x_{\mathrm{intersection}}} y_{\mathrm{rim}}(x_{\mathrm{rim}})\,\mathrm{d}x_{\mathrm{rim}}, \\
A_{\mathrm{rim}} &= \frac{1}{\sqrt{2}}\cos\alpha\sqrt{-\cos2\alpha-\cos2\beta}\cot^2\beta \nonumber \\
\qquad& + \frac{\pi}{2}\cos^2\beta\sin^2\alpha + 
\sin^{-1}\Big[\cot\alpha\cot\beta\Big]\cos\beta\sin^2\alpha.
\end{align}

\subsubsection{Projected Area of the Starspot: II. The Bulge}

We now need to repeat this process for the optimal bulge curve. One may
re-write the expression for $x_{\mathrm{optimal}}(\nu)$ making $\nu$ the
subject:

\begin{align}
\nu(x_{\mathrm{optimal}}) &= \cos^{-1}\Bigg[\frac{\sqrt{2}\sqrt{1-x_{\mathrm{optimal}}^2}}{2-x_{\mathrm{optimal}}+x_{\mathrm{optimal}} \cos2\beta}\Bigg].
\end{align}

Feeding this into the expression for $y_{\mathrm{optimal}}(\nu)$ in order to
obtain $y_{\mathrm{optimal}}(x_{\mathrm{optimal}})$ we obtain the simple
solution:

\begin{align}
y_{\mathrm{optimal}} &= \sqrt{1-x_{\mathrm{optimal}}^2}.
\end{align}

Once again, this result proves the same inuitive result we saw earlier.
Specifically, the optimal bulge curve lies along the projected rim of the star 
itself. The bounded area is given by:

\begin{align}
A_{\mathrm{optimal}} &= 2 \int_{x_{\mathrm{intersection}}}^{x_{\mathrm{optimal}}(\nu=\pi/2)} y_{\mathrm{optimal}}(x_{\mathrm{optimal}})\,\mathrm{d}x_{\mathrm{optimal}}, \\
A_{\mathrm{optimal}} &= \cos^{-1}[\cos\alpha\csc\beta] \nonumber \\
\qquad& - \frac{\cos\alpha}{2} \Bigg[ \sqrt{2}\cot^2\beta \sqrt{-\cos2\alpha-\cos2\beta} \nonumber \\
\qquad& + 2 \sin\alpha\tan\beta\sqrt{-\cot^2\alpha\cot^2\beta + \cos^2\beta\csc^2\alpha} \Bigg].
\end{align}

\subsubsection{Projected Area of the Starspot: Total}

Combining these two results together, we obtain the area of a circular starspot
under case II conditions:

\begin{align}
A_{II}(\alpha,\beta) &= \cos^{-1}[\cos\alpha\csc\beta] \nonumber \\
\qquad& + \sin\alpha \Bigg[ \cos\beta\sin\alpha(\pi-\cos^{-1}[\cot\alpha\cot\beta]) \\
\qquad& - \cos\alpha\tan\beta\sqrt{-\cot^2\alpha\cot^2\beta + \cos^2\beta\csc^2\alpha} \Bigg].
\end{align}

\subsection{Case III}
\subsubsection{Edge Bulge Curve}

For case III, we have $\pi/2<\beta<\pi/2+\alpha$. Here, the centre of the spot
is out-of-view, hidden behind the star. Despite this, a portion of the spot's
surface remains at $z>0$ and thus is still visible. For case II, we defined
an optimal bulge curve which tracked the curve of interest. Similarly, here
we define the ``edge bulge curve'' to the perimeter of the bulge still in
view when case III conditions remain in effect.

The edge bulge curve is much easier to define that the optimal bulge curve. For
the range $\pi/2<\beta<\pi/2+\alpha$, it is simply given by maximizing $\theta$.
Since $\theta$ is bound to be $0<\theta<\alpha$, then 
$\theta_{\mathrm{edge}} = \alpha$. Thus, the parametric equations describing
the edge bulge curve are:

\begin{align}
x_{\mathrm{edge}} &= x_{\mathrm{bulge}}(\theta=\alpha), \nonumber \\
\qquad&= \cos\alpha\sin\beta + \cos\beta\sin\alpha\sin\nu, \\
y_{\mathrm{edge}} &= y_{\mathrm{bulge}}(\theta=\alpha), \nonumber \\
\qquad&= \cos\nu\sin\alpha, \\
z_{\mathrm{edge}} &= z_{\mathrm{bulge}}(\theta=\alpha), \nonumber \\
\qquad&= \cos\alpha\cos\beta - \sin\alpha\sin\beta\sin\nu.
\end{align}

\subsubsection{Boundary of the Edge Bulge Curve}

The edge bulge curve intersects the stellar rim when the quadrature sum of
the $X$ and $Y$ components equals unity. Therefore, we may find the
$\nu$ value of this location, which we dub $\nu_{\mathrm{boundary}}$, by 
solving the following expression for $\nu$:

\begin{align}
x_{\mathrm{edge}}^2 + y_{\mathrm{edge}}^2 = 1,
\end{align}

which yields:

\begin{align}
\cos^2\nu_{\mathrm{boundary}} &= -\frac{ \cos2\alpha+\cos2\beta }{ 2\sin^2\alpha\sin^2\beta }.
\end{align}

Plugging the above into our expressions for $\mathbf{R}_{\mathrm{edge}}$ yields
$\mathbf{R}_{\mathrm{boundary}}$:

\begin{align}
x_{\mathrm{boundary}} &= \cos\alpha\csc\beta, \\
y_{\mathrm{boundary}} &= \frac{\csc\beta}{\sqrt{2}} \sqrt{-\cos2\alpha-\cos2\beta},\\
z_{\mathrm{boundary}} &= 0.
\end{align}

The right-most $X$-point occurs when we cross the $X$-axis i.e. when 
$y_{\mathrm{edge}}=0$. It is trivial to show this occurs for $\nu=\pi/2$ and
correspondingly $x_{\mathrm{edge}}(\nu=\pi/2) = \sin(\beta+\alpha)$.

\subsubsection{Area Bounded by the Edge Bulge Curve}

Taking the expression for $x_{\mathrm{edge}}(\nu)$, we may re-write this to
make $\nu$ the subject via:

\begin{align}
\nu(x_{\mathrm{edge}}) &= \sin^{-1}\Big[\csc\alpha\sec\beta (x_{\mathrm{edge}}-\cos\alpha\tan\beta)\Big].
\end{align}

We may feed this into $y_{\mathrm{edge}}(\nu)$ to obtain 
$y_{\mathrm{edge}}(x_{\mathrm{edge}})$:

\begin{align}
y_{\mathrm{edge}}(x_{\mathrm{edge}}) &= \sin\alpha\sqrt{1-(x\csc\alpha\sec\beta-\cot\alpha\tan\beta)^2}.
\end{align}

Due to the concave nature of the edge bulge curve, the area of the loci of
points on the bulge only is defined by:

\begin{align}
A_{III} &= 2 \Bigg(\int_{x_{\mathrm{boundary}}}^{1} \sqrt{1-x^2}\,\mathrm{d}x\Bigg) \nonumber \\
\qquad& - 2 \Bigg(\int_{x_{\mathrm{boundary}}}^{x_{\mathrm{edge}}(\nu=\pi/2)} y_{\mathrm{edge}}(x_{\mathrm{edge}})\,\mathrm{d}x_{\mathrm{edge}}\Bigg).
\end{align}

Finally, one may express this purely as a function of $\alpha$ and $\beta$:

\begin{align}
A_{III}(\alpha,\beta) &= \frac{\pi}{2} - \sin^{-1}\Big[\cos\alpha\csc\beta\Big] \nonumber \\
\qquad& + \cos\beta\sin^2\alpha\cos^{-1}\Big[-\cot\alpha\cot\beta\Big] \nonumber \\
\qquad& - \cos\alpha \sin\beta\sqrt{1-\cos^2\alpha\csc^2\beta}
\end{align}

\subsection{Cases I \& IV}

\subsubsection{Case I}

For completion, we here briefly derive the expressions for cases I and IV.
Case I has the spot fully in view at some angle $\beta$ where 
$0<\beta<\beta_{\mathrm{crit}}$. The relevant parametric equations are the
rim expressions derived earlier. The area may be found to be:

\begin{align}
A_{I}(\alpha,\beta) &= \frac{1}{2} \int_{\varphi=0}^{2\pi} \Big(x_{\mathrm{rim}} \frac{\partial y_{\mathrm{rim}}}{\partial \varphi} - y_{\mathrm{rim}} \frac{\partial x_{\mathrm{rim}}}{\partial \varphi}\Big)\,\mathrm{d}\varphi \nonumber \\
A_{I}(\alpha,\beta) &= \pi\sin^2\alpha\cos\beta
\end{align}

\subsubsection{Case IV}

Case IV is for $(\pi/2)+\alpha<\beta<\pi$ and corresponds to the spot fully
out-of-view behind the star. The case trivially has an area:

\begin{align}
A_{IV}(\alpha,\beta) &= 0
\end{align}

\section{Modelling the Light Curve}
\label{appendixC}

\subsection{For a Uniform Brightness Star}

\subsubsection{Generalising to a Single Domain Function}

It can be easily shown that these expressions produce the same light curve
profile predicting by \citet{dorren:1987} in the absence of limb darkening
and a black spot.

The expressions for $A_{II}$ and $A_{III}$ possess some similarities in form
and are of course continuous at the point $\beta=\pi/2$. This led us to
investigate if the two equations are equivalent to some simplified form. We
found the following expression describes both $A_{II}$ and $A_{III}$:

\begin{align}
\mathcal{A}(\alpha,\beta) &= \cos^{-1}\Big[\cos\alpha\csc\beta\Big] \nonumber \\
\qquad& + \cos\beta\sin\alpha \Xi - \cos\alpha\sin\beta \Psi,
\end{align}

where

\begin{align}
\Xi &= \sin\alpha\cos^{-1}[-\cot\alpha\cot\beta],
\Psi &= \sqrt{1-\cos^2\alpha\csc^2\beta}.
\end{align}

Encouraged by this, we tried plotting the function in the range 
$0<\beta<\beta_{\mathrm{crit}}$. However, $\mathcal{A}$ becomes complex in this
range. We therefore only considered the real part. It is easy to see by
example that the real part of $\mathcal{A}$ perfectly maps the $A_{I}$ function.

A final success of $\mathcal{A}$ comes from considering the case IV range i.e.
$(\pi/2)+\alpha<\beta<\pi$. Here the real part of $\mathcal{A}$ goes to zero
but the imaginary component gradually increases. Thus, by plotting the real
part of $\mathcal{A}$ only, we can reproduce all four cases with a single
function across the full domain of $0<\beta<\pi$. Thus, we have:

\begin{align}
A(\alpha,\beta) &= \mathbb{R}[\mathcal{A}].
\end{align}

The advantage of using this single-domain function is that we can define an 
analytic Jacobian and Hessian matrices, which are useful in expediting 
regression of photometric data. \macula\ will provide the Jacobian, but not
the Hessian to save computation time (although it may be extended to perform
this function too due its analytic form).

\subsubsection{Model Flux}

For a single-rotating spot on a uniform brightness star, the flux from a star
($F$) can be computed using:

\begin{align}
F(\alpha,\beta) &= (\pi - A) \mathcal{F}_* + A \mathcal{F}_{\mathrm{spot}},
\end{align}

where $\mathcal{F}$ denotes flux-per-unit-area and $A$ is the area of the spot.
Photometric observations are usually normalised to some arbitrary value. A
suitable choice here is the flux from the star in the absence of a starspot
i.e. $F(\alpha=0,\beta)$.

\begin{align}
F_{\mathrm{mod}} &= \frac{F(\alpha,\beta)}{F(\alpha=0,\beta)},\nonumber \\
\qquad&= 1 - \frac{A}{\pi} (1-f_{\mathrm{spot}})
\label{eqn:singleuniformspot}
\end{align}

where $f_{\mathrm{spot}}=\mathcal{F}_{\mathrm{spot}}/\mathcal{F}_*$ and is
the flux-contrast of the spot relative to the star. It may also be thought of
as a proxy for the temperature of the spot. For $N_S$ non-overlapping starspots
labelled $k=1,2,...,N_S-1,N_S$, this can be extended to:

\begin{align}
F_{\mathrm{mod}}(\boldsymbol{\alpha},\boldsymbol{\beta}) &= \frac{ F(\boldsymbol{\alpha},\boldsymbol{\beta}) }{ F(\boldsymbol{\alpha}=\mathbf{0},\boldsymbol{\beta})},
\label{eqn:normalized}
\end{align}

which yields

\begin{align}
F_{\mathrm{mod}}(\boldsymbol{\alpha},\boldsymbol{\beta}) &= 1 - \frac{1}{\pi} \sum_{k=1}^{N_S} A_k (1-f_{\mathrm{spot},k}).
\end{align}

\subsection{For a Limb Darkened Star}
\label{sub:LDspots}

\subsubsection{The Mandel-Agol Cases}

In this work, we will assume that the size of the spot is small relative to
the size of the star. This approximation allows us to easily write down
analytic functions for the light curve and follows on from the work
of \citet{mandel:2002} and \citet{luna:2011}. Specifically, we will use the
small-planet approximation from \citet{mandel:2002} and utilise 4-coefficient
non-linear limb darkening. The flux from a star in the absence of starspots
is therefore modelled via Equation~\ref{eqn:LD4} provided earlier:

\begin{align}
I_*(r) &= 1 - \sum_{n=1}^4 c_n (1-\mu^{n/2}),
\end{align}

where $c_n$ are the limb darkening coefficients, $\mu=\cos\Theta=\sqrt{1-r^2}$,
$0\leq r\leq1$ is the normalised radial coordinate on the disk of the star and
$I_*(r)$ is the specific intensity as a function of $r$, with $I_*(0)=1$.

Limb darkening is present over the entire viewable surface of the star, 
including those portions which are covered in starspots. However, due to the
different temperature and opacity of this surface, the limb darkening
coefficients cannot be assumed to be necessarily the same as that for the
rest of the stellar surface. We therefore consider the specific intensity
of the spot covered surface to be described by limb darkening coefficients
$d_1$, $d_2$, $d_3$ and $d_4$:

\begin{align}
I_{\mathrm{spot}}(r) &= f_{\mathrm{spot}} \Big( 1 - \sum_{n=1}^4 d_n (1-\mu^{n/2}) \Big).
\end{align} 

Note how we assume $f_{\mathrm{spot}}$ is not a function of position on
the star's surface or equivalently the spot's radial angle, $\theta$. In other
words each spot has a uniform temperature, although the temperature may vary
between spots. Also, \S\ref{sub:umbra} describes how umbra/penumbra may be
generated using our model allowing for a more complex profile.

For a single rotating spot, we will denote the model flux as being composed by
the following components:

\begin{align}
F(\alpha,\beta) &= F_{\mathrm{total}} - F_{\mathrm{obscured}} + F_{\mathrm{spot}}.
\end{align}

The obscured and total flux components are computed as one would do so for a
transiting planet model. In this scenario, there are four principal cases, as
shown in Table~\ref{tab:mandelspot}. The table requires we define an angle
at which point the spot no longer covers the centre of the star 
(M3-M9 boundary). When does this occur?

In order to simplify the problem, let us assume it is not 
possible for a spot to both cover the stellar centre and to exceed the angle 
$\beta_{\mathrm{crit}}$. This is equivalent to assuming $0<\alpha<\pi/4$.
For the loci of points along the rim of the starspot, the locus which is
closest to the sky-projected stellar centre has a position 
$\{x,y\}=\{\sin(\beta-\alpha),0\}$. Therefore, when $\sin(\beta-\alpha)>0$,
the spot no longer covers the stellar centre. The spot therefore no longer
covers the stellar centre once $\beta>\alpha$ and this is the critical angle
of interest required for deriving our limb darkening model.

%%%%%%%%%%%%%%%%%%%%%% MANDEL CASES TABLE %%%%%%%%%%%%%%%%%%%%%%%%%%%%%%%%%%%%%%%
%%%%%%%%%%%%%%%%%%%%%%%%%%%%%%%%%%%%%%%%%%%%%%%%%%%%%%%%%%%%%%%%%%%%%%%%%%%%%%%%%
%										%
%%% TABLE OF MANDEL CASES (OF INTEREST)						%
\begin{table*}									%
\caption{List of cases identified by \citet{mandel:2002}. We use the same case	%
classification in this work, but altering the notation.}			%
\centering % used for centering table						%
\begin{tabular}{l c c c} % centered columns (5 columns)				%
\hline\hline %inserts double horizontal lines					%
Case & Analogous Condition for a Planet & Condition for a Spot & $\beta$ Range \\ [0.5ex]%
%heading									%
\hline % inserts single horizontal line						%
M1 & $1+p<S_{P*}<\infty$ & Case IV & $\pi/2+\alpha<\beta<\pi$ \\		%
M2 & $1-p<S_{P*}<1+p$ & Cases II \& III & $\pi/2-\alpha<\beta<\pi/2+\alpha$\\	%
M3 & $p<S_{P*}<1-p$ & Case I & $\alpha<\beta<\pi/2-\alpha$ \\			%
M9 & $0<S_{P*}<p$ & Case I & $0<\beta<\alpha$  \\ [1ex]				%
\hline\hline %inserts single line						%
\end{tabular}									%
\label{tab:mandelspot} % is used to refer this table in the text		%
\end{table*}									%
%										%
%%%%%%%%%%%%%%%%%%%%%%%%%%%%%%%%%%%%%%%%%%%%%%%%%%%%%%%%%%%%%%%%%%%%%%%%%%%%%%%%%

\subsubsection{Case M3}

For case M3, it may be shown (see \citealt{luna:2011}) that:

\begin{align}
F_{\mathrm{total}} &= \int_{r=0}^1 2 r I_*(r)\,\mathrm{d}r, \nonumber \\
\qquad&= 1 - \sum_{n=1}^4 \frac{n c_n}{n+4}.
\end{align}

For the spot, we must compute the flux obscured by its presence. This can be
done by exploiting of the circular symmetry of the limb darkening effect and
integrating the flux over an annulus defined to have an inner radius equal to
the left-most point of the spot and outer radius equal to the right-most point
of the spot.

\begin{align}
F_{\mathrm{obscured},\mathrm{annulus}}^{M3} &= \int_{r=\sin(\beta-\alpha)}^{\sin(\beta+\alpha)} 2 r I_*(r)\,\mathrm{d}r, \nonumber \\
\qquad&= \sum_{n=0}^4 \Big(\frac{4 c_n}{4+n}\Big) \Big[\cos^{\frac{n+4}{2}}(\beta-\alpha) \nonumber \\
\qquad& - \cos^{\frac{n+4}{2}}(\beta+\alpha) \Big].
\end{align}

Note that the solution above has a significantly more compact form that than
acquired for an exomoon in \citet{luna:2011}. We may correct for the fact
this is the flux over the entire annulus by applying:

\begin{align}
F_{\mathrm{obscured}}^{M3} &= \frac{A}{ \pi A_{\mathrm{annulus}} } F_{\mathrm{obscured},\mathrm{annulus}}^{M3}, \nonumber \\
\qquad &= \frac{A}{ \pi[\cos^2(\beta-\alpha)-\cos^2(\beta+\alpha)] } F_{\mathrm{obscured},\mathrm{annulus}}^{M3},
\end{align}

where $A$ and $F_{\mathrm{obscured},\mathrm{annulus}}^{M3}$ have been previously
defined.

Finally, we need to compute $F_{\mathrm{spot}}$, the flux from the spot itself.
As will be the situation for all cases, the derivation for 
$F_{\mathrm{spot}}^{Mx}$ is precisely the same as that as was done for 
$F_{\mathrm{obscured}}^{Mx}$ except that $\{c_1,c_2,c_3,c_4\}\rightarrow
\{d_1,d_2,d_3,d_4\}$ and we multiply the expression by $f_{\mathrm{spot}}$ to
account for the temperature difference:

\begin{align}
F_{\mathrm{spot}}^{Mx} &= f_{\mathrm{spot}} \lim_{\mathbf{c}\rightarrow\mathbf{d}} F_{\mathrm{obscured}}^{Mx},
\end{align}

where the $x$ label emphasises that it is valid for all cases.

\subsubsection{Case M9}

M9 considers the case when the spot overlaps with the centre of the stellar
disc. Here, we must adjust the integration limits of the annulus flux since 
$r\geq0$:

\begin{align}
F_{\mathrm{obscured},\mathrm{annulus}}^{M9} &= \int_{r=0}^{\sin(\beta+\alpha)} 2 r I_*(r)\,\mathrm{d}r, \nonumber \\
\qquad&= \sum_{n=0}^4 \Big(\frac{4 c_n}{4+n}\Big) \Big[1-\cos^{\frac{n+4}{2}}(\beta+\alpha) \Big].
\end{align}

Correcting for the expanded annulus area, we find:

\begin{align}
F_{\mathrm{obscured}}^{M9} &= \frac{A}{ \pi A_{\mathrm{annulus}} } F_{\mathrm{obscured},\mathrm{annulus}}^{M9}, \nonumber \\
\qquad &= \frac{A}{ \pi[1-\cos^2(\beta+\alpha)] } F_{\mathrm{obscured},\mathrm{annulus}}^{M9}.
\end{align}

\subsubsection{Case M2}

M2 considers the case when the spot now hits the stellar limb. Again, we must 
adjust the integration limits of the annulus flux since $r\leq1$:

\begin{align}
F_{\mathrm{obscured},\mathrm{annulus}}^{M2} &= \int_{r=1}^{\sin(\beta-\alpha)} 2 r I_*(r)\,\mathrm{d}r, \nonumber \\
\qquad&= \sum_{n=0}^4 \Big(\frac{4 c_n}{4+n}\Big) \Big[\cos^{\frac{n+4}{2}}(\beta-\alpha) \Big].
\end{align}

Correcting for the expanded annulus area, we find:

\begin{align}
F_{\mathrm{obscured}}^{M2} &= \frac{A}{ \pi A_{\mathrm{annulus}} } F_{\mathrm{obscured},\mathrm{annulus}}^{M2}, \nonumber \\
\qquad &= \frac{A}{ \pi \cos^2(\beta-\alpha) } F_{\mathrm{obscured},\mathrm{annulus}}^{M2}.
\end{align}

\subsubsection{Case M1}

The final case, and the simplest, is when the spot is out-of-view, analogous to
the out-of-transit planet. Here, we have:

\begin{align}
F_{\mathrm{obscured}}^{M1} &= 0.
\end{align}

\subsubsection{Final Expressions}

For a single rotating starspot, satisfying the assumptions made in this work,
we find that the flux from a star with a starspot may be written as:

\begin{align}
F(\alpha,\beta) &= 1 - \sum_{n=0}^4 \Big(\frac{n c_n}{n+4}\Big) \nonumber \\
\qquad& - \frac{A}{\pi} \Bigg[ \Bigg(\sum_{n=0}^4 \frac{4 (c_n-d_n f_{\mathrm{spot}})}{n+4} \frac{ \zeta_{-}^{\frac{n+4}{2}} - \zeta_{+}^{\frac{n+4}{2}} }{ \zeta_{-}^2 - \zeta_{+}^2 } \Bigg) \Bigg],
\label{eqn:singleLDspot}
\end{align}

where

\begin{equation}
\zeta_{-} =
\begin{cases}
1  & \text{if } 0<\beta<\alpha ,\\
\cos(\beta-\alpha)  & \text{if } \alpha<\beta<\frac{\pi}{2}+\alpha ,\\
0 & \text{if } \frac{\pi}{2}+\alpha<\beta<\pi ,
\end{cases}
\end{equation}

and

\begin{equation}
\zeta_{+} =
\begin{cases}
\cos(\beta+\alpha)  & \text{if } 0<\beta<\frac{\pi}{2}-\alpha ,\\
0  & \text{if } \frac{\pi}{2}-\alpha<\beta<\pi ,
\end{cases}
\end{equation}

In the above form, the expressions span two/three domains. A single-domain 
function can be expressed using Heaviside Theta functions, $\mathsf{H}(x)$:

\begin{align}
\zeta_{-} &= \cos(\beta-\alpha) \mathsf{H}(\beta-\alpha) \mathsf{H}(\frac{\pi}{2}-(\beta-\alpha)) + \mathsf{H}(-(\beta-\alpha)), \nonumber \\
\zeta_{+} &= \cos(\beta+\alpha) \mathsf{H}(\beta+\alpha) \mathsf{H}(\frac{\pi}{2}-(\beta+\alpha))  + \mathsf{H}(-(\beta+\alpha)).
\end{align}

Or more generally:

\begin{align}
\zeta(x) &= \cos x \mathsf{H}(x) \mathsf{H}(\frac{\pi}{2}-x) + \mathsf{H}(-x),
\end{align}

where $\zeta_{-} = \zeta(\beta-\alpha)$ and $\zeta_{+} = \zeta(\beta+\alpha)$.

Equation~\ref{eqn:singleLDspot} may be shown to return 
Equation~\ref{eqn:singleuniformspot} if one sets 
$\{c_1,c_2,c_3,c_4\}^T=\{d_1,d_2,d_3,d_4\}^T=\{0,0,0,0\}^T$, as
expected. For Equation~\ref{eqn:singleuniformspot}, we showed how it was trivial 
to generalise the expression to $N_S$ spots, provided one assumes the spots do 
not overlap. The same extension may be used here to yield:

\begin{align}
F(\boldsymbol{\alpha},\boldsymbol{\beta}) &= 1 - \sum_{n=0}^4 \Big(\frac{n c_n}{n+4}\Big) - \sum_{k=1}^{N_S} \frac{A_k}{\pi} \Bigg[ \nonumber \\
\qquad& \Bigg(\sum_{n=0}^4 \frac{4 (c_n-d_n f_{\mathrm{spot}})}{n+4} \frac{ \zeta_{-,k}^{\frac{n+4}{2}} - \zeta_{+,k}^{\frac{n+4}{2}} }{ \zeta_{-,k}^2 - \zeta_{+,k}^2 + \delta_{\zeta_{+,k},\zeta_{-,k}} } \Bigg) \Bigg],
\label{eqn:multiLDspot}
\end{align}

where the expressions for $\zeta_{+/-}$ are trivially generalized to 
$\zeta_{+/-,k}$ by amending $\alpha \rightarrow \alpha_{k}$ and 
$\beta \rightarrow \beta_{k}$. Note that in the above expression we have added
a Kronecker Delta function. This is because for $\beta>\pi/2+\alpha$, the
fraction containing the $\zeta$ terms goes to $0/0$ i.e. undefined. Adding
the Kronecker delta instead causes this to be equal to $0/1=0$ in this special
circumstance and thus adds numerical stability to the function. 

\subsection{Expressing $\beta$ with Physical Parameters}

\subsubsection{Accounting for the Star's Geometry}

So far, we have derived an expression for the flux from a limb-darkened star 
covered in multiple spots of sizes $\boldsymbol{\alpha}$ and instantaneous 
positions $\boldsymbol{\beta}$, as given in Equation~\ref{eqn:multiLDspot}. It 
was shown earlier how $\beta$ could be related to a specific choice of 
apparent longitude, $\tilde{\Lambda}$, and apparent latitude, $\tilde{\Phi}$, 
via Equation~\ref{eqn:beta}:

\begin{align}
\beta(\tilde{\Lambda},\tilde{\Phi}) &= \cos^{-1}[\cos\tilde{\Lambda}\cos\tilde{\Phi}]. \nonumber
\end{align}

As stressed throughout, $\tilde{\Lambda}$ and $\tilde{\Phi}$ are the 
\emph{apparent} longitude and latitude of a starspot. The vector describing the 
Cartesian coordinates of the spot's centre is $\mathbf{R}_{\mathrm{centre}}$ and
so far we have only defined this as a function of $\tilde{\Lambda}$ and 
$\tilde{\Phi}$ i.e. we know 
$\mathbf{R}_{\mathrm{centre}}(\tilde{\Lambda},\tilde{\Phi})$. However, here we
show how the vector can also be expressed as a function of the true longitude
and latitude (i.e. accounting for the star's geometry), 
$\mathbf{R}_{\mathrm{centre}}(\Lambda,\Phi)$. This is crucial since the flux 
from the star is described by the parameters $\alpha$ and $\beta$ only and 
ultimately one wishes to describe the flux as a function of the physical 
parameters and not auxiliary angles.

Consider a frame in which the geometry of the star is such that the rotation
axis has a normal vector given by $\hat{\mathbf{j}}$ i.e. along the 
$\hat{Y}$-axis. In this frame, the apparent longitude and latitude are in
fact equal to the true longitude and latitude, by virtue of definition. In this
frame, which does not account for stellar geometry, we describe the position
vector of the spot's centre with vector $\mathbf{R}_{\mathrm{centre}}'$. Due
to the argument made above, we have:

\begin{align}
\mathbf{R}_{\mathrm{centre}}'(\Lambda,\Phi) &= \mathbf{R}_{\mathrm{centre}}(\tilde{\Lambda}=\Lambda,\tilde{\Phi}=\Phi),
\end{align}

or explicitly

\begin{eqnarray}
 \mathbf{R}_{\mathrm{centre}}' = 
  \begin{bmatrix}
    x_{\mathrm{centre}}' \\
    y_{\mathrm{centre}}' \\
    z_{\mathrm{centre}}'
  \end{bmatrix} =
  \begin{bmatrix}
    \sin\Lambda\cos\Phi \\
    \sin\Phi \\
    \cos\Lambda\cos\Phi
  \end{bmatrix}.
\label{eqn:Rcentredash}
\end{eqnarray}

In order to calculate $\mathbf{R}_{\mathrm{centre}}(\Lambda,\Phi)$, we must
transform the frame to account for the geometry of the star. In other words we
seek to transform 
$\mathbf{R}_{\mathrm{centre}}'\rightarrow\mathbf{R}_{\mathrm{centre}}$.

Euler's rotation theorem states that any large series of three-dimensional 
rotations can be written as a series of just three rotations only. Two 
conventions exist for how these three ``Euler rotations'' may be performed. 
The first is known as ``proper Euler angles'. According to the 
intrinsic/extrinsic rotation equivalences, proper Euler angles are equivalent to 
three combined rotations repeating exactly one axis e.g. 
$\hat{X}$-$\hat{Z}$-$\hat{X}$. The second convention is called the ``Tait-Bryan 
angles'' (also known as the ``Cardan angles'') and these are equivalent to
three composed rotations in different axes e.g. $\hat{X}$-$\hat{Z}$-$\hat{Y}$.

We abstain from choosing a convention for the moment and proceed to consider a
sequential choice of rotations which minimises the degeneracy between the 
various angles. We note that by choosing the first axis to be $\hat{Y}$, we 
can eliminate a redundant angle since we defined an initial configuration with 
the stellar rotation axis aligned to the $\hat{Y}$-axis (i.e. an initial 
$\hat{Y}$ rotation is equivalent to intrinsic stellar rotation). For the sake of 
completeness, we refer to this first rotation as a clockwise rotation 
about the $\hat{Y}$-axis by an angle $\omega_*$.

For the next rotation, it is desirable to include stellar inclination at this
point. A clockwise rotation about $\hat{X}$ by an angle $(\pi/2-I_*)$ 
would correspond to the traditional definition of the stellar inclination angle.
We have now selected the first two rotations, leaving just to the third. If
we follow the proper Euler angles, we will be forced to us a $\hat{Y}$ rotation.
In contrast, the Tait-Bryan convention would require a $\hat{Z}$ rotation. Since
the observer is located down the $\hat{Z}$-axis, a rotation about this axis
cannot change the observed disk-integrated flux. Thus, a rotation about this
axis would be redundant. For this reason, we use the Tait-Bryan convention
and define our Euler rotation scheme as $\hat{Y}$-$\hat{X}$-$\hat{Z}$ leading
to two redundant angles and only one angle of physical interest, $I_*$ (for
completeness we dub the $\hat{Z}$ rotation angle as $\psi_*$). We
therefore define the position of the starspot centre, after applying the
Tait-Bryan rotations, as:

\begin{align}
\mathbf{R}_{\mathrm{centre}} &= \mathbf{M}_{\psi_*} \mathbf{M}_{I_*} \mathbf{M}_{\omega_*} \mathbf{R}_{\mathrm{centre}}'
\end{align}

where the first rotation is a clockwise rotation about the $\hat{Y}$-axis 
by an angle $\omega_*$:

\begin{eqnarray}
 \mathbf{M}_{\omega_*} = 
  \begin{bmatrix}
    \cos\omega_* & 0 & \sin\omega_* \\
    0 & 1 & 0 \\
    -\sin\omega_* & 0 & \cos\omega_*
  \end{bmatrix}.
\label{eqn:Momega}
\end{eqnarray}

The second rotation is a clockwise rotation about the
$\hat{X}$ axis by an angle $(\pi/2-I_*)$.

% NOTE - ISTAR IS A CLOCKWISE ROTATION BY ANGLE (PI/2 - ISTAR)
\begin{eqnarray}
 \mathbf{M}_{I_*} = 
  \begin{bmatrix}
    1 & 0 & 0 \\
    0 & \sin I_* & -\cos I_* \\
    0 & \cos I_* & \sin I_*
  \end{bmatrix}.
\label{eqn:Minc}
\end{eqnarray}

Finally, the third Euler rotation is about the $\hat{Z}$-axis in a clockwise
sense by an angle $\psi_*$.

\begin{eqnarray}
 \mathbf{M}_{\psi_*} = 
  \begin{bmatrix}
    \cos\psi_* & -\sin\psi_* & 0 \\
    \sin\psi_* & \cos\psi_* & 0 \\
    0 & 0 & 1
  \end{bmatrix}.
\label{eqn:Mpsi}
\end{eqnarray}

Recall from Equation~\ref{eqn:beta} that the $\hat{Z}$-component of 
$\mathbf{R}_{\mathrm{centre}}$ directly yields 
$\beta$, via

\begin{align}
\beta &= \cos^{-1}[z_{\mathrm{centre}}]
\end{align}

Applying all three rotations and extracting the $\hat{Z}$-component
allows us to write $\beta$ as a function of the true longitude and latitude:

\begin{align}
\beta &= \cos^{-1}[\cos(\Lambda+\omega_*)\cos\Phi\sin I_* + \cos I_* \sin\Phi].
\end{align}

As discussed earlier, and manifestly evident from the above expression, the
angle $\omega_*$ is fully degenerate with $\Lambda$ and thus may be neglected,
giving us:

\begin{align}
\beta(\Lambda,\Phi,I_*) &= \cos^{-1}[\cos\Lambda\cos\Phi\sin I_* + \cos I_* \sin\Phi].
\label{eqn:finalbeta}
\end{align}

It may be easily seen that this is precisely the same expression as Equation~8 
of \citet{dorren:1987}.

\subsubsection{Accounting for the Star's Rotation}

Stellar rotation causes the a spot's instantaneous longitude to vary as a
function of time. We denote the rotation rate by $\Omega_* = 2\pi/P_*$, where
$P_*$ is the rotational period of the star. Although it may be possible to
envisage spots which migrate in latitude as well as longitude, we here only
consider the simple case of $\dot{\Phi}=\mathrm{d}\Phi/d\mathrm{t}=0$ and
$\dot{\Lambda}=\mathrm{d}\Lambda/d\mathrm{t}=\Omega_{*}$. We may then
decsribe the spot's instantaneous longitude and latitude as a function of
time using:

\begin{align}
\Lambda(t) &= \Lambda(t=t_{\mathrm{ref}}) + \dot{\Lambda}(t-t_{\mathrm{ref},k}) = \Lambda(t=t_{\mathrm{ref}}) + \frac{2\pi (t-t_{\mathrm{ref}})}{P_*}, \nonumber \\
\qquad&= \Lambda_{\mathrm{ref}} + \frac{2\pi (t-t_{\mathrm{ref}})}{P_*}, \\
\Phi(t) &= \Phi(t=t_{\mathrm{ref}}) + \dot{\Phi}(t-t_{\mathrm{ref},k}) = \Phi(t=t_{\mathrm{ref}}), \nonumber \\
\qquad&= \Phi_{\mathrm{ref}}.
\end{align}

\subsection{Differential Rotation}

In general, unique $P_{*,k}$ terms are included to allow for differential
rotation. However, differential rotation is well-described by the following
function:

\begin{align}
P_{*,k} &= \frac{P_{\mathrm{EQ}}}{1 - \kappa_2 \sin^2\Phi_{\mathrm{ref},k} - \kappa_4 \sin^4\Phi_{\mathrm{ref},k}}.
\end{align}

The $\sin^4\Phi_{\mathrm{ref},k}$ is usually only used for Solar rotation 
analysis, but we include it here for cases where a user requires a more 
soPhisticated differential rotation profile. 

\subsection{Starspot Evolution}
\subsubsection{A Linear Model}

There is some debate in the literature regarding how to model the evolution of
a single starspot. \citet{rudiger:2000} show that the theoretical decay rate 
from 2-D modelling of a sunspot is close to linear for the spot area (i.e. a 
square-root rate for $\alpha$). Similarly, \citet{stix:2002} argue that if
the decay of an isolated sunspot is set by the amount of the azimuthal electric
current within the spot, a linear decay in area would result. However,
\citet{petrovay:1997} use a statistical analysis of sunspot data to show that
an ``idealised'' sunspot exhibits parabolic area decay. Additionally,
observations of \citet{martinez:1993} find both linear and parabolic decays.

With our model, the user is free to use any description they desire, but the 
code provided considers a simple linear growth/decay in $\alpha$ model and the 
partial derivatives computed are only valid for said model. 

A linear model has the advantage of being intuitively simple to handle,
making the selection of appropriate boundary conditions and starting points
easier for regression problems. It is also very quick to computationally
evaluate and encapsulates the key physics involved. Our linear model produces a 
linear growth, flat-top and then linear-decay i.e. a trapezoidal profile for the 
spot's evolution. The profile is allowed to asymmetric to reproduce realistic
evolution.

We model starspot growth/decay via the $\alpha$ parameter only i.e. we
consider the flux contrast to be constant. The linear-model has the simple
form:

\begin{align}
\frac{\alpha_k(t_i)}{\alpha_{\mathrm{max},k}} &= \mathcal{I}_k^{-1} [\Delta t_1 \mathsf{H}(\Delta t_1) - \Delta t_2 \mathsf{H}(\Delta t_2)] \nonumber\\
\qquad& - \mathcal{E}_k^{-1} [\Delta t_3 \mathsf{H}(\Delta t_3) - \Delta t_4 \mathsf{H}(\Delta t_4)].
\end{align}

and using

\begin{align}
\Delta t_1 &= t_i-t_{\mathrm{max},k}+\frac{L_k}{2}+\mathcal{I}_k,\\
\Delta t_2 &= t_i-t_{\mathrm{max},k}+\frac{L_k}{2},\\
\Delta t_3 &= t_i-t_{\mathrm{max},k}-\frac{L_k}{2},\\
\Delta t_4 &= t_i-t_{\mathrm{max},k}-\frac{L_k}{2}-\mathcal{E}_k,
\end{align}

where $\mathsf{H}(x)$ is the Heaviside step function, $\alpha_{\mathrm{max},k}$
is the maximum spot-size, $L_k$ is the full-width-full-maximum ``lifetime'' of 
the spot and $\mathcal{I}_k$ \& $\mathcal{E}_k$ are the ingress \& egress
durations of the spot profile.

\subsection{Normalisation and Blended Light}

As discussed earlier, we choose to normalise the flux of spotted star by the
flux which the same star would cause if no spots were present i.e.

\begin{align}
F_{\mathrm{mod}} &= \frac{F(\boldsymbol{\alpha},\boldsymbol{\beta})}{F(\boldsymbol{\alpha}=\mathbf{0},\boldsymbol{\beta})}.
\end{align}

At the time of writing, the most precise and sizable source of photometric time 
series for main sequence stars comes from \emph{Kepler Mission}. With this in
mind, we choose to include a blending factor at this stage to account for
overlapping PSFs, background flux or flux from an associated member in the
system. This is a fairly common occurrence for \emph{Kepler} data and many other
photometric surveys due to the crowded fields observed. Let us consider then
that the total flux observed is changed via 
$F(\boldsymbol{\alpha},\boldsymbol{\beta}) \rightarrow 
(F(\boldsymbol{\alpha},\boldsymbol{\beta}) + F_{\mathrm{blend}})$. Our
normalisation factor must now also be modified if we require that 
$F_{\mathrm{mod}}=1$ for an unspotted star. An appropriate choice is to use
$(F(\boldsymbol{\alpha}=\mathbf{0},\boldsymbol{\beta}) + F_{\mathrm{blend}})$:

\begin{align}
F_{\mathrm{mod}} &= \frac{F(\boldsymbol{\alpha},\boldsymbol{\beta})+F_{\mathrm{blend}}}{F(\boldsymbol{\alpha}=\mathbf{0},\boldsymbol{\beta})+F_{\mathrm{blend}}}.
\end{align}

Using $F_{\mathrm{blend}}$ is cumbersome and a more common approach is to
define a blending factor, relative to the target's flux. 
\citet{kippingtinetti:2010} advocate using 
$B = (F_{\mathrm{source}} + F_{\mathrm{blend}})/F_{\mathrm{source}}$ which we
follow here. This yields:

\begin{align}
F_{\mathrm{mod}} &= \frac{F(\boldsymbol{\alpha},\boldsymbol{\beta})}{B F(\boldsymbol{\alpha}=\mathbf{0},\boldsymbol{\beta})} + \frac{B-1}{B}, \\
B &= \frac{F(\boldsymbol{\alpha}=\mathbf{0},\boldsymbol{\beta}) + F_{\mathrm{blend}}}{F(\boldsymbol{\alpha}=\mathbf{0},\boldsymbol{\beta})}.
\end{align}

\subsection{Allowing for Multiple Time Series}

In many practical cases, we must fit multiple epochs of data which have
different systematics. The most common systematic to be treated is a
baseline parameter, $U$. A common application of this process is using a unique
normalisation factor for each \emph{Kepler} quarter since spacecraft rolls
affect the total flux within a defined aperture. For $M$ data sets (e.g. $M$ 
quarters of data from \emph{Kepler}), each set requires a unique $U_m$ 
parameter. Using a box-car function ($\Pi$), which is a composite of two 
Heaviside Theta functions, one can reproduce the desired behaviour:

\begin{align}
&F_{\mathrm{mod}} = \sum_{m=1}^M U_m \Pi_m \Bigg(\frac{F(\boldsymbol{\alpha},\boldsymbol{\beta})}{B_m F(\boldsymbol{\alpha}=\mathbf{0},\boldsymbol{\beta})} + \frac{B_m-1}{B_m}\Bigg), \\
&\Pi_m(t;T_{\mathrm{start},m},T_{\mathrm{end},m}) = \mathsf{H}(t-T_{\mathrm{start},m}) - \mathsf{H}(t-T_{\mathrm{end},m}),
\end{align}

where it is understood that 
$T_{\mathrm{start},m+1}\geq T_{\mathrm{end},m}>T_{\mathrm{start},m}$.  Note
that we have also assumed that each data set has a unique blending factor.
For \emph{Kepler} data, it is typical for each quarter to have a unique $B$
factor from spacecraft motion altering the PSF overlaps.

Consider we have two data sets separated by $N$ rotation periods where
$N\gg1$. Further assume that the time span of data sets 1 and 2 are
shorter than the spot lifetime of all spots i.e. 
$(T_{\mathrm{end},m} - T_{\mathrm{start},m})<L_k$ for all $k$ and $m$. In this
case, may one wish to treat the spots in data set 1 as independent of data set
2. This can be implemented by making use of the starspot evolution equations.
Specifically, one wishes to impose box-car spots (unchanging during each
data set) with cut-offs in-between the two data sets. So the $k^{\mathrm{th}}$ 
starspot of the $m^{\mathrm{th}}$ data set will have
$\alpha_{m,k}(t;\alpha_{\mathrm{max},m,k},t_{\mathrm{max},m,k},L_{m,k},s_{m,k})$ 
take the form $s_{m,k}\rightarrow\infty$ and 
$t_{\mathrm{max},m,k}=(T_{\mathrm{end},m}-T_{\mathrm{start},m})/2$ and 
$L_{m,k}=(T_{\mathrm{end},m} - T_{\mathrm{start},m})$.

\section{Partial Derivatives}
\label{appendixD}

\subsection{Motivation}

One of the major benefits of writing our expression as a single-domain function
is that one can consider writing down a set of a single-domain partial 
derivatives. Partial derivatives are highly useful in optimisation problems 
where frequently the Jacobian matrix is computed to expedite a regression 
problem.

\subsection{Partial Derivatives of the Likelihood Function}

The commonly used Gaussian noise likelihood has the form:

\begin{align}
\mathcal{L}(\mathbf{\Theta}) &= \prod_{i=1}^N \frac{1}{\sqrt{2\pi\sigma_i^2}} \exp\Big[-\frac{(F_{\mathrm{obs},i} - F_{\mathrm{mod},i}(\mathbf{\Theta}))^2}{2\sigma_i^2}\Big].
\end{align}

Taking the partial derivative of the log likelihood with respect to parameter
$\Theta_j$ yields:

\begin{align}
\frac{\partial\log\mathcal{L}}{\partial\Theta_j} &= -\sum_{i=1}^N (r_i/\sigma_i) \frac{\partial(r_i/\sigma_i)}{\partial\Theta_j},\nonumber\\
\qquad&= \sum_{i=1}^N \Big( \frac{r_i}{\sigma_i^2} \Big) \Big(\frac{\partial F_{\mathrm{mod},i}}{\partial\Theta_j}\Big),
\end{align}

where $r_i = (F_{\mathrm{obs},i} - F_{\mathrm{mod},i})$. Also note that in
the above, and what follows throughout, that any partial derivatives taken with
respect to $\Theta_j$ implicitly means that all other parameters are held 
constant except $\Theta_j$. In other words, for a set of parameters $\Theta_l$ 
where $l=1,2,...L-1,L$, we use the notation that the partial derivative of some
quantity $X$ follows

\begin{align}
\frac{\partial X}{\partial \Theta_j} &= \Bigg(\frac{\partial X}{\partial \Theta_j}\Bigg)_{\Theta_l,l\neq j}.
\end{align}

The outstanding problem is to derive $\partial F_{\mathrm{mod},i}/\partial\Theta_j$, which we deal
with in the next subsection. We point out that any reasonable likelihood 
function, even if non-Gaussian, will still require 
$\partial F_{\mathrm{mod},i}/\partial\Theta_j$. For this reason, in the provided 
code \macula, we do not provide the partial derivatives of a Gaussian 
likelihood function directly but instead provide the partial derivatives of
the model flux instead, $\partial F_{\mathrm{mod},i}/\partial\Theta_j$. In this
way, the results from \macula\ are more general and hopefully of greater
use to typical observers.

\subsection{Partial Derivatives of the Model Flux}

Recall our final expression for the model flux, evaluated for the 
$i^{\mathrm{th}}$ data point:

\begin{align}
F_{\mathrm{mod},i} &= \sum_{m=1}^M U_m \Pi_{m,i} \Bigg(\frac{F_i(\boldsymbol{\alpha},\boldsymbol{\beta})}{B_m F_i(\boldsymbol{\alpha}=\mathbf{0},\boldsymbol{\beta})} + \frac{B_m-1}{B_m}\Bigg),
\label{eqn:Fmodfull}
\end{align}

which may be written as

\begin{align}
F_{\mathrm{mod},i} &= \sum_{m=1}^M \tilde{F}_{\mathrm{mod},m,i}.
\label{eqn:Fmodtilde}
\end{align}

So one may easily see that

\begin{align}
\frac{ \partial F_{\mathrm{mod},i} }{ \partial\Theta_j } &= \sum_{m=1}^M \frac{ \partial \tilde{F}_{\mathrm{mod},m,i} }{ \partial\Theta_j }.
\end{align}

The $\tilde{F}_{\mathrm{mod},m,i}$ function now requires partial derivatives.
We adopt the assumption that $\Pi_{m,i}$ is not a function of any of the
$\Theta_j$ parameters. This is perfectly reasonable as the function is only a
function of $T_{\mathrm{start},m}$ and $T_{\mathrm{end},m}$, which the user
would define rather than fit for. Using this assumption, 
$\partial \Pi_{m,i}/\partial\Theta_j=0$ for all $i,j,m$. Using the
replacement (purely to save space) that 
$\mathbb{F}_i=F_i(\boldsymbol{\alpha},\boldsymbol{\beta})$ and 
$\mathbb{F}_{0,i}=F_i(\boldsymbol{\alpha}=\mathbf{0},\boldsymbol{\beta})$, one may now 
show:

\begin{align}
\frac{ \partial \tilde{F}_{\mathrm{mod},m,i} }{\partial \Theta_j} &= \frac{\Pi_{m,i}}{\mathbb{F}_{0,i}^2 B_m^2} \Bigg( \mathbb{F}_0 B_m \Big(\mathbb{F}_i+\mathbb{F}_{0,i} (B_m-1)\Big) \frac{\partial U_m}{\partial \Theta_j} \nonumber \\
\qquad& + U_m \Big( B_m \mathbb{F}_{0,i} \frac{\partial\mathbb{F}_i}{\partial\Theta_j}
- B_m \mathbb{F}_i \frac{\partial\mathbb{F}_{0,i}}{\partial\Theta_j} \nonumber \\
\qquad& + \mathbb{F}_{0,i} (\mathbb{F}_{0,i}-\mathbb{F}_i) \frac{\partial B_m}{\partial\Theta_j} \Big) \Bigg).
\end{align}

The above expression shows that the partial derivatives of the model flux
can be expressed as a function of four other partial derivatives (which in turn
may be broken down into other partial derivatives).

$\mathbb{F}_{0,i}$ and in particular $\mathbb{F}_{i}$ are functionally
dependent upon many $\Theta_j$ parameters but $U_m$ and $B_m$ do not.
Rather, they represent a fitted parameter and have no other dependencies. We
therefore have:

\begin{equation}
\frac{\partial U_m}{\partial \Theta_j} =
\begin{cases}
0  & \text{if } \Theta_j \neq U_m ,\\
1 & \text{if } \Theta_j = U_m ,
\end{cases}
\end{equation}

and

\begin{equation}
\frac{\partial B_m}{\partial \Theta_j} =
\begin{cases}
0  & \text{if } \Theta_j \neq B_m ,\\
1 & \text{if } \Theta_j = B_m .
\end{cases}
\end{equation}

With these expressions the only remaining partial derivatives to find
are those of $\mathbb{F}_i$ and $\mathbb{F}_{0,i}$. In fact, since
$\mathbb{F}_{0,i}$ is defined as simply a special case version of 
$\mathbb{F}_i$ then we only require solving the partial derivatives of
$\mathbb{F}_{i}$ or equivalently $F_i(\boldsymbol{\alpha},\boldsymbol{\beta})$.

\subsection{Partial Derivatives of the Flux w.r.t. Limb Darkening}

The $F_i(\boldsymbol{\alpha},\boldsymbol{\beta})$ function is fully expressed as:

\begin{align}
F_i(\boldsymbol{\alpha},\boldsymbol{\beta}) &= 1 - \sum_{n=0}^4 \Big(\frac{n c_n}{n+4}\Big) - \sum_{k=1}^{N_S} \frac{A_{k,i}}{\pi} \Bigg[ \nonumber \\
\qquad& \Bigg(\sum_{n=0}^4 \frac{4 (c_n-d_n f_{\mathrm{spot},k})}{n+4} \frac{ \zeta_{-,k,i}^{\frac{n+4}{2}} - \zeta_{+,k,i}^{\frac{n+4}{2}} }{ \zeta_{-,k,i}^2 - \zeta_{+,k,i}^2 + \delta_{\zeta_{+,k,i},\zeta_{-,k,i}} } \Bigg) \Bigg]. \nonumber
\end{align}

For an unspotted star, $A_{k,i}=0$ for all $k$ \& $i$ and so one may write:

\begin{align}
&F_i(\boldsymbol{\alpha},\boldsymbol{\beta}) = F_i(\boldsymbol{\alpha}=\mathbf{0},\boldsymbol{\beta}) - Q_i \\
&F_i(\boldsymbol{\alpha}=\mathbf{0},\boldsymbol{\beta}) = 1 - \sum_{n=0}^4 \Big(\frac{n c_n}{n+4}\Big) \\
&Q_i = \sum_{k=1}^{N_S} q_{k,i} \\
&q_{k,i} = \frac{A_{k,i}}{\pi} \Bigg(\sum_{n=0}^4 \frac{4 (c_n-d_n f_{\mathrm{spot},k})}{n+4} \frac{ \zeta_{-,k,i}^{\frac{n+4}{2}} - \zeta_{+,k,i}^{\frac{n+4}{2}} }{ \zeta_{-,k,i}^2 - \zeta_{+,k,i}^2 + \delta_{\zeta_{+,k,i},\zeta_{-,k,i}} } \Bigg).
\end{align}

This allows us to write that

\begin{align}
\frac{\partial F_i(\boldsymbol{\alpha},\boldsymbol{\beta})}{\partial \Theta_j} &= \frac{\partial F_i(\boldsymbol{\alpha}=\mathbf{0},\boldsymbol{\beta})}{\partial \Theta_j} - \frac{\partial Q_i}{\partial \Theta_j} \nonumber \\
\qquad&= \frac{\partial F_i(\boldsymbol{\alpha}=\mathbf{0},\boldsymbol{\beta})}{\partial \Theta_j} - \sum_{k=1}^{N_S} \frac{\partial q_{k,i}}{\partial \Theta_j}.
\end{align}

It is easy to show that

\begin{equation}
\frac{\partial F_i(\boldsymbol{\alpha}=\mathbf{0},\boldsymbol{\beta})}{\Theta_j} =
\begin{cases}
-\frac{1}{5}	& \text{if } \Theta_j = c_1 ,\\
-\frac{1}{3}	& \text{if } \Theta_j = c_2, \\
-\frac{3}{7}	& \text{if } \Theta_j = c_3, \\
-\frac{1}{2}	& \text{if } \Theta_j = c_4, \\
0 		& \text{otherwise}.
\end{cases}
\label{eqn:LDderivs}
\end{equation}

\subsection{Partial Derivatives of $q_{k,i}$}

The outstanding problem is now to find the partial derivatives of $q_{k,i}$ with 
respect to $\Theta_j$. $q_{k,i}$ is defined as:

\begin{align}
q_{k,i} &= \frac{A_{k,i}}{\pi} \Bigg(\sum_{n=0}^4 \frac{4 (c_n-d_n f_{\mathrm{spot},k})}{n+4} \frac{ \zeta_{-,k,i}^{\frac{n+4}{2}} - \zeta_{+,k,i}^{\frac{n+4}{2}} }{ \zeta_{-,k,i}^2 - \zeta_{+,k,i}^2 + \delta_{\zeta_{+,k,i},\zeta_{-,k,i}} } \Bigg) \nonumber.
\end{align}

We therefore proceed to derive the full four-coefficient partial derivatives, 
which we start by re-writing:

\begin{align}
q_{k,i} &= \frac{A_{k,i}}{\pi} \sum_{n=0}^4 w_{n,k,i} \\
w_{n,k,i} &= \frac{4 (c_n-d_n f_{\mathrm{spot},k})}{n+4} \frac{ \zeta_{-,k,i}^{\frac{n+4}{2}} - \zeta_{+,k,i}^{\frac{n+4}{2}} }{ \zeta_{-,k,i}^2 - \zeta_{+,k,i}^2 + \delta_{\zeta_{+,k,i},\zeta_{-,k,i}} }.
\end{align}

For the complex function $\mathcal{A}(\alpha,\beta)$, the only derivatives of
interest are with respect to $\alpha$ and $\beta$ since $\mathcal{A}$ is
functionally dependent on these terms alone. It may easily shown and numerically
verified that:

\begin{align}
\Bigg(\frac{\partial \mathbb{R}[\mathcal{A}] }{\partial\alpha}\Bigg)_{\beta} &= \mathbb{R}\Bigg[ \Bigg(\frac{\partial \mathcal{A} }{\partial\alpha}\Bigg)_{\beta} \Bigg],\\
\Bigg(\frac{\partial \mathbb{R}[\mathcal{A}] }{\partial\beta}\Bigg)_{\alpha} &= \mathbb{R}\Bigg[ \Bigg(\frac{\partial \mathcal{A} }{\partial\beta}\Bigg)_{\alpha} \Bigg].
\end{align}

Since all other partial derivatives of $A$ can be expressed using the chain
rule as a combination of the above two forms, then partial derivatives of
$A$ can be derived for all $\Theta_j$ using this simple trick. This allows
us to write:

\begin{align}
\frac{\partial q_{k,i}}{\partial \Theta_j} &= \frac{\mathbb{R}[\mathcal{A}_{k,i}]}{\pi} \sum_{n=0}^4 \frac{\partial w_{n,k,i}}{\partial\Theta_j} + \frac{1}{\pi}\mathbb{R}\Big[\frac{\partial\mathcal{A}_{k,i}}{\partial\Theta_j}\Big] \sum_{n=0}^4 w_{n,k,i}.
\end{align}

With the above, one can see the outstanding problem is to find partial 
derivatives of $\mathcal{A}_{k,i}$ \& $w_{n,k,i}$ with respect to $\Theta_j$.

\subsection{Partial Derivatives of $\mathcal{A}_{k,i}$}

$\mathcal{A}_{k,i}$ is a function of $\alpha_{k,i}$ and $\beta_{k,i}$ only. 
Whilst these two terms will be functions of other parameters themselves, they 
offer the obvious starting point for a derivation of $\mathcal{A}_{k,i}$'s 
partial derivatives. The partial derivatives with respect to $\alpha_{k,i}$ and 
$\beta_{k,i}$ are easily shown to be given by:

\begin{align}
\Bigg(\frac{\partial \mathcal{A}_{k,i}}{\partial \alpha_{k,i}}\Bigg)_{\beta_{k,i}} &= -\sin\alpha_{k,i} \sin\beta_{k,i} \epsilon_{k,i} + 2\cos\alpha_{k,i}\cos\beta_{k,i}\Xi_{k,i},\\
\Bigg(\frac{\partial\mathcal{A}_{k,i}}{\partial\beta_{k,i}}\Bigg)_{\alpha_{k,i}} &= 0.5 \cos\alpha_{k,i}\cos\beta_{k,i} \epsilon_{k,i} - \sin\alpha_{k,i}\sin\beta_{k,i}\Xi_{k,i}.
\label{eqn:Aderivs}
\end{align}

where we use

\begin{align}
\epsilon_{k,i} &= \frac{\csc^2\beta_{k,i}(\cos2\alpha_{k,i}+\cos2\beta_{k,i})}{\Psi_{k,i}}.
\end{align}

One may now employ the chain rule to write:

\begin{align}
\Bigg(\frac{\partial \mathcal{A}_{k,i}}{\partial \Theta_j}\Bigg)_{\Theta_l,l\neq j} &= \Bigg(\frac{\partial \mathcal{A}_{k,i}}{\partial \alpha_{k,i}}\Bigg)_{\beta_{k,i}} \Bigg(\frac{\partial \alpha_{k,i}}{\partial \Theta_j}\Bigg)_{\Theta_l,l\neq j} \nonumber \\
\qquad& + \Bigg(\frac{\partial \mathcal{A}_{k,i}}{\partial \beta_{k,i}}\Bigg)_{\alpha_{k,i}} \Bigg(\frac{\partial \beta_{k,i}}{\partial \Theta_j}\Bigg)_{\Theta_l,l\neq j},
\end{align}

where we temporarily re-include the implicit notation to make the expression
less ambiguous. Partial derivatives of $\alpha_{k,i}$ \& $\beta_{k,i}$ with 
respect to $\Theta_j$ will be provided later. 

\subsection{Partial Derivatives of $w_{n,k,i}$}

$w_{n,k,i}$ is fully expressed as:

\begin{align}
w_{n,k,i} &= \frac{4 (c_n-d_n f_{\mathrm{spot},k})}{n+4} \Upsilon_{n,k,i}, \\
\Upsilon_{n,k,i} &= \frac{ \zeta_{-,k,i}^{\frac{n+4}{2}} - \zeta_{+,k,i}^{\frac{n+4}{2}} }{ \zeta_{-,k,i}^2 - \zeta_{+,k,i}^2 + \delta_{\zeta_{+,k,i},\zeta_{-,k,i}} }
\end{align}

We first turn our attention to taking the partial derivatives of $\Upsilon_{n,k,i}$ 
with respect to $\Theta_j$. We note that that:

\begin{align}
\Bigg(\frac{\partial\delta_{\zeta_{+,k,i},\zeta_{-,k,i}}}{\partial\alpha_{k,i}}\Bigg)_{\beta_{k,i}} &= 0, \\
\Bigg(\frac{\partial\delta_{\zeta_{+,k,i},\zeta_{-,k,i}}}{\partial\beta_{k,i}}\Bigg)_{\alpha_{k,i}} &= 0
\end{align}

which via the chain rule imply:

\begin{align}
\frac{\partial\delta_{\zeta_{+,k,i},\zeta_{-,k,i}}}{\partial\Theta_j} &= 0\,\,\,\forall \{i,j,k\}
\end{align}

With this simplification, we find:

\begin{align}
\frac{\partial\Upsilon_{n,k,i}}{\Theta_j} &= \Bigg[\frac{1}{\zeta_{-,k,i}^2-\zeta_{+,k,i}^2+\delta_{\zeta_{+,k,i},\zeta{-,k,i}}}\Bigg] \nonumber\\
\qquad& \times \Bigg[ \Bigg(\frac{n+4}{2}\Bigg) \Bigg( \zeta_{-,k,i}^{\frac{n+2}{2}} \frac{\partial\zeta_{-,k,i}}{\partial\Theta_j} - \zeta_{+,k,i}^{\frac{n+2}{2}} \frac{\partial\zeta_{+,k,i}}{\partial\Theta_j} \Bigg) \nonumber \\
\qquad& - 2\Upsilon_{n,k,i} \Bigg( \frac{\partial\zeta_{-,k,i}}{\partial\Theta_j} - \frac{\partial\zeta_{+,k,i}}{\partial\Theta_j} \Bigg) \Bigg].
\label{eqn:Upsilonderivs}
\end{align} 

The partial derivatives of $\zeta_{-,k,i}$ are given by:

\begin{align}
\Bigg(\frac{\partial\zeta_{-,k,i}}{\partial\alpha_{k,i}}\Bigg)_{\beta_{k,i}} &= \delta[-(\beta_{k,i}-\alpha_{k,i})] \nonumber \\
\qquad& + \frac{\mathsf{H}(\frac{\pi}{2}-(\beta_{k,i}-\alpha_{k,i})) \mathsf{H}(\beta_{k,i}-\alpha_{k,i})}{\csc(\beta_{k,i}-\alpha_{k,i})} \nonumber \\
\qquad& + \frac{2\delta[\pi - 2(\beta_{k,i}-\alpha_{k,i})] \mathsf{H}(\beta_{k,i}-\alpha_{k,i})}{\sec(\beta_{k,i}-\alpha_{k,i})} \nonumber \\
\qquad& - \frac{\delta[\beta_{k,i}-\alpha_{k,i}] \mathsf{H}(\frac{\pi}{2}-(\beta_{k,i}-\alpha_{k,i}))}{\sec(\beta_{k,i}-\alpha_{k,i})}, \\
\Bigg(\frac{\partial\zeta_{-,k,i}}{\partial\beta_{k,i}}\Bigg)_{\alpha_{k,i}} &= -\Bigg(\frac{\partial\zeta_{-,k,i}}{\partial\alpha_{k,i}}\Bigg)_{\beta_{k,i}},
\label{eqn:zetanegderivs}
\end{align}

and of $\zeta_{+,k,i}$

\begin{align}
\Bigg(\frac{\partial\zeta_{+,k,i}}{\partial\alpha_{k,i}}\Bigg)_{\beta_{k,i}} &= -\frac{2\delta[\pi-2(\beta_{k,i}+\alpha_{k,i})]}{\sec(\beta_{k,i}+\alpha_{k,i})} \nonumber \\
\qquad& - \frac{ \mathsf{H}(\beta_{k,i}+\alpha_{k,i}) \mathsf{H}(\frac{\pi}{2}-(\beta_{k,i}+\alpha_{k,i})) }{ \csc(\beta_{k,i}+\alpha_{k,i}) }, \\
\Bigg(\frac{\partial\zeta_{+,k,i}}{\partial\beta_{k,i}}\Bigg)_{\alpha_{k,i}} &= \Bigg(\frac{\partial\zeta_{+,k,i}}{\partial\alpha_{k,i}}\Bigg)_{\beta_{k,i}}.
\label{eqn:zetaposderivs}
\end{align}

In practice, the $\delta(x)$ functions always yield zero unless
$x=0$. Since they are a function of a continuous variable, namely time, the 
probability that the time will precisely yield a non-zero $\delta$ function is
infinitesimal. For this purpose, they are simply ignored in the \macula\ code.
The latter relations lead to a simplification of the chain rule
expansion of the partial derivatives with respect to $\Theta_j$:

\begin{align}
\frac{\partial \zeta_{-,k,i}}{\partial \Theta_j} &= \Bigg(\frac{\partial \zeta_{-,k,i}}{\partial \alpha_{k,i}}\Bigg)_{\beta_{k,i}} \Bigg[ \frac{\partial \alpha_{k,i}}{\partial \Theta_j} - \frac{\partial \beta_{k,i}}{\partial \Theta_j} \Bigg],\\
\frac{\partial \zeta_{+,k,i}}{\partial \Theta_j} &= \Bigg(\frac{\partial \zeta_{+,k,i}}{\partial \alpha_{k,i}}\Bigg)_{\beta_{k,i}} \Bigg[ \frac{\partial \alpha_{k,i}}{\partial \Theta_j} + \frac{\partial \beta_{k,i}}{\partial \Theta_j} \Bigg].
\end{align}

Finally, the partial derivatives of $w_{n,k,i}$ are:

\begin{align}
\frac{\partial w_{n,k,i}}{\Theta_j} &= \Bigg( \frac{4}{n+4} \Bigg) \Bigg( \Upsilon_{n,k,i} \frac{\partial c_n}{\partial\Theta_j} + (c_n-d_n f_{\mathrm{spot},k}) \frac{\partial\Upsilon_{n,k,i}}{\partial\Theta_j} \nonumber \\
\qquad&  - d_n \Upsilon_{n,k,i} \frac{\partial f_{\mathrm{spot},k}}{\partial\Theta_j} - f_{\mathrm{spot},k} \Upsilon_{n,k,i} \frac{\partial d_n}{\partial\Theta_j} \Bigg).
\end{align}

Since the partial derivatives of $\Upsilon_{n,k,i}$ have been dealt with above,
this leaves us to comment on the partial derivatives of $c_n$, $d_n$ and 
$f_{\mathrm{spot},k}$. These represent fitted parameters (or perhaps fixed)
and thus one may trivially evaluate their derivatives to be

\begin{equation}
\frac{\partial c_n}{\partial \Theta_j} =
\begin{cases}
0  & \text{if } \Theta_j \neq c_n ,\\
1 & \text{if } \Theta_j = c_n ,
\end{cases}
\end{equation}

\begin{equation}
\frac{\partial d_n}{\partial \Theta_j} =
\begin{cases}
0  & \text{if } \Theta_j \neq d_n ,\\
1 & \text{if } \Theta_j = d_n ,
\end{cases}
\end{equation}

\begin{equation}
\frac{\partial f_{\mathrm{spot},k}}{\partial \Theta_j} =
\begin{cases}
0  & \text{if } \Theta_j \neq f_{\mathrm{spot},k} ,\\
1 & \text{if } \Theta_j = f_{\mathrm{spot},k} .
\end{cases}
\end{equation}

\subsection{Partial Derivatives of $\beta_{k,i}$}

The only partial derivatives now missing are those of $\alpha_{k,i}$ and 
$\beta_{k,i}$ with respect to the fitted parameters, $\Theta_j$. $\beta_{k,i}$ 
is defined as:

\begin{align}
\beta_{k,i} &= \cos^{-1}\Big[\cos I_* \sin\Phi_{k,i} + \sin I_* \cos\Phi_{k,i}\cos\Lambda_{k,i} \Big], \nonumber \\
\Lambda_{k,i} &= \Lambda_{\mathrm{ref},k} + \frac{2\pi (t_i-t_{\mathrm{ref},k})}{P_{*,k}}, \nonumber \\
\Phi_{k,i} &= \Phi_{\mathrm{ref},k}. \nonumber
\end{align}

Accounting for differential rotation, the longitude evolution is described by:

\begin{align}
\Lambda_{k,i} &= \Lambda_{\mathrm{ref},k} + \frac{2\pi (t_i-t_{\mathrm{ref},k})}{P_{\mathrm{EQ}}} (1 - \kappa_2 \sin^2\Phi_{\mathrm{ref},k} - \kappa_4 \sin^4\Phi_{\mathrm{ref},k}).
\end{align}

Now the partial derivatives yield:

\begin{align}
\frac{\partial\beta_{k,i}}{\partial I_*} &= \frac{\sin\Phi_{\mathrm{ref},k}\sin I_* - \cos\Lambda_{k,i}\cos\Phi_{\mathrm{ref},k}\cos I_*}{\sin\beta_{k,i}}, \\
\frac{\partial\beta_{k,i}}{\partial P_{\mathrm{EQ}}} &= - \frac{2\pi (t_i-t_{\mathrm{ref},k})}{P_{\mathrm{EQ}} P_{*,k}} \frac{\cos\Phi_{\mathrm{ref},k}\sin\Lambda_{k,i}\sin I_*}{\sin\beta_{k,i}}, \\
\frac{\partial\beta_{k,i}}{\partial \kappa_2} &= - \frac{2\pi (t_i-t_{\mathrm{ref},k})}{P_{\mathrm{EQ}}} \frac{ \sin^2\Phi_{\mathrm{ref},k}\cos\Phi_{\mathrm{ref},k}\sin\Lambda_{k,i}\sin I_* }{\sin\beta_{k,i}}, \\
\frac{\partial\beta_{k,i}}{\partial \kappa_4} &= - \frac{2\pi (t_i-t_{\mathrm{ref},k})}{P_{\mathrm{EQ}}} \frac{ \sin^4\Phi_{\mathrm{ref},k}\cos\Phi_{\mathrm{ref},k}\sin\Lambda_{k,i}\sin I_* }{\sin\beta_{k,i}}, \\
\frac{\partial\beta_{k,i}}{\partial \Phi_{\mathrm{ref},k}} &= \csc\beta_{k,i}\sin I_*\sin\Phi_{\mathrm{ref},k} \Big( \cos\Lambda_{k,i} \nonumber \\
\qquad& - \frac{2\pi (t_i-t_{\mathrm{ref},k})}{P_{\mathrm{EQ}}} \sin\Lambda_{k,i} [2\kappa_2\cos^2\Phi_{\mathrm{ref},k}+\kappa_4\sin^2(2\Phi_{\mathrm{ref},k})] \Big) \nonumber \\
\qquad& - \csc\beta_{k,i} \cos I_* \cos\Phi_{\mathrm{ref},k}, \\
\frac{\partial\beta_{k,i}}{\partial \Lambda_{\mathrm{ref},k}} &= \frac{ \sin I_*\cos\Phi_{\mathrm{ref},k}\sin\Lambda_{k,i} }{ \sin\beta_{k,i} }.
\end{align}

Aside from the above, the remainder of the partial derivatives satisfy:

\begin{align}
\frac{\partial\beta_{k,i}}{\partial \Theta_j} &= 0 \text{ if } \Theta_j\neq I_*,P_{\mathrm{EQ}},\kappa_2,\kappa_4,\Phi_{\mathrm{ref},k},\Lambda_{\mathrm{ref},k}.
\end{align}

\subsection{Partial Derivatives of $\alpha_{k,i}$}

The $k^{\mathrm{th}}$ starspot evolves via Equation~\ref{eqn:evolution}. The
 partial derivatives are found to be:

\begin{align}
\frac{\partial\alpha_{k,i}}{\partial\alpha_{\mathrm{max},k}} &= \frac{\alpha_{k,i}}{\alpha_{\mathrm{max},k}},  \\
% d(alpha)/d(tmax)
\frac{\partial\alpha_{k,i}}{\partial t_{\mathrm{max},k}} &= -\frac{\alpha_{\mathrm{max},k}}{\mathcal{I}_k} (\mathsf{H}(\Delta t_1) - \mathsf{H}(\Delta t_2)) \nonumber\\
\qquad& + \frac{\alpha_{\mathrm{max},k}}{\mathcal{E}_k} (\mathsf{H}(\Delta t_3) - \mathsf{H}(\Delta t_4)),\\
% d(alpha)/d(L)
\frac{\partial\alpha_{k,i}}{\partial L_k} &= \frac{\alpha_{\mathrm{max},k}}{2\mathcal{I}_k} (\mathsf{H}(\Delta t_1) - \mathsf{H}(\Delta t_2)) \nonumber \\
\qquad& + \frac{\alpha_{\mathrm{max},k}}{2 \mathcal{E}_k} (\mathsf{H}(\Delta t_3) - \mathsf{H}(\Delta t_4)) , \\
% d(alpha)/d(I)
\frac{\partial\alpha_{k,i}}{\partial \mathcal{I}_k} &=-\Big(\frac{\alpha_{\mathrm{max},k} (\Delta t_1 + \Delta t_2)}{2 \mathcal{I}^2}\Big) ( \mathsf{H}(\Delta t_1) - \mathsf{H}(\Delta t_2) ),\\
% d(alpha)/d(E)
\frac{\partial\alpha_{k,i}}{\partial \mathcal{E}_k} &= \Big(\frac{\alpha_{\mathrm{max},k} (\Delta t_3 + \Delta t_4)}{2 \mathcal{E}^2}\Big) ( \mathsf{H}(\Delta t_3) - \mathsf{H}(\Delta t_4) ).\\
\end{align}

Aside from the above, the remainder of the partial derivatives satisfy:

\begin{align}
\frac{\partial\alpha_k}{\partial \Theta_j} &= 0 \text{ if } \Theta_j\neq \alpha_{\mathrm{max},k},t_{\mathrm{max},k},L_k,s_k.
\end{align}

Finally, it is necessary to define a reference longitude, 
$\Lambda_{\mathrm{ref},k}$. A convenient choice is to define it as the 
longitude at the instant $t=t_{\mathrm{max},k}$, which is the default
assumption of \macula.

\section{Manipulating Partial Derivatives}
\label{appendixE}

\subsection{Alternative Limb Darkening Laws}

In this model, we have adopted the four-coefficient limb darkening proposed
by \citet{claret:2000}. Several, but by no means all, alternative limb darkening 
laws can be adopted by re-parametrising the four-coefficient model presented in 
this work. In this subsection, we discuss four examples: i) the quadratic law
\citep{kopal:1950} ii) the linear law \citep{russell:1912} iii) the 
three-coefficient law \citep{sing:2009} iv) the square-root law 
\citep{diaz:1992}. In each case, we show how one may use the results from 
\macula\ to obtain the partial derivatives for regression purposes.
In what follows, we assume that the spot and the star have distinct limb
darkening coefficients.

\subsubsection{Quadratic Law}

The quadratic law, first proposed by \citet{kopal:1950}, is perhaps the most
commonly adopted model in the exoplanet literature. Recall from 
Equation~\ref{eqn:LD4} that the four-coefficient limb darkening law
is described by

\begin{align}
I_*(r) &= 1 - \sum_{n=1}^4 c_n (1-\mu^{n/2}). \nonumber
\end{align}

In contrast, the quadratic law is described by

\begin{align}
I_*(r) &= 1 - u_1 (1-\mu) - u_2 (1-\mu)^2.
\end{align}

By comparing the coefficients relative the four-coefficient model, one may show
that the quadratic law may be reproduced by setting:

\begin{align}
c_1 &= 0, \nonumber \\
c_2 &= u_1 + 2 u_2, \nonumber \\
c_3 &= 0, \nonumber \\
c_4 &= -u_2.
\end{align}

The quadratic model is popular in the exoplanet community when one wishes to
fit for the limb darkening parameters. The reason for this is two-fold. Firstly,
photometric data are rarely precise enough to regress a unique solution for all
four coefficients of the non-linear limb darkening law and so using the
quadratic model reduces the number of free parameters by two yet preserves 
curvature in the intensity profile of the star. Secondly, the two quadratic
coefficients, $u_1$ and $u_2$, have well-described priors by imposing that
the intensity profile is monotonic and everywhere positive. \citet{carter:2009}
show that these conditions impose

\begin{align}
u_1>0, \nonumber \\
0<u_1+u_2<1.
\end{align}

\citet{hek:2012} point out that a sensible upper-bound on $u_1$ may be imposed 
from inspection of typical coefficient tables presented in works such as 
\citet{claret:2000}. A typical choice is $u_1<2$ for Sun-like stars. With this 
upper-bound one may re-define $u_{1+2} = u_1 + u_2$ and regress the parameters 
$\{u_1,u_{1+2}\}$ subject to the uniform priors:

\begin{align}
0<u_1<2,\nonumber \\
0<u_{1+2}<1. 
\end{align}

The four-coefficient model can be set to these parameters using:

\begin{align}
c_1 &= 0, \nonumber \\
c_2 &= 2u_{1+2} - u_1, \nonumber \\
c_3 &= 0, \nonumber \\
c_4 &= u_1-u_{1+2}.
\end{align}

In the previous section, we have derived 
$\partial F_{\mathrm{mod}}/\partial c_n$ for $n=1,2,3,4$. We now require
$\partial F_{\mathrm{mod}}/\partial u_1$ and 
$\partial F_{\mathrm{mod}}/\partial u_{1+2}$. Firstly, one may show:

\begin{align}
u_1 &= c_2 + 2 c_4, \nonumber \\
u_{1+2} &= c_2 + c_4.
\end{align}

It is therefore trivial to write:

\begin{align}
\frac{\partial F_{\mathrm{mod}} }{\partial u_1} &= \frac{\partial F_{\mathrm{mod}} }{\partial c_2} + 2\frac{\partial F_{\mathrm{mod}} }{\partial c_4},\nonumber \\
\frac{\partial F_{\mathrm{mod}} }{\partial u_{1+2}} &= \frac{\partial F_{\mathrm{mod}} }{\partial c_2} + \frac{\partial F_{\mathrm{mod}} }{\partial c_4}.
\end{align}

For the starspot's limb darkening, the same argument may be made to show:

\begin{align}
\frac{\partial F_{\mathrm{mod}} }{\partial v_1} &= \frac{\partial F_{\mathrm{mod}} }{\partial d_2} + 2\frac{\partial F_{\mathrm{mod}} }{\partial d_4},\nonumber \\
\frac{\partial F_{\mathrm{mod}} }{\partial v_{1+2}} &= \frac{\partial F_{\mathrm{mod}} }{\partial d_2} + \frac{\partial F_{\mathrm{mod}} }{\partial d_4},
\end{align}

where we define $d_1=d_3=0$ and

\begin{align}
v_1 &= d_2 + 2 d_4, \nonumber \\
v_{1+2} &= d_2 + d_4.
\end{align}

\subsubsection{Linear Law}

The linear limb darkening law, which can be traced back to \citet{russell:1912}, 
is expressed as:

\begin{align}
I_*(r) &= 1 - u_L (1-\mu).
\end{align}

It is therefore trivial to see that this is identical to the quadratic law
where $u_1=u_L$ and $u_2=0$. Relative to the four coefficient model, we have
$\{c_1,c_2,c_3,c_4\}=\{0,u_L,0,0\}$. In such a model then, one may simply use:

\begin{align}
\frac{\partial F_{\mathrm{mod}} }{\partial u_L} &= \frac{\partial F_{\mathrm{mod}} }{\partial c_2}.
\end{align}

As before, this can be easily applied to the starspot's limb darkening too via

\begin{align}
\frac{\partial F_{\mathrm{mod}} }{\partial u_L} &= \frac{\partial F_{\mathrm{mod}} }{\partial c_2}.
\end{align}

where we define $v_L = d_2$ and $d_1=d_3=d_4=0$.

\subsubsection{Three-Coefficient Law}

The three-coefficient law, proposed by \citet{sing:2009}, is described by:

\begin{align}
I_*(r) &= 1 - c_2 (1-\mu) - c_3 (1-\mu^{3/2}) - c_4 (1-\mu^2),
\end{align}

which is precisely the same as the four-coefficient law in the limit 
$c_1\rightarrow0$. For this reason, the partial derivatives are unchanged from
before and one may ignore the partial derivative with respect to $c_1$ \& $d_1$.

\subsubsection{Square-Root Law}

The three-coefficient law, proposed by \citet{diaz:1992}, is described by:

\begin{align}
I_*(r) &= 1 - c_1 (1-\mu^{1/2}) - c_2 (1-\mu),
\end{align}

which is again identical to the four-coefficient law in the limit of
$c_3\rightarrow0$ and $c_4\rightarrow0$. The same applies, of course, for
the spot's limb darkening profile and thus the partial derivatives are
trivially obtained.

\subsection{Allowing for Global Parameters}
\subsubsection{Principle}

Practically speaking, it is common to consider a subset of the $\mathbf{\Theta}$
parameters to be equal to some global term. For example, rather than regressing
for $N_S$ unique spot contrast fluxes, $f_{\mathrm{spot},k}$, one may wish to
enforce the condition that all spots have the same flux contrast (i.e. 
temperature). The advantage of implementing such a condition is that one 
reduces the number of free parameters in the regression by $(N_S-1)$. 

In such a case, one requires the partial derivatives of the model flux with
respect to this new global parameter, rather than the individual terms. Since
the individual partial derivatives have already been calculated and are
directly returned by the \macula\ code, it is highly advantageous if we
can phrase the partial derivatives of this new global parameter as a function
of the individual terms. In this subsection, we provide a method for 
accomplishing this.

The model flux is a function of parameters $\mathbf{\Theta}$ i.e. 
$F_{\mathrm{mod}}(\mathbf{\Theta})$. For $L$ model parameters, one may write out 
the differential as:

\begin{align}
\delta F_{\mathrm{mod}} &= \sum_{l=1}^L \frac{\partial F_{\mathrm{mod}}}{\partial \Theta_l} \delta \Theta_l, \nonumber \\
\qquad &= \frac{\partial F_{\mathrm{mod}}}{\partial \Theta_1} \delta \Theta_1 + \frac{\partial F_{\mathrm{mod}}}{\partial \Theta_2} \delta \Theta_2 +... \frac{\partial F_{\mathrm{mod}}}{\partial \Theta_L} \delta \Theta_{L}.
\label{eqn:G1}
\end{align}

Now consider that a subset of the $\mathbf{\Theta}$ parameter vector is set to 
be equal to some global parameter, $G$. Let this subset run from parameter 1 to
$L'$ i.e. $\Theta_1=\Theta_2=...=\Theta_L=G$ where $G$ is some global parameter.
The differential now becomes:

\begin{align}
\delta F_{\mathrm{mod}} &= \frac{\partial F_{\mathrm{mod}}}{\partial G} \delta G + \sum_{l=L'}^L \frac{\partial F_{\mathrm{mod}}}{\partial \Theta_l} \delta \Theta_l.
\label{eqn:G2}
\end{align}

And so by equivalence of Equations~\ref{eqn:G1}\&\ref{eqn:G2}, one can see that:

\begin{align}
\frac{\partial F_{\mathrm{mod}}}{\partial G} \delta G &= \sum_{l=1}^{L'} \frac{\partial F_{\mathrm{mod}}}{\partial \Theta_l} \delta \Theta_l,
\end{align}

And finally this yields:

\begin{align}
\frac{\partial F_{\mathrm{mod}}}{\partial G} &= \lim_{\Theta_1,\Theta_2,...,\Theta_{L'}\rightarrow G} \Bigg[ \sum_{l=1}^{L'} \frac{\partial F_{\mathrm{mod}}}{\partial \Theta_l}\Bigg].
\label{eqn:globallaw}
\end{align}

\subsubsection{Common Examples}

As we cited earlier, a common application of Equation~\ref{eqn:globallaw} is
to $N_S$ individual spot contrast values, $f_{\mathrm{spot},k}$ to be equal to 
some global spot contrast term, $g_{\mathrm{spot}}$. The partial derivative
of the model flux with respect to this new global flux contrast new may
be expressed, using Equation~\ref{eqn:globallaw}, as:

\begin{align}
\frac{\partial F_{\mathrm{mod}}}{\partial g_{\mathrm{spot}}} &= \lim_{f_{\mathrm{spot,k}}\rightarrow g_{\mathrm{spot}}} \Bigg[ \sum_{k=1}^{N_S} \frac{\partial F_{\mathrm{mod}}}{\partial f_{\mathrm{spot},k}} \Bigg].
\end{align}

Another example is to enforce a global blending factor, $C$, rather than 
individual values, $B_m$:

\begin{align}
\frac{\partial F_{\mathrm{mod}}}{\partial C} &= \lim_{B_m\rightarrow C} \Bigg[ \sum_{m=1}^{M} \frac{\partial F_{\mathrm{mod}}}{\partial B_m} \Bigg],
\end{align}

Finally, one may wish to set the spot's limb darkening parameters to be
equal to the star's limb darkening parameters i.e. 
$\mathbf{c}=\mathbf{d}=\mathbf{b}$ where $\mathbf{b}$ is the global limb
darkening parameters in vector-form.

\begin{align}
\frac{\partial F_{\mathrm{mod}}}{\partial b_n} &= \lim_{c_n\rightarrow b_n} \lim_{d_n\rightarrow b_n} \Bigg[\frac{\partial F_{\mathrm{mod}}}{\partial c_n} + \frac{\partial F_{\mathrm{mod}}}{\partial d_n}\Bigg].
\end{align} 

\section{Partial Derivatives with Respect to Time}
\label{appendixF}

Recall from Equation~\ref{eqn:Fmodfull} and Equation~\ref{eqn:Fmodtilde}
that the $m^{\mathrm{th}}$ component of the model flux is given by

\begin{align}
\tilde{F}_{\mathrm{mod},m,i} &=  \Bigg( \frac{U_m \Pi_{m,i} \mathbb{F}_i}{B_m \mathbb{F}_0} + \frac{(B_m-1) U_m \Pi_{m,i}}{B_m} \Bigg).
\end{align}

Taking the partial derivative of the above with respect to time yields

\begin{align}
\frac{ \partial \tilde{F}_{\mathrm{mod},m,i} }{\partial t_i} &= \Bigg[ \frac{U_m \Pi_{m,i}}{B_m \mathbb{F}_0} \frac{\partial \mathbb{F}_i}{\partial t_i} \nonumber \\
\qquad& + \Bigg( \frac{U_m \mathbb{F}_i}{B_m \mathbb{F}_0} + \frac{U_m (B_m-1)}{B_m}\Bigg) \frac{\partial \Pi_{m,i}}{\partial t_i} \Bigg].
\end{align}

The partial derivatives of the box-car function, $\Pi_{m,i}$, is simply two 
Dirac Delta functions and thus may be neglected in what follows i.e.

\begin{align}
\frac{ \partial \tilde{F}_{\mathrm{mod},m,i} }{\partial t_i} &= \frac{U_m \Pi_{m,i}}{B_m \mathbb{F}_0} \frac{\partial \mathbb{F}_i}{\partial t_i}.
\end{align}

Since $\mathbb{F}_i = \mathbb{F}_0 - Q_i$ and $\mathbb{F}_0$ has no time
dependency, then $\partial \mathbb{F}_i/\partial t_i = 
-\partial Q_i/\partial t_i$ giving

\begin{align}
\frac{\partial \mathbb{F}_i}{\partial t_i} &= - \sum_{k=1}^{N_S} \frac{\partial q_{k,i}}{\partial t_i}.
\end{align}

The $q_{k,i}$ partial derivative may be expressed via

\begin{align}
q_{k,i} &= \frac{A_{k,i}}{\pi} \sum_{n=0}^4 w_{n,k,i}, \nonumber \\
\frac{\partial q_{k,i}}{\partial t_i} &= \frac{\mathbb{R}[\mathcal{A}_{k,i}]}{\pi} \sum_{n=0}^4 \frac{\partial w_{n,k,i}}{\partial t_i}
+ \frac{1}{\pi} \mathbb{R}\Big[\frac{\partial \mathcal{A}_{k,i}}{\partial t_i}\Big] \sum_{n=0}^4 w_{n,k,i}.
\end{align}

For the partial derivatives of $\mathcal{A}_{k,i}$, we can use the same chain 
rule trick as was used earlier:

\begin{align}
\Bigg(\frac{\partial \mathcal{A}_{k,i}}{\partial t_i}\Bigg)_{\Theta_l\,\forall\,l} &= \Bigg(\frac{\partial \mathcal{A}_{k,i}}{\partial \alpha_{k,i}}\Bigg)_{\beta_{k,i}} \Bigg(\frac{\partial \alpha_{k,i}}{\partial t_i}\Bigg)_{\Theta_l\,\forall\,l} \nonumber \\
\qquad& + \Bigg(\frac{\partial \mathcal{A}_{k,i}}{\partial \beta_{k,i}}\Bigg)_{\alpha_{k,i}} \Bigg(\frac{\partial \beta_{k,i}}{\partial t_i}\Bigg)_{\Theta_l\,\forall\,l},
\end{align}

where the partial derivatives of $\mathcal{A}_{k,i}$ with respect to 
$\alpha_{k,i}$ and $\beta_{k,i}$ are given in Equations~\ref{eqn:Aderivs}. Let 
us leave aside the issue of the partial derivatives of $\alpha_{k,i}$ and 
$\beta_{k,i}$ for the moment and focus on those of $w_{n,k,i}$:

\begin{align}
w_{n,k,i} &= \frac{4 (c_n-d_n f_{\mathrm{spot},k})}{n+4} \Upsilon_{n,k,i}, \nonumber \\
\frac{\partial w_{n,k,i}}{\partial t_i} &= \frac{4 (c_n-d_n f_{\mathrm{spot},k})}{n+4} \frac{\partial \Upsilon_{n,k,i}}{\partial t_i}.
\end{align}

Partial derivatives of $\Upsilon_{n,k,i}$ with respect to $\Theta_l$ have already 
been calculated earlier in Equation~\ref{eqn:Upsilonderivs}, in terms of the 
derivatives of $\zeta_{-,k,i}$ and $\zeta_{+,k,i}$. This result is easily 
modified to be with respect to time:

\begin{align}
\frac{\partial\Upsilon_{n,k,i}}{\partial t_i} &= \Bigg[\frac{1}{\zeta_{-,k,i}^2-\zeta_{+,k,i}^2+\delta_{\zeta_{+,k,i},\zeta{-,k,i}}}\Bigg] \nonumber\\
\qquad& \times \Bigg[ \Bigg(\frac{n+4}{2}\Bigg) \Bigg( \zeta_{-,k,i}^{\frac{n+2}{2}} \frac{\partial\zeta_{-,k,i}}{\partial t_i} - \zeta_{+,k,i}^{\frac{n+2}{2}} \frac{\partial\zeta_{+,k,i}}{\partial t_i} \Bigg) \nonumber \\
\qquad& - 2\Upsilon_{n,k,i} \Bigg( \frac{\partial\zeta_{-,k,i}}{\partial t_i} - \frac{\partial\zeta_{+,k,i}}{\partial t_i} \Bigg) \Bigg].
\end{align} 

Those terms have also had their partial derivatives computed
wth respect to $\alpha_{k,i}$ and $\beta_{k,i}$, which lead to the chain rule 
relation:

\begin{align}
\frac{\partial \zeta_{-,k,i}}{\partial t_i} &= \Bigg(\frac{\partial \zeta_{-,k,i}}{\partial \alpha_{k,i}}\Bigg)_{\beta_{k,i}} \Bigg[ \frac{\partial \alpha_{k,i}}{\partial t_i} - \frac{\partial \beta_{k,i}}{\partial t_i} \Bigg],\\
\frac{\partial \zeta_{+,k,i}}{\partial t_i} &= \Bigg(\frac{\partial \zeta_{+,k,i}}{\partial \alpha_{k,i}}\Bigg)_{\beta_{k,i}} \Bigg[ \frac{\partial \alpha_{k,i}}{\partial t_i} + \frac{\partial \beta_{k,i}}{\partial t_i} \Bigg].
\end{align}

The partial derivatives of $\zeta_{-,k,i}$ and $\zeta_{+,k,i}$ with respect to
$\alpha_{k,i}$ have already been calculated in Equation~\ref{eqn:zetanegderivs}
and Equation~\ref{eqn:zetaposderivs} respectively. The outstanding task is
now to compute the partial derivatives of $\alpha_{k,i}$ and $\beta_{k,i}$ with
respect to time. It is easy to show that the $\alpha_{k,i}$ partial derivative 
is given by:

\begin{align}
\frac{\partial \alpha_{k,i}}{\partial t_i} &= \frac{\alpha_{\mathrm{max},k}}{\mathcal{I}_k} (\mathsf{H}(\Delta t_1) - \mathsf{H}(\Delta t_2) ) \nonumber \\
\qquad& - \frac{\alpha_{\mathrm{max},k}}{\mathcal{E}_k} (\mathsf{H}(\Delta t_3) - \mathsf{H}(\Delta t_4) ),
\end{align}

and that of $\beta_{k,i}$ by:

\begin{align}
\frac{\partial \beta_{k,i}}{\partial t_i} &= \frac{2 \pi}{P_{*,k}} \frac{ \sin I_* \cos \Phi_{\mathrm{ref},k} \sin \Lambda_{k,i} }{ \sqrt{ 1 - (\sin I_*\cos\Phi_{\mathrm{ref},k}\cos\Lambda_{k,i} + \cos I_*\sin\Phi_{\mathrm{ref},k})^2 } }.
\end{align}

\bsp

\label{lastpage}


\begin{thebibliography}{99}
\bibitem[\protect\citeauthoryear{Aigrain et al.}{2012}]{aigrain:2012} 
Aigrain, S., Pont, F. \& Zucker, S. 2012, MNRAS, 419, 3147
\bibitem[\protect\citeauthoryear{Basri et al.}{2011}]{basri:2011} 
Basri, G. et al. 2011, AJ, 141, 20
\bibitem[\protect\citeauthoryear{Barnes}{2009}]{barnes:2009} 
Barnes, S. 2009, IAU Symposium 258, 345
\bibitem[\protect\citeauthoryear{Batalha et al.}{2012}]{batalha:2012} 
Batalha, N. M. et al. 2012, ApJS, submitted (astro-ph:1202.5852)
\bibitem[\protect\citeauthoryear{Berdyugina}{2005}]{berdyugina:2005} 
Berdyugina S. V. 2005, LRSP, 2, 8
\bibitem[\protect\citeauthoryear{Boisse et al.}{2012}]{boisse:2012} 
Boisse, I., Bonfils, X. \& Santos, N. C. 2012, A\&A, accepted 
(astro-ph:1206.5493)
\bibitem[\protect\citeauthoryear{Budding}{1977}]{budding:1977} 
Budding, E. 1977, Ap\&SS, 48, 207
\bibitem[\protect\citeauthoryear{Cantiello \& Braithwaite}{2011}]{cantiello:2011}
Cantiello, M. \& Braithwaite, J. 2011, A\&A, 534, 140
\bibitem[\protect\citeauthoryear{Carter et al.}{2009}]{carter:2009} 
Carter, J. A., Winn, J. N., Gilliland, R. \& Holman, M. J. 2009, ApJ, 696, 241
\bibitem[\protect\citeauthoryear{Carter \& Winn}{2010}]{carter:2010} 
Carter, J. A. \& Winn, J. N. 2010, ApJ, 716, 850
\bibitem[\protect\citeauthoryear{Charbonneau et al.}{2000}]{charbonneau:2000} 
Charbonneau, D., Brown, T. M., Latham, D. W. \& Mayor, M. 2000, ApJ, 529, 45
\bibitem[\protect\citeauthoryear{Claret}{2000}]{claret:2000} 
Claret, A. 2000, A\&A, 363, 1081
\bibitem[\protect\citeauthoryear{Claret}{2011}]{claret:2011} 
Claret, A. \& Bloemen, S. 2011, A\&A, 529, 75
\bibitem[\protect\citeauthoryear{Collier-Cameron et al.}{1994}]{cameron:1994} 
Collier-Cameron, A. \& Unruh, Y. C. 1994, MNRAS, 269, 814
\bibitem[\protect\citeauthoryear{Collier-Cameron et al.}{1997}]{cameron:1997} 
Collier-Cameron, A. 1997, MNRAS, 287, 556
\bibitem[\protect\citeauthoryear{Croll et al.}{2006a}]{starspotz:2006} 
Croll, B. et al. 2006a, in Bulletin of the American Astronomical Society,
Vol. 38, Bulletin of the American Astronomical Society, 1217
\bibitem[\protect\citeauthoryear{Croll et al.}{2006b}]{croll:2006} 
Croll, B. et al. 2006b, ApJ, 648, 607
\bibitem[\protect\citeauthoryear{Czela et al.}{2009}]{czesla:2009} 
Czesla, S., Huber, K. F., Wolter, U., Schroter, S. \& Schmitt, J. H. M. M.
2009, A\&A, 505, 1277
\bibitem[\protect\citeauthoryear{Degroote et al.}{2011}]{degroote:2011} 
Degroote, P. et al. 2011, A\&A, 536, 82
\bibitem[\protect\citeauthoryear{D\'esert et al.}{2009}]{desert:2009} 
D\'esert, J.-M., Sing, D., Vidal-Madjar, A., H\'ebrard, G., Ehrenreich, D., 
Lecavelier Des Etangs, A., Parmentier, V., Ferlet, R. \& Henry, G. W. 2009,
A\&A, 526, 12
\bibitem[\protect\citeauthoryear{D\'esert et al.}{2011}]{desert:2011} 
D\'esert, J.-M. et al. 2011, ApJS, 197, 14
\bibitem[\protect\citeauthoryear{D\'iaz-Cordov\'es \& Gim\'enez}{1992}]{diaz:1992} 
D\'iaz-Cordov\'es, J. \& Gim\'enez, A. 1992, A\&A, 259, 227
\bibitem[\protect\citeauthoryear{Dorren}{1987}]{dorren:1987} Dorren, J. D., 
1987, ApJ, 320, 756
\bibitem[\protect\citeauthoryear{Feroz et al.}{2008}]{feroz:2008} 
Feroz, F. \& Hobson, M. P. 2008, MNRAS, 384, 449
\bibitem[\protect\citeauthoryear{Feroz et al.}{2009}]{feroz:2009} 
Feroz, F., Hobson, M. P. \& Bridges, M. 2009, MNRAS, 398, 1601
\bibitem[\protect\citeauthoryear{Henry et al.}{2000}]{henry:2000} 
Henry, G. W., Marcy, G. W., Butler, P. R. \& Vogt, S. S. 2000, ApJ, 529, 41
\bibitem[\protect\citeauthoryear{Hirano et al.}{2012}]{hirano:2012}
Hirano, T., Sanchis-Ojeda, R., Takeda, Y., Narita, N., Winn, J. N., Taruya, A. 
\& Suto, Y. 2012, ApJ, accepted (astro-ph:1205.3233)
\bibitem[\protect\citeauthoryear{Kipping \& Tinetti}{2010}]{kippingtinetti:2010} 
Kipping, D. M. \& Tinetti, G. 2010, MNRAS, 407, 2589
\bibitem[\protect\citeauthoryear{Kipping}{2011}]{luna:2011} Kipping, D. M.
2011, MNRAS, 416, 689
\bibitem[\protect\citeauthoryear{Kipping et al.}{2012}]{hek:2012} Kipping, 
D. M., Bakos, G. \'A., Buchhave, L., Nesvorn\'y, D. \& Schmitt, A.
2012, ApJ, 750, 115
\bibitem[\protect\citeauthoryear{Kopal}{1950}]{kopal:1950} 
Kopal, Z. Harvard Coll. Obser. Circ., 454, 1
\bibitem[\protect\citeauthoryear{Kron}{1947}]{kron:1947} Kron, G, E. 1947,
PASP, 59, 261
\bibitem[\protect\citeauthoryear{Mandel \& Agol}{2002}]{mandel:2002} Mandel, K.
\& Agol, E. 2002, ApJ, 580, 171
\bibitem[\protect\citeauthoryear{Mart\'inez et al.}{1993}]{martinez:1993}
Mart\'inez, P. V, Moreno, I. F \& Vázquez, M. 1993, Astron. Astrophys, 274, 521
\bibitem[\protect\citeauthoryear{McLaughlin}{1924}]{mclaughlin:1924} 
McLaughlin, D. B. 1924, ApJ, 60, 22
\bibitem[\protect\citeauthoryear{Mukherjee et al.}{2006}]{mukherjee:2006}
Mukherjee P., Parkinson D. \& Liddle A. R., 2006, ApJ, 638, L51
\bibitem[\protect\citeauthoryear{Mullan}{1974}]{mullan:1974} Mullan, D. J.
1974, ApJ, 192, 149
\bibitem[\protect\citeauthoryear{Nutzman et al.}{2011}]{nutzman:2011} 
Nutzman, P. A., Fabrycky, D. C. \& Fortney, J. J. 2011, ApJ, 740, 10
\bibitem[\protect\citeauthoryear{P\'al}{2012}]{pal:2012} P\'al, A. 2012,
MNRAS, 420, 1630
\bibitem[\protect\citeauthoryear{Petrovay \& Van Driel-Gesztelyi}{1997}]
{petrovay:1997} Petrovay K. \& Van Driel-Gesztelyi, L. 1997, Sol. Phys., 176, 
249
\bibitem[\protect\citeauthoryear{Rabus et al.}{2009}]{rabus:2009}
Rabus, M. et al. 2009, A\&A, 494, 391
\bibitem[\protect\citeauthoryear{Rossiter}{1924}]{rossiter:1924} 
Rossiter, R. A. 1924, ApJ, 60, 15
\bibitem[\protect\citeauthoryear{Rucinski et al.}{2004}]{rucinski:2004} 
Rucinski, S. M. et al. 2004, PASP, 116, 1093
\bibitem[\protect\citeauthoryear{R\"{u}diger \& Kitchatinov}{2000}]{rudiger:2000} 
R\"{u}diger, G. \& Kitchatinov, L. L., 2000, Astronomische Nachrichten, 321, 75
\bibitem[\protect\citeauthoryear{Russell}{1912}]{russell:1912} 
Russell, H. N. 1912, ApJ, 36, 54
\bibitem[\protect\citeauthoryear{Sanchis-Ojeda et al.}{2011}]{sanchis:2011} 
Sanchis-Ojeda, R., Winn, J. N., Holman, M. J., Carter, J. A., Osip, D. J.
\& Fuentes, C. I. 2011, ApJ, 733, 127
\bibitem[\protect\citeauthoryear{Sanchis-Ojeda et al.}{2012}]{sanchis:2012} 
Sanchis-Ojeda, R. et al. 2012, Nature, 487, 449
\bibitem[\protect\citeauthoryear{Skilling}{2004}]{skilling:2004} 
Skilling, J. 2004, in Fischer R., Preuss R., Toussaint U. V., eds,
American Institute of Physics Conference Series Nested Sampling. pp 395–405
\bibitem[\protect\citeauthoryear{Sing et al.}{2009}]{sing:2009} 
Sing, D. K., D\'esert, J.-M., Lecavelier Des Etangs, A., Ballester, G. E., 
Vidal-Madjar, A., Parmentier, V., Hebrard, G. \& Henry, G. W. 2009,
A\&A, 505, 891
\bibitem[\protect\citeauthoryear{Slawson et al.}{2011}]{slawson:2011} 
Slawson, R. W. et al. 2011, ApJ, 142, 160
\bibitem[\protect\citeauthoryear{Skumanich}{1972}]{skumanich:1972} 
Skumanich, A. 1972, ApJ, 171, 565
\bibitem[\protect\citeauthoryear{Stix}{2002}]{stix:2002} 
Stix, M. 2002, Astronomische Nachrichten, 323, 178
\bibitem[\protect\citeauthoryear{Strassmeier}{1999}]{strassmeier:1999} 
Strassmeier, K. G. 1999, A\&A, 347, 225
\bibitem[\protect\citeauthoryear{Tuominen et al.}{2002}]{tuominen:2002} 
Tuominen, I., Berdyugina, S. V. \& Korpi, M. J. 2002, AN, 323, 367
\bibitem[\protect\citeauthoryear{Walker et al.}{2007}]{walker:2007} 
Walker, G. A. H. et al. 2007, ApJ, 659, 1611
\bibitem[\protect\citeauthoryear{van Leeuwen et al.}{1997}]{leeuwen:1997} van 
Leeuwen, F., Evans, D. W., Grenon, M., Grossmann, V., Mignard, F. \& Perryman, 
M. A. C. 1997, A\&A, 323, 61
\bibitem[\protect\citeauthoryear{Vogt}{1975}]{vogt:1975} Vogt, S. S.
1975, ApJ, 199, 418
\bibitem[\protect\citeauthoryear{Vogt \& Penrod}{1983}]{vogt:1983} Vogt, S. S.
\& Penrod, G. D. 1983, PASP, 95, 565
\end{thebibliography}
\end{document}